\documentclass[]{aa}
\let\bibfont\relax
\usepackage{natbib}
\usepackage{latexsym}
\usepackage{graphicx}
\usepackage{amssymb,dcolumn, rotating, savefnmark, lscape, multirow,longtable}
\usepackage{txfonts}
\bibpunct{(}{)}{;}{a}{}{,} 
\renewcommand{\d}{\textrm{d}}

\newcommand{\ap}{$\sim$}
\newcommand{\Lx}{$L_\mathrm{X}$ }

\begin{document}
\title{Dust amorphization in protoplanetary disks}

\author{A.~M. Glauser \inst{1,2} \and M. G\"udel \inst{1} \and D.~M. Watson \inst{3} \and T. Henning \inst{4}  \and A.~A. Schegerer \inst{5} \and S. Wolf \inst{6} \and M. Audard \inst{7,8} \and C. Baldovin-Saavedra \inst{7,8}}

\offprints{A. M. Glauser, \email{glauser@astro.phys.ethz.ch}}

\institute{Institute of Astronomy, ETH Zurich, 8093 Zurich,
Switzerland \and UK Astronomy Technology Centre, Blackford Hill,
Edinburgh EH9 3HJ, United Kingdom \and University of Rochester,
Department of Physics and Astronomy, Rochester, NY, USA  \and Max Planck Institute for Astronomy,
K\"onigstuhl 17, 69117 Heidelberg, Germany \and
Helmholtz Zentrum M\"unchen, German Research Center for
Environmental Health, Ingolst\"adter Landstra{\ss}e 1, 85758
Neuherberg, Germany \and University of Kiel,
Institute for Theoretical Physics and Astrophysics, Leibnizstr. 15,
24098 Kiel, Germany \and Observatoire de Gen\`eve, University of
Geneva, Ch. de Maillettes 51, 1290 Sauverny, Switzerland \and ISDC
Data Center for Astrophysics, University of Geneva, Ch. d'Ecogia 16,
1290 Versoix, Switzerland}

\date{Received ... / Accepted ...}

\abstract{}{High-energy irradiation of the circumstellar material
might impact the structure and the composition of a protoplanetary
disk and hence the process of planet formation. In this paper, we
present a study on the possible influence of the stellar
irradiation, indicated by X-ray emission, on the crystalline
structure of the circumstellar dust.}{The dust crystallinity is
measured for 42 class II T~Tauri stars in the Taurus star-forming
region using a decomposition fit of the 10~$\mu$m silicate feature,
measured with the \textsc{Spitzer} IRS instrument. Since the sample includes objects with disks of various evolutionary stages, we further confine the target selection, using the age of the objects as a selection parameter.}{We correlate the
X-ray luminosity and the X-ray hardness of the central object with
the crystalline mass fraction of the circumstellar dust and find a significant
anti-correlation for 20 objects within an age range of approx. 1 to 4.5~Myr. We
postulate that X-rays represent the stellar activity and
consequently the energetic ions of the stellar winds which interact
with the circumstellar disk. We show that the fluxes around 1~AU and
ion energies of the present solar wind are sufficient to amorphize
the upper layer of dust grains very efficiently, leading to an
observable reduction of the crystalline mass fraction of the
circumstellar, sub-micron sized dust. This effect could also erase other relations
between crystallinity and disk/star parameters such as age or
spectral type.}{}


\maketitle

\section{Introduction}
The evolution of the dust in a protoplanetary disk is one of the key
subjects in the overall research on mechanisms of planet formation.
As we now know, dust in young circumstellar disks differs
significantly from the dust in the interstellar medium (ISM). There
is evidence for grain growth from the typical ISM and sedimentation
in the vertical direction (or dust settling) of a disk (see various
references such as \citealt{Rodmann:2006}, \citealt{Sicilia:2007},
\citealt{Furlan:2006}). While the dust grains in the ISM and in
molecular clouds are amorphous, protoplanetary disks contain an
increased content of crystalline silicates \citep{Bouwman:2008}.
This transition of the dust grain structure during the star-forming
process is poorly understood and could be important in the later
scenario of planet formation.

However, as many authors have pointed out (e.g.
\citealt{Watson:2009}, \citealt{Sicilia:2007}, \citealt{Boekel:2005},
\citealt{Schegerer:2006}), no definitive connection have been found so far between the
properties of the disk or of the central object and the crystalline
mass fraction of the dust disk. From the point of view of standard
dust processing scenarios for protoplanetary disks, this conclusion
is surprising. We expect the crystallization of the dust grains to
occur due to thermal annealing or evaporation and recondensation processes either close
to the star or within accretion shocks \citep{Henning:2008}. Radial
mixing may transport the crystalline grains to more distant regions
\citep[e.g.,][]{Gail:2004}. Therefore, we expect an evolutionary
trend for the crystalline mass fraction and/or correlations with
stellar parameters such as the bolometric luminosity, the
photospheric temperature, the accretion rate or the disk/star mass
ratio. The fact that no such relation has been found raises the
question on alternative mechanisms, controlling the process of
crystallization. \citet{Kessler:2006} and \citet{Watson:2009} suggested that the
crystallizing process might be dominated by the impact of X-ray
irradiation which destroys the crystalline structure of the dust
grains.

Young stars are very strong sources of X-rays. A typical T~Tauri
star emits between $10^{29}$ and $10^{31}$~erg~s$^{-1}$ in the soft
($0.1-10$~keV) X-ray band, i.e., $2-4$ orders of magnitude more than
the Sun (see \citealt{Guedel:2004} for a review of stellar X-ray
radiation). The radiation is thought to be mostly coronal,
originating from hot ($1-20$ million K), magnetically trapped plasma
above the stellar photosphere, in analogy to the solar coronal X-ray
radiation.

There is little doubt that X-rays have some impact on the gas and
dust in circumstellar disks, at least relatively close to the star
and at the disk surface. For example, \cite{Igea:1999}, \citet{Glassgold:2004} or
\citet{Ercolano:2008} computed detailed models for radial distances between 0.5-10~AU that indicate
efficient ionization of circumstellar disks by X-rays and also
heating of the gaseous surface layers to several thousands of
Kelvin. Complicated chemical networks are a consequence (e.g.,
\citealt{Semenov:2004} or \citealt{Ilgner:2006}). A
direct evidence for these processes is suggested from the presence
of strong line radiation of [Ne\,{\sc ii}] at 12.8~$\mu$m detected
by {\sc Spitzer} in many T~Tauri stars (e.g.,
\citealt{Pascucci:2007}, \citealt {Herczeg:2007} or
\citealt{Lahuis:2007}) which in some cases may be triggered by
shocks \citep{Boekel:2009}. This transition requires ionization and
heating of the ambient gas to several 1000~K \citep{Glassgold:2007}.

Magnetic energy release events, so-called flares, occurring in the
same stellar coronae can increase the X-ray output up to hundreds of
times, but as we know from solar observations, such events are also
accompanied by high-energy electrons, protons and ions ejected from
the Sun. \citet{Feigelson:2002} speculated that the expected
elevated proton flux around T~Tauri stars leads to isotopic
anomalies in solids in the accretion disk, as suggested from
measurements of meteoritic composition for our early 
solar system \citep{Caffee:1987}.

The destructive impact of high-energy irradiation on crystalline
structures by ions has been demonstrated in laboratory measurements
by, e.g., \citet{Jaeger:2003}, \citet{Bringa:2007},
\citet{Demyk:2001}, \citet{Carrez:2002}, mainly for low energetic
cosmic rays ($E\ga$50~keV) but rarely also for lower energies
typical of stellar winds.

In this study we look in particular for correlations between the
crystalline mass fraction and stellar properties related to X-ray
emission by deriving these parameters for T~Tauri stars in the
Taurus-Auriga star formation region. We present in
Sect.~\ref{si:datasample} the target sample and some aspects of the
data reduction, describe in Sect.~\ref{si:decomposition} the
methodology of measuring the crystalline mass fraction based on
decomposition fits to the 10~$\mu$m silicate feature, present the
derived values in Sect.~\ref{si:results} and bring them into context
with X-ray parameters in Sect.~\ref{si:discussion}. Our conclusions
are presented in Sect.~\ref{si:conclusions}.

\section{Data sample and data reduction}\label{si:datasample}

We use the sample of objects in common to two recent surveys of the
Taurus-Auriga star-forming region. The first survey was obtained by
the {\sc Spitzer} InfraRed Spectrograph (IRS) and was published
previously by \citet{Furlan:2006} and further analyzed by
\citet{Watson:2009}. The second survey was performed in the X-ray
range with XMM-Newton as described by \citet{Guedel:2007}. While the
former survey provides information on the dust properties of the
circumstellar disks, the latter allows the investigation of stellar X-rays. We focus only on the disk surrounded (Class II, as listed in \citealt{Guedel:2007})
T~Tauri stars which appear in both surveys and show significant
emission in the 10~$\mu$m silicate feature. Table~\ref{si:objects}
provides an overview of the objects used for this study and their
properties derived in the framework of the XMM-Newton survey.

\begin{table}
\caption{Object sample and stellar properties (X-ray luminosity
\Lx, hardness $H$, spectral type, photosphere temperature
$T_\star$ and stellar age) published in
\citet{Guedel:2007}}\label{si:objects}

\begin{minipage}[t]{\columnwidth} \centering
\renewcommand{\footnoterule}{}
\newcolumntype{+}{D{.}{.}{-1}}

\begin{tabular}{l@{}c + l l +}
\hline \hline Name & $L_\mathrm{X}$\footnote{From Table 6 in
\citet{Guedel:2007}}\saveFN\Gue &
\multicolumn{1}{c}{$H$\footnote{Hardness,
see definition in Eq.~(\ref{si:hardness})}}&\multicolumn{1}{c}{Spect\footnote{From Table 9 in \citet{Guedel:2007}}\saveFN\Gub}&\multicolumn{1}{c}{$T_\star$\useFN\Gub}&\multicolumn{1}{c}{Age\footnote{From Table 10 in \citet{Guedel:2007}}\saveFN\Guc}\\
& [10$^{30}$ erg/s]&&&\multicolumn{1}{c}{[K]}&\multicolumn{1}{c}{[My]}\\
\hline
04187+1927      &   0.91    &   0.57    &   M0  &   3850    &   -   \\
04303+2240      &   4.99    &   0.76    &   M0.5    &   3700    &   0.5 \\
04385+2550      &   0.40    &   2.26    &   M0  &   3850    &   -   \\
AA Tau          &   1.24    &   1.98    &   K7  &   4060    &   2.4 \\
BP Tau          &   1.36    &   0.68    &   K7  &   4060    &   1.9    \\
CI Tau          &   0.19    &   1.62    &   K7  &   4060    &   2.2    \\
CoKu Tau/3      &   5.83    &   0.68    &   M1  &   3705    &   0.9    \\
CW Tau          &   2.84    &   -   &   K3  &   4730    &   7.0    \\
CY Tau          &   0.13    &   0.46    &   M1.5    &   3632    &   1.5    \\
CZ Tau          &   0.42    &   0.43    &   M3  &   3415    &   2.1 \\
DD Tau          &   0.09    &   2.55    &   M3  &   3412    &   4.5    \\
DK Tau          &   0.91    &   0.75    &   K7  &   4060    &   1.3    \\
DN Tau          &   1.15    &   1.01    &   M0  &   3850    &   1.1    \\
FM Tau          &   0.53    &   1.34    &   M0  &   3850    &   2.8    \\
FO Tau          &   0.06    &   1.13    &   M2  &   3556    &   1.5    \\
FQ Tau          &   0.12    &   0.36    &   M3  &   3416    &   2.8    \\
FS Tau          &   3.21    &   0.73    &   M0  &   3876    &   3.1    \\
FV Tau          &   0.53    &   0.50    &   K5  &   4395    &   4.8    \\
FX Tau          &   0.50    &   0.52    &   M1  &   3720    &   0.9 \\
FZ Tau          &   0.64    &   0.87    &   M0  &   3850    &   1.1    \\
GH Tau          &   0.11    &   0.98    &   M1.5    &   3631    &   2.0    \\
GI Tau          &   0.83    &   0.90    &   K7  &   4060    &   1.8    \\
GK Tau          &   1.47    &   0.96    &   K7  &   4060    &   1.2    \\
GN Tau          &   0.78    &   1.18    &   M2.5    &   3488    &   1.0    \\
GO Tau          &   0.25    &   1.09    &   M0  &   3850    &   3.8    \\
Haro 6-13       &   0.80    &   1.99    &   M0  &   3800    &   0.6    \\
Haro 6-28       &   0.25    &   0.75    &   M2  &   3556    &   10  \\
HK Tau          &   0.08    &   4.19    &   M0.5    &   3778    &   1.8    \\
HO Tau          &   0.05    &   0.08    &   M0.5    &   3778    &   9.1    \\
HP Tau          &   2.54    &   1.02    &   K3  &   4730    &   6.9 \\
IQ Tau          &   0.41    &   1.17    &   M0.5    &   3778    &   1.1    \\
IS Tau          &   0.66    &   0.48    &   K7  &   3999    &   4.2    \\
IT Tau          &   6.47    &   2.05    &   K2  &   4900    &   4.8    \\
MHO-3           &   0.46    &   1.26    &   K7  &   4060    &   2   \\
RY Tau          &   5.50    &   1.41    &   K1  &   5080    &   2.1    \\
UZ Tau/e        &   0.89    &   0.49    &   M1  &   3705    &   2.3    \\
V410 Anon 13    &   0.01    &   0.68    &   M5.8    &   3024    &   -   \\
V710 Tau        &   1.37    &   0.76    &   M0.5    &   3778    &   1.7    \\
V773 Tau        &   9.46    &   1.08    &   K2  &   4898    &   6.4    \\
V807 Tau        &   1.05    &   0.52    &   K7  &   3999    &   1.5    \\
V955 Tau        &   1.62    &   0.46    &   K5  &   4395    &   6.7 \\
XZ Tau          &   0.96    &   1.05    &   M2  &   3561    &   4.6    \\
\hline
\end{tabular}
\end{minipage}
\end{table}

The {\sc Spitzer} IRS data were obtained during the observing
campaigns 3, 4 and 12 using mainly the low resolution channel. Two
exposures per object were obtained in different nod positions
allowing the subtraction of the background by subtracting the two
spectra, $F_{\nu,1}(\lambda)$ and $F_{\nu,2}(\lambda)$, from each
other. The fluxes are then averaged over the two observations and
the difference of the two spectra
$d(\lambda)=F_{\nu,1}(\lambda)-F_{\nu,2}(\lambda)$ is used to
estimate the uncertainties of the observation. However, $d(\lambda)$
is not the correct error estimator; although it does contain the
errors due to statistical fluctuations, it is also sensitive to
systematic effects such as the truncation of the flux by
misalignment of one of the two observations. Therefore, we make a
first order correction of the uncertainty by subtracting the
averaged difference
\begin{equation}
\tilde{d}(\lambda)=d(\lambda)-\langle d(\lambda)\rangle_{\lambda\in
[7,14\mu\mathrm{m}]}
\end{equation}
where the average is taken over the wavelength range which is
relevant for the fit procedure used later for the 10~$\mu$m-silicate
feature. This correction brings $\tilde{d}(\lambda)$ closer to the
truly random noise contributions. However, $\tilde{d}(\lambda)$ can
underestimate the purely statistical fluctuations (noise) by chance
at various wavelengths. But assuming that the noise remains similar
across a narrow window in wavelength, we can obtain its statistical
value from the fluctuations of $\tilde{d}(\lambda)$ itself. We use
$N=11$ data points around each wavelength bin $\lambda_i$ to derive
the measurement uncertainties $\sigma$ by computing a geometrical
sum of their average and standard deviation of the systematic
differences $\tilde{d}(\lambda)$:
\begin{eqnarray}
\sigma^2(\lambda_i)&=&\sum_{j\in
\mathbb{A}}{\frac{(\tilde{d}(\lambda_j)-\langle\tilde{d}(\lambda_k)\rangle_{k\in\mathbb{A}})^2}{N-1}}+\left(\langle\tilde{d}(\lambda_k)\rangle_{k\in\mathbb{A}}\right)^2\nonumber\\
&&\textrm{with }\mathbb{A}=[i-5,i+5]
\end{eqnarray}

The X-ray data were obtained from the XMM-Newton Extended Survey of
the Taurus Molecular Cloud (XEST) which consisted of 28 exposures in
total, spread over the whole Taurus region. Parameters used for this
work are the stellar  X-ray luminosity \Lx and further, we
defined the hardness $H$ of the stellar X-ray emission as the ratio
between the hard and soft luminosity components. This allows
to measure the relative contribution of different energy bands to
the X-ray radiation; the hard and soft bands comprise the
$1-10$~keV and the $0.3-1$~keV ranges, respectively:
\begin{equation}\label{si:hardness}
H=\frac{L_\mathrm{X}(1~\mathrm{keV} < E <
10~\mathrm{keV})}{L_\mathrm{X}(0.3~\mathrm{keV}<E<1~\mathrm{keV})}\quad
.
\end{equation}
As a few objects such as DH~Tau showed excessive emission during the
observation (as described by \citealt{Telleschi:2007b}),
their value for \Lx does not correspond to the actual
averaged luminosity which is the parameter of interest for the
present study. Therefore, these objects have been discarded from our
sample.

The last column of Table~\ref{si:objects} lists the stellar age (from \citealt{Guedel:2007}). These values were derived from $T_{\rm eff}$ and $L_*$ as given in the 
literature and provide best-estimate values for which systematic 
uncertainties are difficult to provide. Although G\"udel et al. indicated
conservative age uncertainties (based on a variety of literature
values for $T_{\rm eff}$ and $L_*$) of factors of 2--3, these are 
extreme values, and most ages are - within the framework of one
set of evolutionary tracks (\citealt{Siess:2000} in \citealt{Guedel:2007}) -
much more narrowly confined.  As for the original parameters used for age
determination, $T_{\rm eff}$ and $L_*$, \citet{White:2001}
(used in the XEST study) estimate uncertainties in $L_*$ of 0.1--0.3 dex
for their sample. \citet{Hartigan:2003} (also used in the XEST study)
estimate errors in $L_*$ from the disagreement of ages between components
of binaries, amounting to typically 0.1--0.2 dex. A similar uncertainty
(0.11 dex) for $L_*$ has also been given  by \citet{Kenyon:1995}.
A scatter of 0.2-0.3 dex is furthermore found when comparing values
from various authors. We thus conclude that most of our ages (primarily
determined by $L_*$ and much less by $T_{\rm eff}$) are accurate to within
a factor of 1.5--2.

The range of the stellar ages spreads from 0.5 to 10~Myr. Although our sample consists of Class II T~Tauri stars only, this wide range indicates that the selected objects can be categorized into three evolutionary groups: Very young objects ($\lesssim$1~Myr) which are likely to be in a transitional phase from embedded to disk-only geometries; typical Class II objects with a pure disk geometry; and older objects ($\gtrsim$5~Myr) which are more mature Class II sources and are likely to have changed characteristics compared to their younger counterparts. These groups are not separated sharply given the age uncertainties. Also, we note that age is not the only parameter determining the evolutionary state of the disk. Therefore we continue to study the full sample and will investigate the object selection later in this paper (see Sect.~\ref{si:Correlations}).

\section{The decomposition of the 10~$\mu$m silicate
feature}\label{si:decomposition}
\subsection{Modeling of the Emission Profile}
\begin{figure}[!h]\begin{center}
\resizebox{\hsize}{!}{\includegraphics{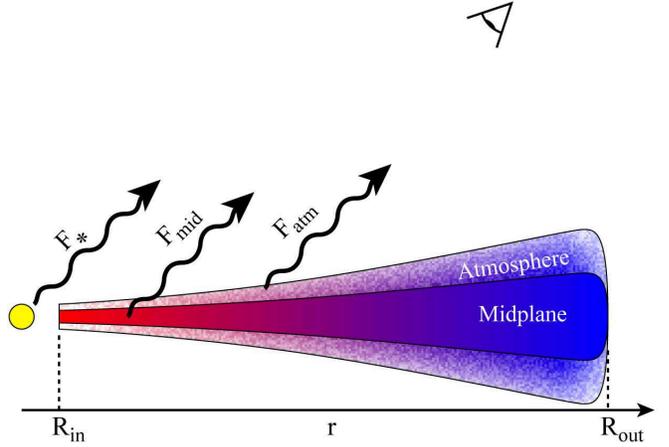}}
\caption{Circumstellar disk model describing the observable total
flux which is a superposition of the stellar light, $F_\star$, the
continuum emission from the disk midplane, $F_\mathrm{mid}$, and the
emission from the optically thin disk atmosphere,
$F_\mathrm{atm}$.}\label{si:diskmodel}
\end{center}\end{figure}

Our model describes the total observable flux with three components
based on the two-layer temperature distribution (TLTD) method
introduced by \citet{Juhasz:2009}: The emission from the stellar
photosphere, a continuum emission from the opaque disk midplane and
emission from the optically thin disk atmosphere consisting of
thermal emission from dust grains of different mineralogy and
temperature. The atmosphere is transparent with respect to the
continuum emission of the disk midplane. Figure~\ref{si:diskmodel}
shows a sketch of the disk model.

The total observable flux is given by
\begin{equation}
F_{\nu,\mathrm{tot}}=F_{\nu,\star}+F_{\nu,\mathrm{mid}}+F_{\nu,\mathrm{atm}}\quad
.
\end{equation}
We describe the flux of the stellar photosphere approximately with the
emission of a blackbody with a temperature $T_\star$:
\begin{equation}
F_{\nu,\star}=C_\star\frac{2 h \nu^3}{c^2}\frac{1}{e^{h \nu / k
T_\star}-1}\equiv C_\star\cdot B_\nu (T_\star)\quad .
\end{equation}
$C_\star$ is the normalization factor used for the later fit and
$B_\nu (T)$ defines the Planck function at temperature $T$. To model
the flux from the midplane and the atmosphere of the disk, we set up
two simple continuum emission profiles, one of which we use directly
for the midplane while the other is multiplied by dust grain
absorption coefficients $\kappa_{\nu,i}$ for modeling the flux from
the atmosphere. These two continuum emission profiles are
constructed as a superposition of blackbody spectra taking a radial
temperature distribution $T(r)$ and geometrical aspects of the disk
into account, assuming axisymmetry. Hence, the fluxes can be written
as
\begin{eqnarray}
F_{\nu,\mathrm{mid}}=\tilde{C_0}&\cdot&\int_{R_\mathrm{mid,in}}^{R_\mathrm{mid,out}}
B_\nu (T_\mathrm{mid}(r))\cdot r\cdot\d r\label{si:fluxmid1}\\
F_{\nu,\mathrm{atm}}=\sum_{i=1}^N \tilde{C_i}\cdot
\kappa_{\nu,i}&\cdot&\int_{R_\mathrm{atm,in}}^{R_\mathrm{atm,out}}
B_\nu (T_\mathrm{atm}(r))\cdot r\cdot\d r\label{si:fluxatm1}
\end{eqnarray}
where $r$ is the radial distance to the central object,
$R_\mathrm{atm/mid,in/out}$ the inner and outer radii of the disk
midplane and atmosphere, respectively (see Fig.~\ref{si:diskmodel}),
$\tilde{C_i}$ are the normalization factors of the $N$ dust species
$i$ (in our case we use $N=10$, see
Sect.~\ref{emissionprofiles}) and $\tilde{C_0}$ of the disk midplane
emission, respectively. We assume the temperature distributions to
follow a simple power law with
\begin{eqnarray}
\frac{T_\mathrm{mid}(r)}{T_\mathrm{mid,max}}&=&\left(
\frac{r}{R_\mathrm{mid,in}}\right)^{q_\mathrm{mid}}\\
\frac{T_\mathrm{atm}(r)}{T_\mathrm{atm,max}}&=&\left(
\frac{r}{R_\mathrm{atm,in}}\right)^{q_\mathrm{atm}}
\end{eqnarray}
which implies that all grains in the atmosphere or in the disk mid-plane follow the same temperature profile, regardless their size and chemical composition. This allows a substitution of $r$ with $T$ and
Eqs.~(\ref{si:fluxmid1}) and (\ref{si:fluxatm1}) can be rewritten as
\begin{eqnarray}
F_{\nu,\mathrm{mid}}=\frac{C_0}{q_\mathrm{mid}} &\cdot&
\int_{T_\mathrm{mid,max}}^{T_\mathrm{mid,min}}
B_\nu(T)\cdot T^{(2-q_\mathrm{mid})/q_\mathrm{mid}}~\d T\label{si:fluxmid2}\\
F_{\nu,\mathrm{atm}}=\sum_{i=1}^N \frac{C_i\cdot
\kappa_{\nu,i}}{q_\mathrm{atm}}&\cdot&
\int_{T_\mathrm{atm,max}}^{T_\mathrm{atm,min}} B_\nu(T)\cdot
T^{(2-q_\mathrm{atm})/q_\mathrm{atm}}~\d T\label{si:fluxatm2}
\end{eqnarray}
where $C_i$ are the new normalization factors used for the later
fit. In this fitting approach, $C_\star$,$C_0$, $\dots$, $C_N$,
$q_\mathrm{mid}$, $q_\mathrm{atm}$, $T_\star$, $T_\mathrm{mid,max}$,
$T_\mathrm{mid,min}$, $T_\mathrm{atm,max}$ and $T_\mathrm{atm,min}$
are fitting parameters. As we are only interested in a wavelength
range between \ap 7~$\mu$m and 14~$\mu$m (see
Sect.~\ref{si:validity} for a discussion), we can eliminate several
of these parameters:
\begin{itemize}
\item $T_\star$ can be derived from the literature. We used the
values summarized in Table~\ref{si:objects}.
\item We chose to set $T_\mathrm{mid,min}$ and $T_\mathrm{atm,min}$ to 10~K to account for
the contribution of the cooler region of the disk which is
irrelevant for the mid-infrared regime.
\item Figure~\ref{si:T_maxvar} shows the impact of $T_\mathrm{mid,max}$ on
the shape of the flux function given by Eq.~(\ref{si:fluxmid2}) by
keeping $q_\mathrm{mid}$ constant.
\begin{figure}[!h]\begin{center}
\resizebox{\hsize}{!}{\includegraphics{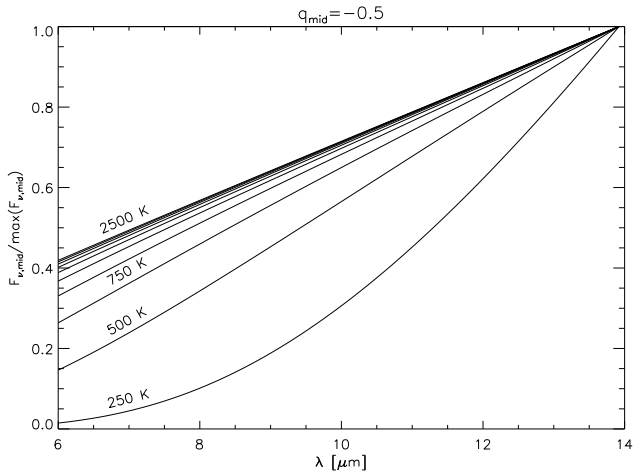}}
\caption{Normalized flux as expressed by Eq.~(\ref{si:fluxmid2}) for
$q=-0.5$ and $T_\mathrm{mid,max}=250$~K, 500~K, $\dots$,
2500~K.}\label{si:T_maxvar}
\end{center}\end{figure}
It is obvious that $T_\mathrm{mid,max}$ has no significant influence
on the shape of the flux function for values
$T_\mathrm{mid,max}>1000$~K, which is true for various values of
$q_\mathrm{mid}$. Further, \citet{DAlessio:1998} showed that the
inner disks of classical T~Tauri stars (CTTS) reach temperatures
around $T_\mathrm{sub}=1800-2000$~K due to the dust sublimation.
Therefore, we set without loss of generality
$T_\mathrm{mid,max}=1800$~K. With the same argumentation we set
$T_\mathrm{atm,max}=1800$~K.
\item First trials of fitting simulated spectra showed that the fit
is not sensitive to $q_\mathrm{atm}$ for any physically
meaningful value (e.g., $0\ge q_\mathrm{atm}\ge -2$) as the
influence on the shape of the flux function in
Eq.~(\ref{si:fluxatm2}) is fully dominated by the dust emission
profile. We therefore set $q_\mathrm{atm}=q_\mathrm{mid}\equiv q$.
\end{itemize}
Consequently, the remaining fit parameters are $C_\star$, $C_0$,
$\dots$, $C_N$ and $q$. For a given $q$ and more than $N+2$ data
points, $C_\star$, $C_0$, $\dots$, $C_N$ can be derived analytically
according to a non-negative least-square fit (see e.g.
\citealt{Lawson:1974}) and further, a unique solution exists.
Therefore, we can search efficiently for a global minimum of
$\chi^2(q)$.

To derive the components $C_i$ and $q$ of the fit function and their
errors for a given observation, we compute 100 spectra adding
Monte-Carlo simulated, normally distributed noise to the original
data; the noise distribution is based on the spectral errors of the
original data. Each of these spectra is then fitted with a
least-square method approach as we take the median values (the
distributions tend to be very asymmetric) of the resulting
parameters to obtain the final parameters and their errors
(1$\sigma$ range of the frequency distribution). 

\subsection{Emission profiles}\label{emissionprofiles}

We use emission profiles analogous to \citet{Schegerer:2006} and
\citet{Bouwman:2008}: For the amorphous silicates we use profiles
calculated for homogeneous, compact, and spherical grains applying
Mie theory. For the crystalline silicates we use emission profiles
calculated for inhomogeneous spheres according to the distribution of
hollow spheres (DHS, \citealt{Min:2005}). We start from the complex
refractive indices $n_i$ for silicate material $i$. The result is
the dimensionless absorption efficiency $Q_{i}$, which is used for
calculating the mass absorption coefficient $\kappa_{m,i} = Q_{i}\pi
a^2/(4/3\pi a^3 \rho)$ where $a$ is the particle radius and $\rho$
the material density. We fit the 10~$\mu$m silicate feature with
amorphous silicates with the stoichiometries of olivine
(MgFeSiO$_4$) and pyroxene (MgFe$[$SiO$_3]_2$) and the crystalline
silicates forsterite (Mg$_2$SiO$_4$), enstatite (MgSiO$_3$) and
quartz (SiO$_2$). Table~\ref{si:profiles} summarizes the dust
species considered here.
\begin{table}[!h]
\caption{Dust species used for the decomposition fit of the 10~$\mu$m
feature fit.}\label{si:profiles}
\begin{minipage}[t]{\columnwidth}
\renewcommand{\footnoterule}{}
\begin{center}
\begin{tabular}{l l c l}
\hline \hline
Name&Stoichiometry&Structure\footnote{a=amorphous, c=crystalline}&Reference\\
\hline
Olivine&MgFeSiO$_4$&a&\citealt{Dorschner:1995}\\
Pyroxene&MgFe$[$SiO$_3]_2$&a&\citealt{Dorschner:1995}\\
Forsterite&Mg$_2$SiO$_4$&c&\citealt{Servoin:1973}\\
Enstatite&MgSiO$_3$&c&\citealt{Jaeger:1998}\\
Quartz&SiO$_2$&c&\citealt{Spitzer:1960}\\
\hline
\end{tabular}
\end{center}
\end{minipage}
\end{table}

\begin{figure}[!h]
\begin{center}
\includegraphics[width=9cm]{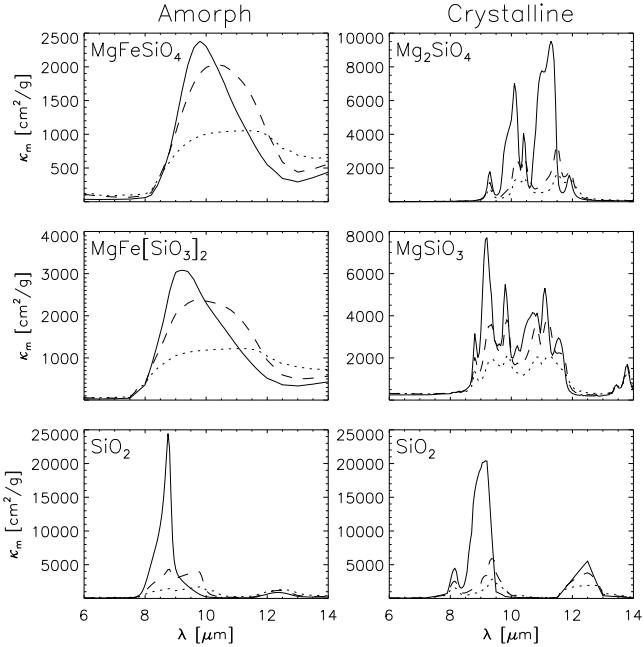}
\caption{Mass absorption coefficients of grains with radii
0.1~$\mu$m (solid line), 1.26~$\mu$m (dashed line) and 2.5~$\mu$m
(dotted lines). Plot after
\citet{Schegerer:2006}.}\label{si:opacities}
\end{center}
\end{figure}

Figure~\ref{si:opacities} shows a reproduction of Fig.~3 from
\citet{Schegerer:2006} where the mass absorption coefficients for
different grain sizes and grain compositions are shown.

In this work, we use only grains of sizes 0.1~$\mu$m and 1.26~$\mu$m
as this is sufficient to fit the data reasonably well. Further, the fit with only two grain sizes has been confirmed to be valid by
\citet{Bouwman:2001} and \citet{Schegerer:2006}. The 10~$\mu$m
silicate feature does not put constraints on silicate particles larger
than about 5~$\mu$m in general. Therefore, discussing the
crystallinity implies that we understand it as a fraction of
sub- or micron sized grains. We do not fit features from PAH molecules.
The latter produce emission lines at 7.7~$\mu$m, 8.6~$\mu$m,
11.2~$\mu$m and 12.8~$\mu$m within the spectral range of our
interest (see, e.g., \citealt{Geers:2006}). Rather, we concentrate
only on the silicates described above and perform the fit
with 5 different silicates of 2 different grain sizes each (consequently, $N=10$
different fit components). Therefore, we exclude
wavelength regions in the spectra that clearly show PAH emission
features .

\section{Results}\label{si:results}
\begin{figure}[!h]\begin{center}
\includegraphics[width=8cm]{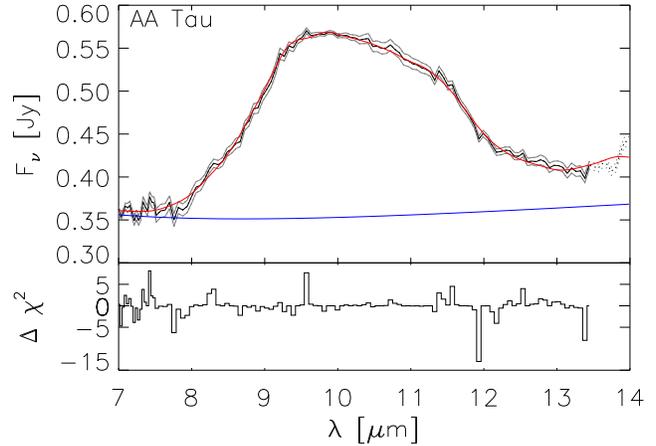}
\newline
\caption{IRS spectrum (black line) and fit (red line) including the
corresponding uncertainties (gray lines) of AA~Tau as an example of
the full data sample. The continuum background used for the fit
function is shown in blue. In the lower part of the figure, the
resulting $\Delta \chi^2$, multiplied by the sign of the deviation, is
shown.}\label{si:fitplot}
\end{center}\end{figure}

We list the fit results, i.e. the
relative mass fractions of the minerals as well as the derived
values for $q_\mathrm{disk}$, in
Appendix~\ref{si:fitresult}. Table~\ref{si:fitresult2} lists values of the crystallinity $\Gamma$
which is the sum of all relative mass fractions of the crystalline
components regardless of their size. The
reduced $\chi^2_\mathrm{red}$ values are listed in the third column,
calculated from the averaged fit parameters. In
Fig.~\ref{si:fitplot} we present the spectrum and resulting fit
functions for the example of AA~Tau; the complete sample is shown in
Appendix~B.

Most of the spectra are fitted reasonably well with a reduced
$\chi^2_\mathrm{red}$ in the range of 1-3. The fit of a few spectra such as
IRAS~04303+2240, FV~Tau, Haro 6-13, MHO-3, and RY~Tau show
systematic discrepancies between the data and the fit, resulting in
a large $\chi^2_\mathrm{red}$. In most of these cases, the data
show a high signal-to-noise ratio and consequently, the error bars
of the data are small. Our fit model is too incomplete to describe
these object accurately. In the particular case of IRAS~04303+2240,
the spectrum shows a large variety of emission features which we are
not able to describe with the selected minerals. We did not intend
to increase the complexity of the model to avoid including too many
degree of freedom for the bulk of the sample with lower
signal-to-noise ratios. We do not use these poorly fitted objects
for our further studies.

\begin{table}
\renewcommand{\footnoterule}{}
\newcolumntype{+}{D{.}{.}{-1}}
\caption{Derived quantities for the decomposition analysis of the
10~$\mu$m silicate feature. The values in brackets correspond to the
border values of the 1$\sigma$ range. The last two columns
list the values of the crystallinity for the cold
($\Gamma_\mathrm{c}$) and warm ($\Gamma_\mathrm{w}$) components as
derived by \citet{Sargent:2009}.}\label{si:fitresult2}

\begin{minipage}[t]{\columnwidth} \centering

\begin{tabular}{l +@{ }r@{ }+@{ }r@{$\pm$}l@{ }r@{$\pm$}l}
\hline \hline Name        &   \multicolumn{2}{c}{$\Gamma$
[\%]}&\multicolumn{1}{c}{$\chi^2_\mathrm{red}$}&\multicolumn{2}{@{
}c@{
}}{$\Gamma_\mathrm{c}$\footnote{From \citet{Sargent:2009}}\saveFN\Saa [\%]}            &   \multicolumn{2}{@{ }c@{ }}{$\Gamma_\mathrm{w}$\useFN\Saa [\%]}            \\
\hline \\
04187+1927  &   33.4    &($ 32.6    ,   33.6    $)& 2.7                                 \\
04303+2240  &   33.6    &($ 32.8    ,   34.4    $)& 14.3                                    \\
04385+2550  &   1.7 &($ 1.5 ,   1.9 $)& 1.8                                 \\
AA Tau      &   15.6    &($ 14.4    ,   16.9    $)& 1.3 &   25  &   21  &   5   &   5   \\
BP Tau      &   8.8 &($ 8.1 ,   9.1 $)& 0.7 &   15  &   5   &   18  &   6   \\
CI Tau      &   15.3    &($ 14.4    ,   16.3    $)& 1.7 &   8   &   13  &   6   &   4   \\
CoKu Tau/3  &   22.7    &($ 21.4    ,   24.2    $)& 1.8 &   18  &   7   &   11  &   4   \\
CW Tau      &   20.7    &($ 20.1    ,   26.4    $)& 2.3 &   75  &   27  &   6   &   9   \\
CY Tau      &   37.0    &($ 32.7    ,   43.1    $)& 1.0 &   77  &   52  &   30  &   27  \\
CZ Tau      &   2.2 &($ 2.2 ,   2.3 $)& 0.8                                 \\
DD Tau      &   17.6    &($ 16.6    ,   18.7    $)& 2.2 &   19  &   13  &   20  &   10  \\
DK Tau      &   17.3    &($ 16.7    ,   17.6    $)& 2.2 &   56  &   14  &   14  &   3   \\
DN Tau      &   44.3    &($ 41.2    ,   47.2    $)& 1.1 &   7   &   17  &   50  &   27  \\
FM Tau      &   1.2 &($ 1.0 ,   1.3 $)& 1.2 &   13  &   9   &   4   &   5   \\
FO Tau      &   24.7    &($ 19.1    ,   34.5    $)& 0.4 &   14  &   7   &   17  &   11  \\
FQ Tau      &   50.7    &($ 50.1    ,   52.3    $)& 1.1 &   53  &   153 &   14  &   8   \\
FS Tau      &   1.1 &($ 1.0 ,   1.2 $)& 1.3 &   9   &   10  &   7   &   8   \\
FV Tau      &   1.5 &($ 1.1 ,   2.0 $)& 4.9 &   6   &   7   &   8   &   5   \\
FX Tau      &   4.6 &($ 3.9 ,   5.4 $)& 1.9 &   15  &   6   &   14  &   4   \\
FZ Tau      &   43.1    &($ 42.1    ,   44.1    $)& 2.7 &   49  &   28  &   22  &   7   \\
GH Tau      &   34.9    &($ 31.8    ,   39.0    $)& 1.4 &   19  &   11  &   12  &   8   \\
GI Tau      &   10.3    &($ 10.0    ,   11.4    $)& 0.5 &   13  &   7   &   3   &   4   \\
GK Tau      &   7.7 &($ 7.6 ,   7.8 $)& 1.9 &   26  &   9   &   8   &   3   \\
GN Tau      &   17.2    &($ 16.4    ,   18.9    $)& 2.2 &   64  &   19  &   12  &   4   \\
GO Tau      &   16.1    &($ 13.5    ,   19.6    $)& 1.1 &   14  &   12  &   9   &   6   \\
Haro 6-13   &   1.4 &($ 1.4 ,   1.5 $)& 3.8                                 \\
Haro 6-28   &   30.3    &($ 30.1    ,   31.3    $)& 1.0 &   18  &   11  &   45  &   13  \\
HK Tau      &   40.8    &($ 37.3    ,   44.5    $)& 1.6 &   8   &   7   &   24  &   8   \\
HO Tau      &   5.0 &($ 4.4 ,   5.5 $)& 1.2 &   17  &   23  &   14  &   7   \\
HP Tau      &   4.1 &($ 4.0 ,   4.4 $)& 1.6 &   12  &   24  &   4   &   4   \\
IQ Tau      &   6.9 &($ 6.4 ,   7.3 $)& 1.5 &   7   &   31  &   9   &   5   \\
IS Tau      &   29.0    &($ 26.6    ,   31.7    $)& 1.7 &   100 &   45  &   14  &   4   \\
IT Tau      &   32.3    &($ 28.3    ,   40.1    $)& 1.7 &   8   &   11  &   63  &   45  \\
MHO-3       &   1.5 &($ 1.4 ,   1.6 $)& 13.6                                    \\
RY Tau      &   7.0 &($ 6.7 ,   7.5 $)& 6.6                                 \\
UZ Tau/e    &   11.6    &($ 9.4 ,   11.6    $)& 1.1 &   5   &   10  &   2   &   4   \\
V410 Anon   13&   59.1    &($ 54.6    ,   61.2    $)& 1.6 &   6   &   7   &   43  &   13  \\
V710 Tau    &   16.2    &($ 15.3    ,   19.0    $)& 1.4 &   37  &   21  &   23  &   10  \\
V773 Tau    &   22.4    &($ 22.4    ,   26.0    $)& 2.7                                 \\
V807 Tau    &   32.1    &($ 28.6    ,   37.1    $)& 1.7 &   15  &   7   &   8   &   18  \\
V955 Tau    &   36.5    &($ 33.8    ,   38.7    $)& 1.9 &   21  &   9   &   25  &   8   \\
XZ Tau      &   23.5    &($ 22.0    ,   25.4    $)& 1.8 &   4   &   14  &   8   &   12  \\

\hline
\end{tabular}
\end{minipage}
\end{table}
\section{Discussion}\label{si:discussion}
\subsection{Disk model validity}\label{si:validity}
The fit method presented here for decomposing the 10~$\mu$m silicate
feature takes advantage of a continuous temperature distribution.
Therefore, this model is more realistic than previous methods using
polynomials \citep{Bouwman:2001}, single temperature blackbody
functions \citep{Meeus:2003} or two temperature fits (e.g.
\citealt{Boekel:2005} or \citealt{Sargent:2009}). For a
systematic comparison between the methods, see \citet{Juhasz:2009}.

We conclude from the spectral fits that the continuous temperature
distribution of the disk is a robust model for approximating the
continuum emission. We are able to adjust the background by just one
geometrical parameter, i.e. the slope $q$ of the radial temperature
distribution, which shows the strength and the simplicity of this
method. Although previous studies pointed out that the exact shape
of the background function has little effect on the relative
composition of the dust mineralogy, the application of a more
physical model is appropriate and avoids wrong conclusions about
dust temperatures from a single blackbody approach. Further,
using the TLTD method to describe the silicate emission features of the
disk atmosphere allows to model the contribution to the 10~$\mu$m
flux by grains of varying temperatures. As \citet{Juhasz:2009}
pointed out, this is crucial to reduce the systematic uncertainty of
the decomposition.

On the other hand, it has to be mentioned that the temperature profile is assumed to be independent of the dust grain size and composition, which is an evident simplification, especially at the disk surface. Further, the TLTD method in its presented form does not account for radially dependent distributions of the individual dust species. As suggested by \citet{Juhasz:2009}, to
optimize the validity of the applied method, we confined the wavelength range to the 10~$\mu$m feature only and did not include
longer wavelengths. It would be very interesting to extend the
wavelength range to the full Spitzer IRS spectral coverage; a
radially dependent distribution of the dust composition will be
implemented in the TLTD method in a future study.

Most objects of our sample have been studied likewise by
\citet{Sargent:2009} (SA09), using a two temperature decomposition
fit (2T fit). Although this method appears to be less realistic, it probes
different regions of the disk and is worth to compare with our
results. The last two columns of Table~\ref{si:fitresult2} list the
accumulated values for all crystalline mass fractions for the warm
and the cold fit components, respectively. The errors have been
calculated using standard error propagation of the uncertainties given in Table~6 of SA09. The method used by SA09 to derive the uncertainties of the fitting parameters is dissimilar from our work and the values for the uncertainties as presented by SA09 are very conservative. The consequence is that large errors appear in the propagated values for the crystallinity as shown in Table~\ref{si:fitresult2}. Therefore, we expect many objects to have overlaps between the 1~sigma range of our $\Gamma$-values and the values derived by
SA09. Indeed, out of 34 common objects, 23 show an overlap with
either the warm or the cold component. This number increases to 28 objects when considering our results with a 2 sigma uncertainty. However, such an ad hoc
comparison is not very valuable as a real comparison of the derived
crystallinity is not possible: Basically, we should compare our
results with the warm dust component of SA09 only as this
corresponds more likely to the 10~$\mu$m feature. However, the
largest fraction of the mass contribution for the two temperature
fit is contained in the cold component ($\sim$100~-~200~K) which does not contribute
to the 10~$\mu$m flux significantly. 

\subsection{Correlations between crystallinity and X-ray emission}\label{si:Correlations}

We compare stellar high-energy properties with structural characteristics of the dust disk. For this purpose, we correlate the X-ray luminosity \Lx and the product of \Lx and Hardness $H$ with the total crystalline mass fraction $\Gamma$. Using the full sample of the remaining 37 objects shows that these quantities do not correlate. As discussed in Sect.~\ref{si:datasample}, our sample consists of objects of a wide spread in age and therefore it is likely that different evolutionary stages are present even if we are studying class II objects only. As mentioned in Sect.~\ref{si:datasample}, to increase the uniformity of the evolutionary epoch, we selected the objects according their age and divided them into three groups; very young, intermediate and older objects. Unfortunately, our sample size decreases to 34 as we lose another three objects (IRAS~04187+1927, IRAS~04385+2550 and V410~Anon~13) for which no age determination could be found in the literature. 

\begin{figure}[!h]\begin{center}
\begin{tabular}{l l c r}
\multicolumn{4}{r}{\includegraphics[width=8.8cm]{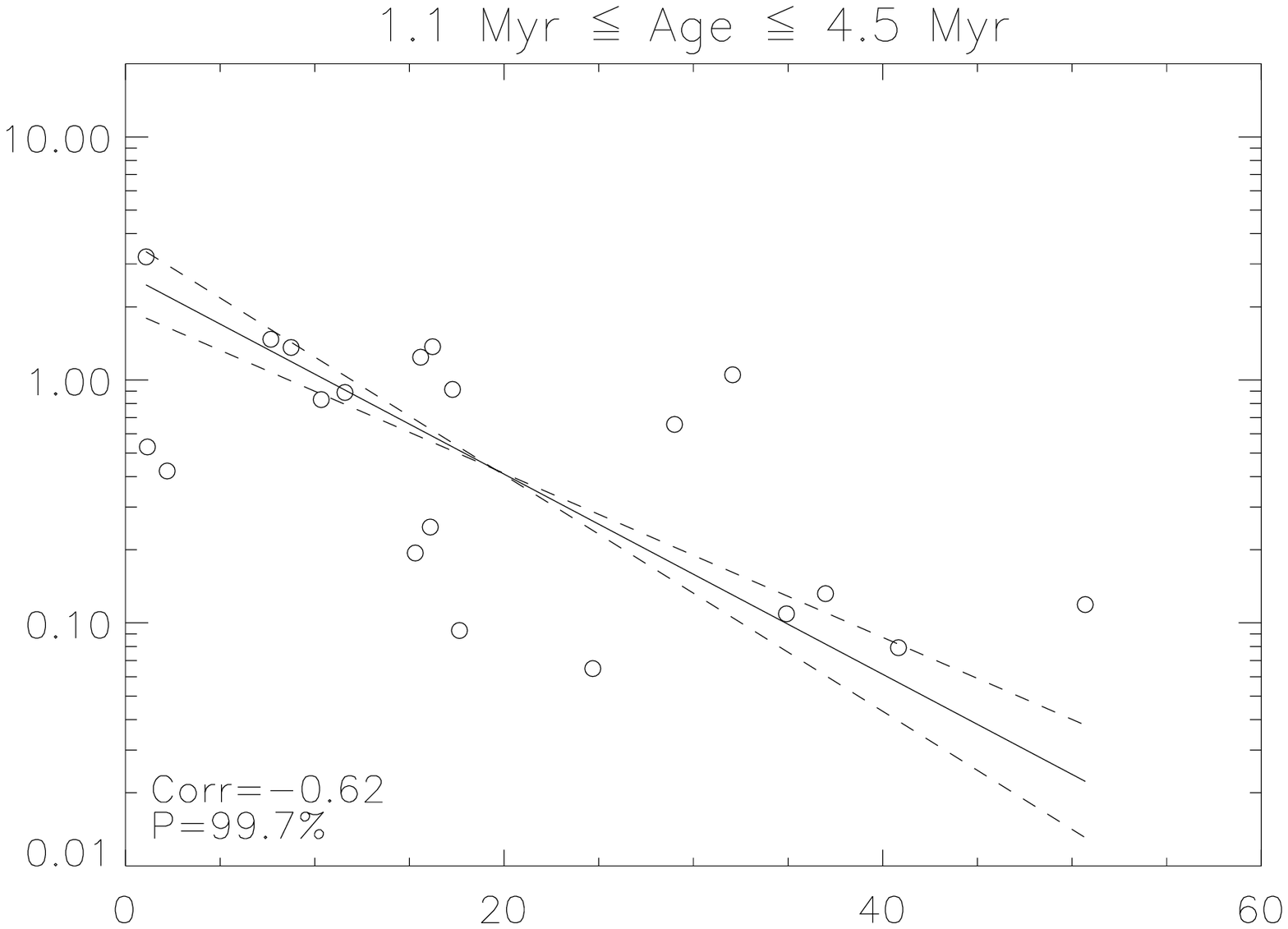}}\\
\\
&&&\\
 &\includegraphics[width=3.6cm]{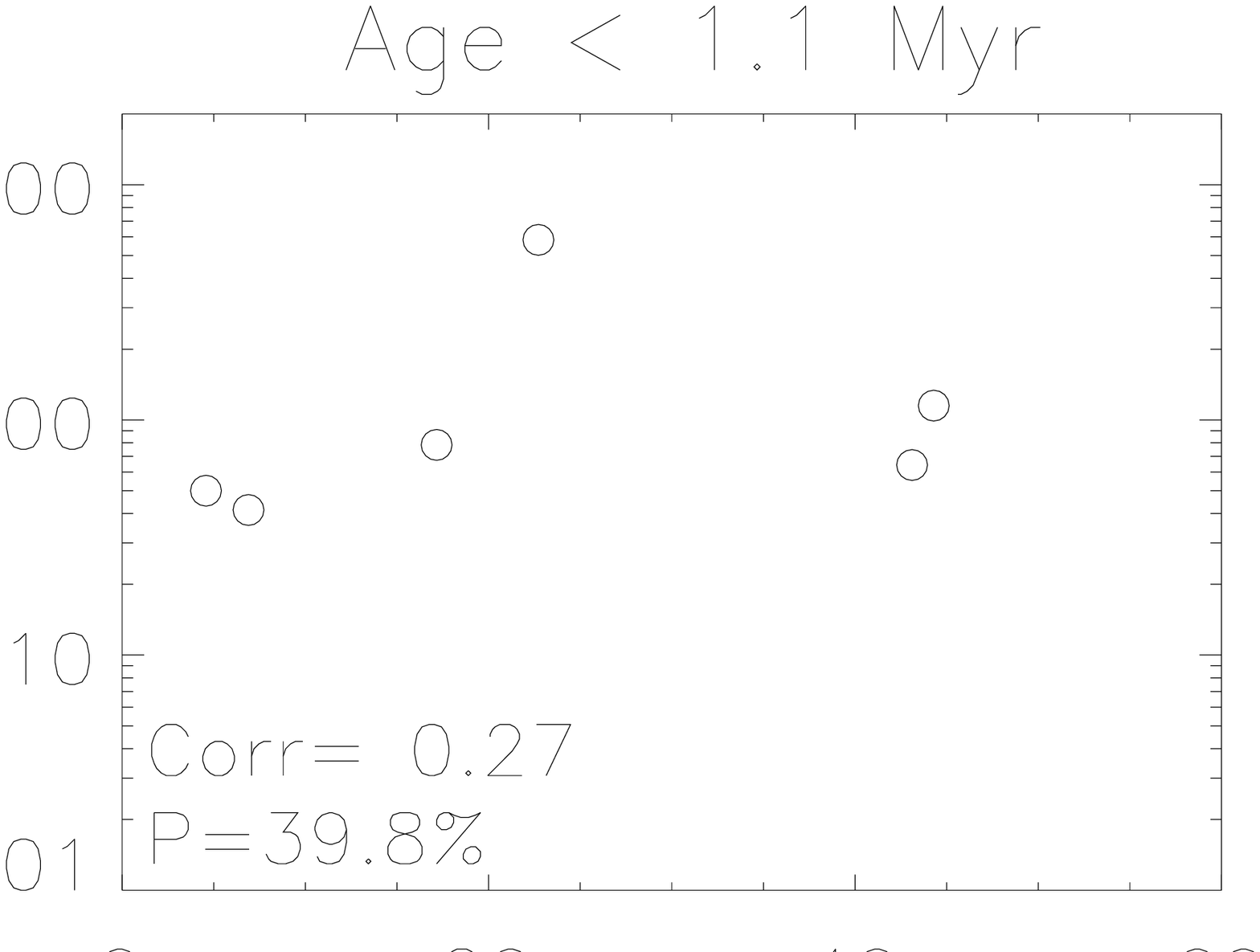}&&
\includegraphics[width=3.6cm]{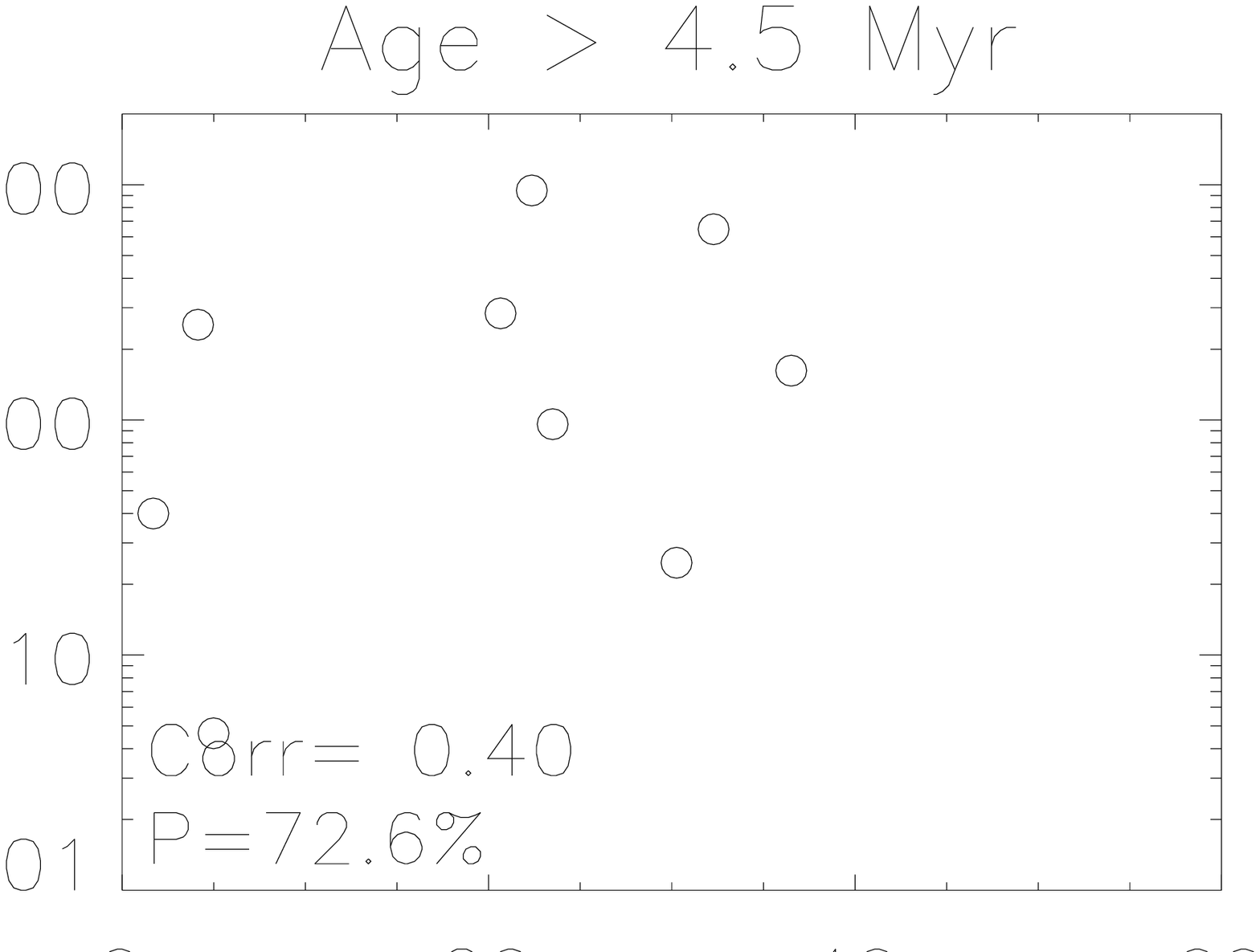}\\
&&&\\
\end{tabular}
\newline
\caption{X-ray luminosity \Lx vs. total crystalline mass
fraction $\Gamma$ for objects of an age within $1.1-4.5$~Myr (top
panel), for objects younger than 1.1~Myr (bottom left panel), and
older than 4.5~Myr (bottom right panel). The lines in the top panel
represent the OLS bisector regression (solid) with the uncertainties
for the slope (dashed).}\label{si:Lx_vs_Crist_cut}
\end{center}\end{figure}

We found a significant correlation between \Lx and $\Gamma$ when selecting the objects in the intermediate range (\ap 1~Myr to \ap 5~Myr). Figure~\ref{si:Lx_vs_Crist_cut} shows the correlation plots for the three groups while the age limits of 1.1~Myr and 4.5~Myr have been set to optimize the correlation performance with the number of objects used for the intermediate group (see below). We measure a correlation coefficient of -0.62 and a significance of 99.7\% for a correlation.

We calculate the regression lines $L_\mathrm{X}(\Gamma)$ according the ordinary least
square (OLS) bisector method described by \citet{Isobe:1990} to
treat the two variables symmetrically while we use the logarithmic
value of \Lx. The dashed lines in
Fig.~\ref{si:Lx_vs_Crist_cut} correspond to lines with slopes adding
and subtracting the regression slope uncertainty as calculated
according to Table~1 of \citet{Isobe:1990}.

We note here that measurement errors (or errors derived from measurements) 
will not be used in the derivation of regression lines. If measurement
errors are a minor contribution to the scatter around a regression line,
then the deviation of a point from the best-fit line is dominated by
systematic processes not considered here; using measurement errors as
weights for such points will introduce arbitrary bias that is 
unrelated to the actual scatter of the points. This problem has been 
discussed in detail by Isobe et al. (1990) and Feigelson \& Babu (1992). 
Is the scatter dominated by unknown, additional processes in our case?  
The uncertainties in $L_{\rm X}$ of our sample were discussed in G\"udel et al. (2007) with the result that the 
errors resulting from the X-ray measurement process and the spectral fit
procedures are usually smaller than variations due to intrinsic X-ray 
variability on time scales of hours to days. The uncertainty introduced
by X-ray variability usually corresponds to a factor of $\sim$2
between maximum and minimum values. As can be seen in our correlation plots
discussed here, such variations are smaller than the scatter around
the regression lines. We conclude that other factors not considered here
dominate the scatter, and errors from the measurement or spectral
fit procedures are inappropriate to use here. We therefore use
{\it unweighted} regression (Isobe et al. 1990). 

\begin{figure}[!b]\begin{center}
\includegraphics[width=8.8cm]{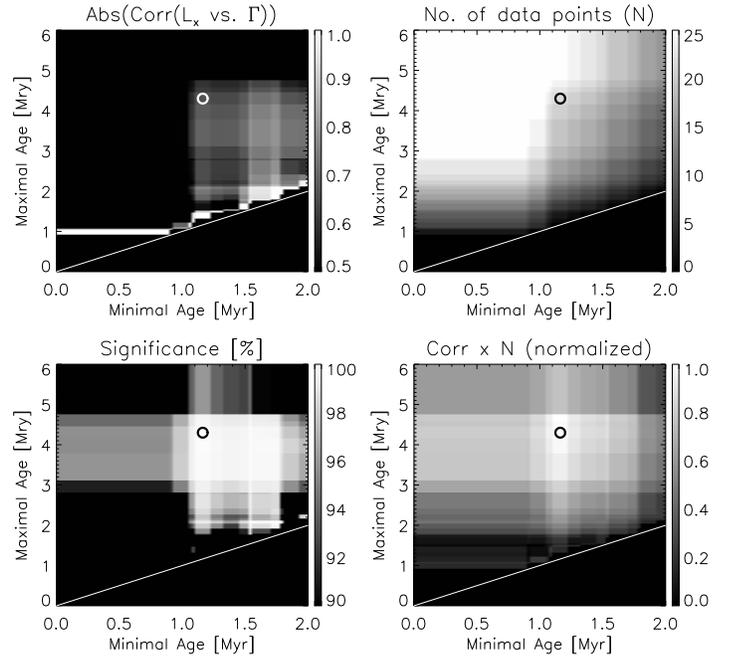}
\caption{Maps of the correlation coefficient (absolute value, upper left), the number of selected data points (upper right), the significance (lower left) and the product of the correlation coefficient and the number of selected datapoint (lower right) for the \Lx vs. $\Gamma$ correlation taking only objects into consideration which have a stellar age in between the minimum and the maximum age as indicated by the two axes of the maps. The circles indicate our selection of the stellar age for which the correlation coefficient and the statistics has been optimized. The straight lines indicate where the minimum and maximum age limitations are identical.}\label{si:agecut2d}
\end{center}\end{figure}

To define the limits of the stellar age for which the
correlation works best, we varied the lower and upper limit and
calculated the correlation coefficient of the intermediate group. Figure~\ref{si:agecut2d} shows a map of the correlation coefficient and the correlation significance as a function of the minimum and maximum stellar age used for selecting the sample. We see that the minimum and maximum stellar ages for which the correlation is still present, is not sharply defined: It varies between 1-2~Myr for the minimum and 2-4.5~Myr for the maximum age, respectively. This is in agreement with the age uncertainties discussed in Sect.~\ref{si:datasample}. Beside the goodness of the fit, we also considered the statistics in terms of the number of selected objects which is shown in the upper right panel of Fig.~\ref{si:agecut2d}. Since we tried to optimize the selection in terms of correlation and statistical sample, we show in the lower right panel of the same figure the product of the correlation coefficient and the number of objects. 
We decided to use objects of age older than 1.1~Myr and younger than 4.5~Myr and 20 objects satisfied this condition while 14 objects fall outside the borders. We emphasize that these
age limits are somewhat arbitrary within the typical age uncertainties
adopted for our stellar ages (see Sect. 2) and correspond to an optimum
choice based on the limited number of objects in our sample. The selected
age interval should be interpreted as essentially containing CTTS of
typical ages in Taurus ($\approx 1-5$~Myr). We marked the final selection with a circle in Fig.~\ref{si:agecut2d}.

Since we selected the sample according the goodness of decompositional fit of the 10~$\mu$m feature using $\chi^2_\mathrm{red}\le3$ as a criterium, we investigated how this cut effects the correlation. For this purpose we added all objects to the sample within the optimum age limits but higher $\chi^2_\mathrm{red}$, in particular we added MHO-3 and RY~Tau. The correlation coefficient of this enlarged sample was found to be -0.63 with a probability of 99.6\%. This is very comparable to the original result and we conclude that the applied threshold for $\chi^2_\mathrm{red}$ has no influence on the systematics of our study.

\begin{figure}[!t]\begin{center}
\begin{tabular}{l l c r}
\multicolumn{4}{r}{\includegraphics[width=8.8cm]{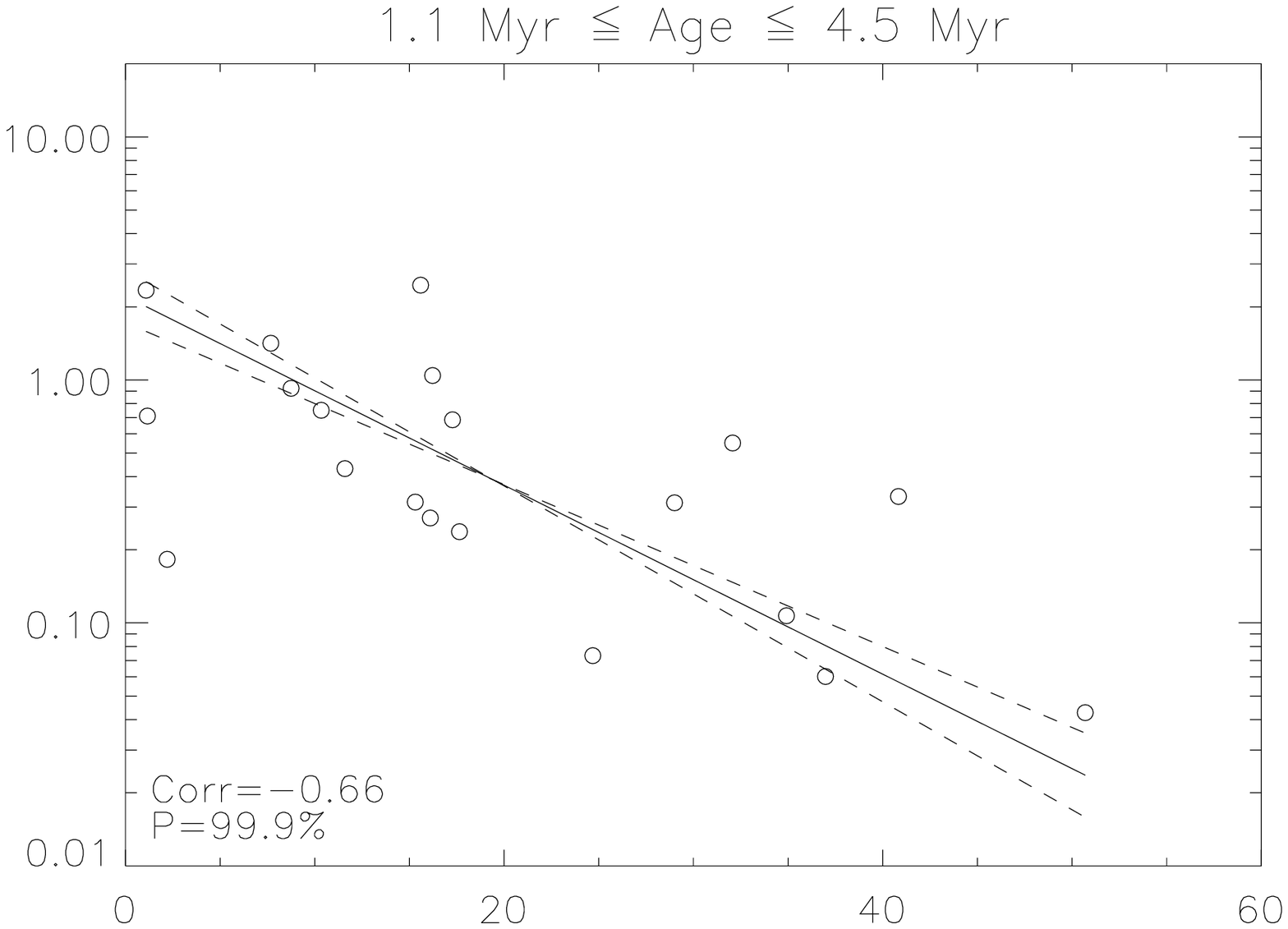}}\\
\\
&&&\\
&\includegraphics[width=3.6cm]{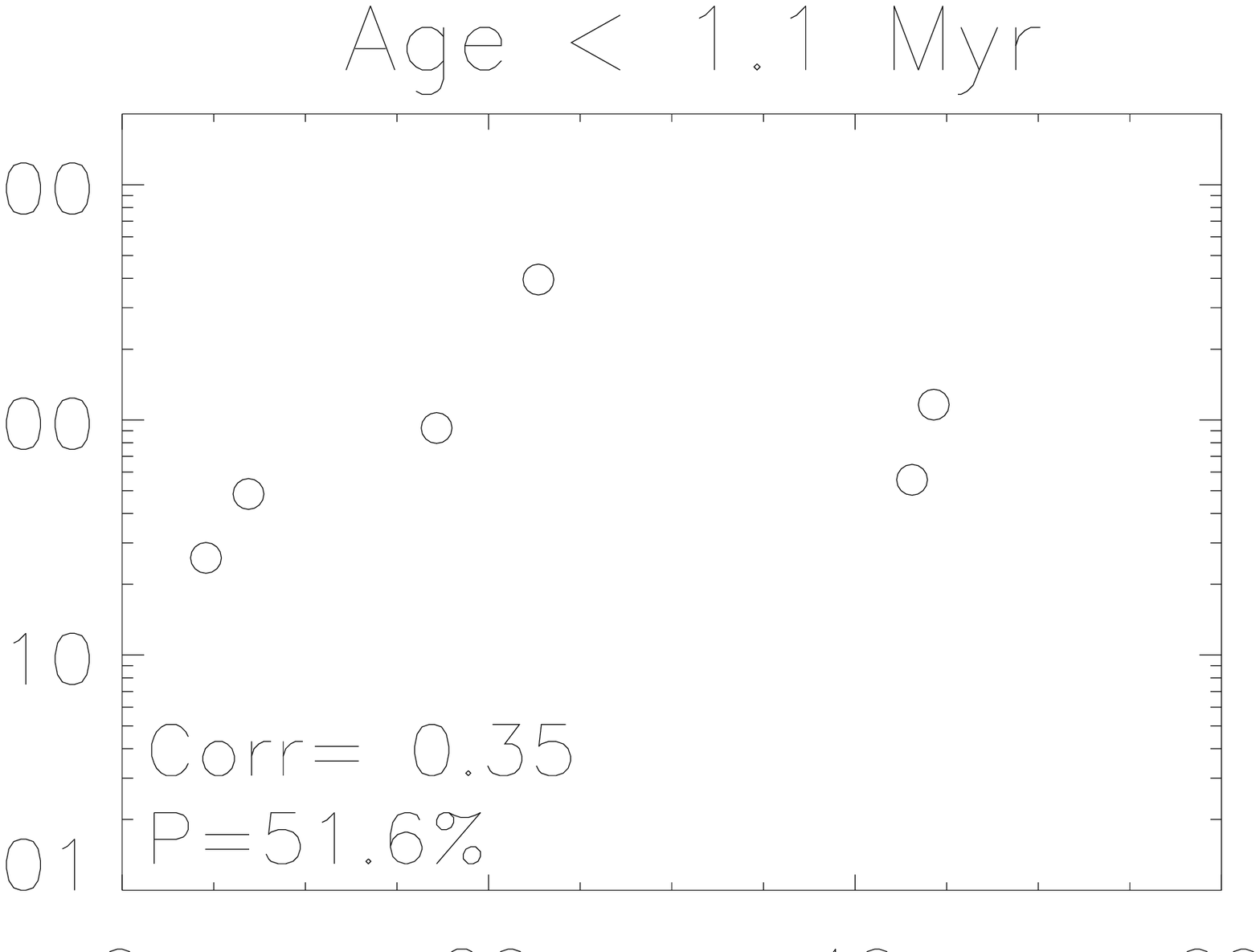}&&
\includegraphics[width=3.6cm]{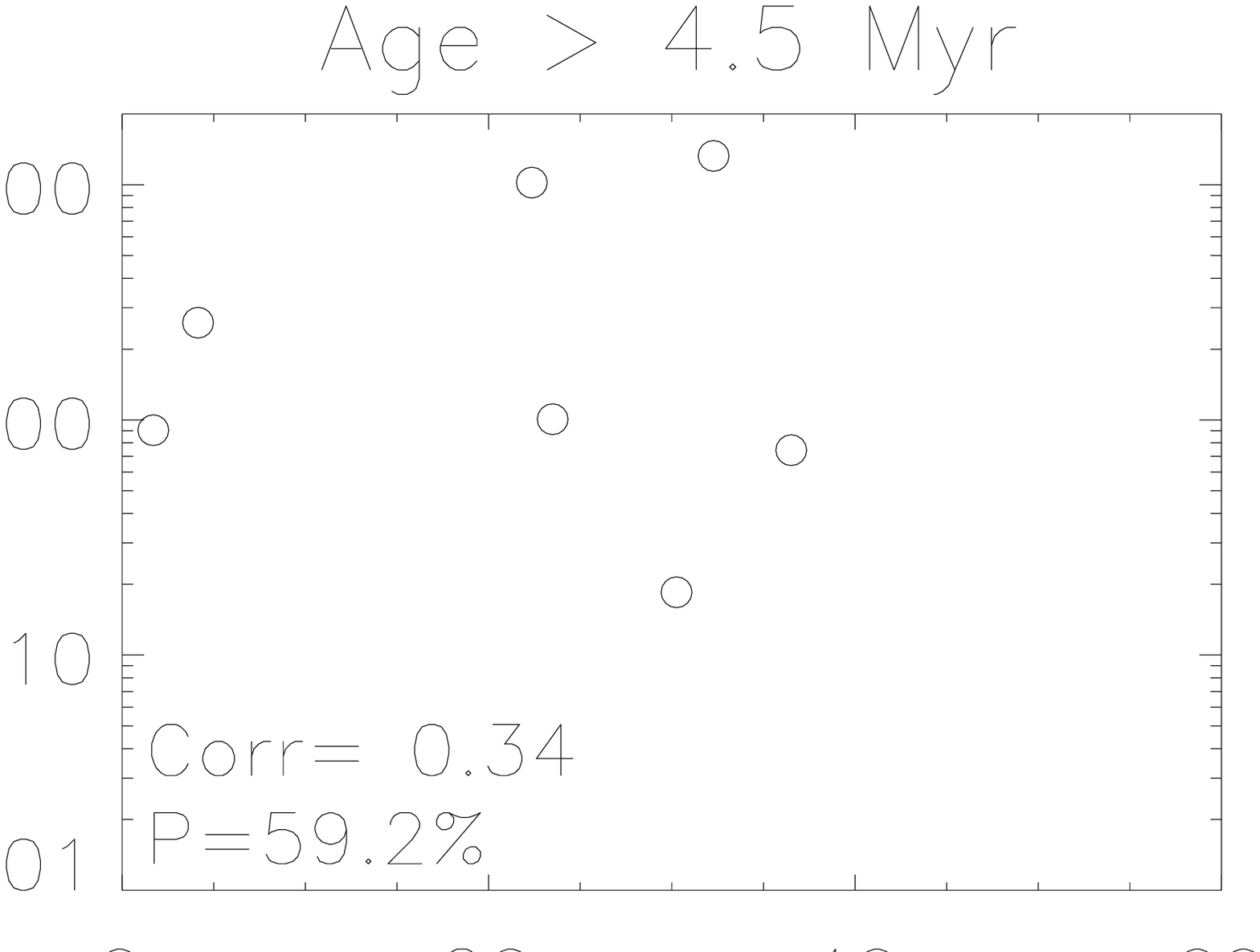}\\
&&&\\
\end{tabular}
\newline
\caption{Product of the X-ray luminosity \Lx and the
X-ray hardness $H$ vs. total crystalline mass fraction $\Gamma$ for
objects of an age within $1.1-4.5$~Myr (top panel), for objects
younger than 1.1~Myr (bottom left panel), and older than 4.5~Myr
(bottom right panel).}\label{si:LxH_vs_Crist_cut}
\end{center}\end{figure}

Finally, we plot in Fig.~\ref{si:LxH_vs_Crist_cut} the product of
\Lx and $H$ against $\Gamma$,
using the same object selection as before.
We measure a correlation coefficient of -0.66 with a significance of 99.9\%. The product of \Lx and $H$ may represent the deposited energy of
the X-rays in the disk. It is remarkable that the correlation for the product of \Lx and the hardness correlate better with the crystalline mass fraction than the X-ray luminosity alone. We investigated the optimum age selection limits as before, and got for both the minimum and maximum age identical values as derived for the \Lx vs. $\Gamma$ correlation.

We repeated this study with the values of the crystalline mass fraction derived by SA09 to discuss the dependencies of the correlation on the applied method. Therefore, we correlated the crystalline mass fractions $\Gamma_\mathrm{c}$ and $\Gamma_\mathrm{w}$ (see Sect.~\ref{si:validity} and Table~\ref{si:fitresult2}) with \Lx. Since the reduced $\chi^2_\mathrm{red}$ in SA09 are systematically high, we had to increase the limit to $\chi^2_\mathrm{red}\le 5$ to calculate the correlation with a sufficient sample. With this approach we use a sample of 20 common objects of which 14 fall into the age range of 1.1-4.5~Myr. We found that $\Gamma_\mathrm{w}$ correlates weakly with \Lx (corr=-0.54, $P$=95.6~\%) and strongly with $H\cdot L_\mathrm{X}$ (corr=-0.65, $P$=98.9~\%). $\Gamma_\mathrm{c}$ does not correlate with any of these parameters at all. Further, repeating the investigation of the age range selection showed a comparable picture to the valid range (1.1-1.5~Myr for the lower limit and 1.5-6 Myr for the upper limit).

This result confirms three aspects of our study: First, the correlation seems to be independent of the applied fitting method and is therefore not biased by the systematics of the decomposition approach. Second, our TLTD fit refers to the 10~$\mu$m feature, for which
we found a correlation between crystallinity and \Lx. In 
the 2T fit (as used by SA09), the 10~$\mu$m feature is emitted by the warm component,
and a corresponding correlation is indeed found.
Our focus on the 10~$\mu$m silicate feature is therefore justified
a posteriori as the cold component from the 2T fit does 
not show a correlation. And third, we see in both data sets that the correlation for the combined product of hardness and X-ray luminosity is better than for the X-ray luminosity alone.

\subsection{Is the dust amorphized by the stellar wind?}

It is a remarkable result that an anticorrelation between the X-ray
emission and the crystalline mass fraction is found; so far, no
correlation between properties of the central object and the
crystalline structure of the dust disk has been reported from
observations (e.g., \citealt{Sicilia:2007}).

However, the observed anticorrelation demands an indirect
explanation because X-rays of the observed energies carry too little
momentum to damage the crystalline structure of the dust. Their
energy will rather be absorbed by the electrons, which temporarily
leads to ionization of lattice atoms and finally to the production
of phonons and a resulting increase of the temperature within the
grain.

Although the processes of X-ray emission are still controversial for
T~Tauri objects, it seems clear that they mostly originate in
magnetic coronae analogous to the solar corona, although stellar
coronae are much more X-ray luminous and hotter. Therefore, we
assume that the X-ray emission of our sample of T~Tauri stars is
related to the high energy processes in the stellar corona that also
lead to a solar-like wind composed of ions, electrons and neutrons.
We show in the following that the flux of the solar wind and the
particle energies would be sufficient to amorphize dust grains
efficiently in a protoplanetary disk at a radius of 1~AU and that
this process induces significant changes in the emission profile of
the dust grains. Although we may observe the silicates at larger
distances from their central object, the choice of 1~AU is based on
the availability of data from the solar wind and allows for an
extrapolation to larger distances. We use a simple disk model where
homogeneous dust grains at the disk surface are directly exposed to
the stellar wind and where recrystallization effects are neglected.

\subsubsection{Stellar wind properties}

We first derive the particle fluxes and energies from data from the
solar wind as this is the only object for which direct and reliable
measurements of these parameters are available. We know from the
enrichment of spallation products in meteorites that the young Sun
had a proton flux several orders of magnitude higher than at present
\citep{Caffee:1987}. Unfortunately, the energies which lead to
nuclear spallation are in the order of $E>10~$MeV while the energies
of interest for amorphization are $E<1~$MeV (see below and, e.g.,
\citealt{Jaeger:2003}). The processes producing these energies might
be different and consequently it may not be correct to scale
particle fluxes with the well measured soft X-ray flux for younger
stars. Therefore, we proceed with present solar particle fluxes and
energies and use its amorphization potential as a lower limit.

We used the database from OMNIWeb \citep{King:2005} to determine the
present proton density of the solar wind plasma, its velocities and
proton to He-ion ratio near the Earth. We averaged the full dataset
over the years 1963 until 2008 and obtained a density of
$\sim$7.2~protons/cm$^3$, a wind speed of $\sim$450~km/s and a
He-ion to proton ratio of $\sim$4.4~\%. Uncertainties of these
quantities are up to 30~\%. This leads to a mean proton flux of
3.2$\cdot 10^8$~protons/(cm$^2\cdot$s) at energies around 1~keV and
to a He-ion flux of 1.4$\cdot 10^7$~ions/(cm$^2\cdot$s) at energies
around 4~keV.

\subsubsection{Calculation of dust amorphization}

We use these values to compare them with the ion dose required to
amorphize a dust grain. For this purpose we calculate the number of
displacements of lattice atoms per incident ion using the SRIM-2008
software \citep{ziegler:2008}. SRIM is a collection of software packages which calculate many features of the transport of ions in matter such as ion stopping, range and straggling distributions in multilayer targets of any material. Further, it allows the calculation of ion implantation including damages to solid targets by atom displacement, sputtering and transmission in mixed gas/solid targets. Table~\ref{si:SRIM} summarizes the
calculation performed for two minerals, pyroxene and enstatite,
respectively, irradiated by protons and He-ions at various energies.
With the SRIM software we determined the penetration depth (range)
of the ion and the number of displaced lattice atoms (due to elastic
scattering on the atom's nuclei) per incident ion.

\begin{table*}
\newcolumntype{+}{D{.}{.}{-1}}
\caption{Ion-penetration depth (range), number of displaced lattice
atoms per ion $N_{\mathrm{Disp.}}/\mathrm{ion}$, and the minimum
required dose for full amorphization of the upper layer (of
thickness equal to the ion penetration depth) of pyroxene and
enstatite by proton and He-ion irradiation at various energies. For
this calculation, mineral data from the webmineral data base and the
calculation software SRIM-2008 \citep{ziegler:2008} have been used.
For the calculation of the dose, a minimal displacement per lattice
atom (dpa) of 2.5 have been assumed to observe
amorphization.}\label{si:SRIM}
\begin{center}

\begin{tabular}{l|+@{ }+|+@{ }+@{ }+|+@{ }+@{ }+}
\hline \hline
\multirow{3}{*}{\begin{sideways}Ion\end{sideways}}&\multicolumn{1}{c}{Stellar Wind}    &\multicolumn{1}{c|}{Particle}      &   \multicolumn{3}{c}{Pyroxene, $\rho=1.2\cdot10^{23}\frac{\mathrm{atoms}}{\mathrm{cm}^3}$}      &   \multicolumn{3}{c|}{Enstatite, $\rho=9.6\cdot10^{22}\frac{\mathrm{atoms}}{\mathrm{cm}^3}$}\\
&\multicolumn{1}{c}{Speed}&\multicolumn{1}{c|}{Energy}&\multicolumn{1}{c}{Range}&\multicolumn{1}{c}{\multirow{2}{*}{$\frac{N_{\mathrm{Disp.}}}{\textrm{ion}}$}}&\multicolumn{1}{c|}{Dose}&\multicolumn{1}{c}{Range}&\multicolumn{1}{c}{\multirow{2}{*}{$\frac{N_{\mathrm{Disp.}}}{\textrm{ion}}$}}&\multicolumn{1}{c}{Dose}\\
&\multicolumn{1}{c}{[km/s]}&\multicolumn{1}{c|}{[keV]}&\multicolumn{1}{c}{[nm]}&&\multicolumn{1}{c|}{$\cdot10^{16}~\frac{\mathrm{ions}}{\mathrm{cm}^2}$}&\multicolumn{1}{c}{[nm]}&&\multicolumn{1}{c}{$\cdot10^{16}~\frac{\mathrm{ions}}{\mathrm{cm}^2}$}\\
\hline \multirow{8}{*}{\begin{sideways}p$^+$\end{sideways}}
&   300 &   0.5 &   8.3 &   0.8 &   30.7    &   9.4 &   0.8 &   28.2    \\
&   400 &   0.8 &   12  &   1.5 &   23.7    &   10.7    &   1.5 &   17.1    \\
&   450 &   1.1 &   15.8    &   2.1 &   22.3    &   17.3    &   2.1 &   19.8    \\
&   500 &   1.3 &   17.7    &   2.4 &   21.8    &   19.6    &   2.5 &   18.8    \\
&   600 &   1.9 &   24.4    &   3.3 &   21.9    &   26.6    &   3.3 &   19.3    \\
&   700 &   2.6 &   31.7    &   4   &   23.5    &   35.5    &   4.1 &   20.8    \\
&   800 &   3.3 &   38.8    &   4.7 &   24.4    &   43  &   4.8 &   21.5    \\
&   1384    &   10.0    &   97.2    &   8.2 &   35.1    &   107.1   &   8.1 &   31.7    \\
&   3094    &   50.0    &   332.7   &   12.6    &   78.2    &   364.5   &   12.4    &   70.6    \\
\hline \multirow{7}{*}{\begin{sideways}He\end{sideways}}
&   300 &   1.9 &   14.4    &   20.1    &   2.1 &   15.5    &   19.4    &   1.9 \\
&   400 &   3.3 &   23.4    &   31.1    &   2.2 &   26.3    &   31.1    &   2.0 \\
&   450 &   4.2 &   29.1    &   36.9    &   2.3 &   33.6    &   36.6    &   2.2 \\
&   500 &   5.2 &   35.7    &   42.6    &   2.5 &   40.1    &   42.3    &   2.3 \\
&   600 &   7.5 &   50.5    &   53.3    &   2.8 &   56.5    &   52.8    &   2.6 \\
&   700 &   10.2    &   67.5    &   63.8    &   3.1 &   74.3    &   62.7    &   2.8 \\
&   800 &   13.3    &   84.9    &   72.7    &   3.5 &   95.5    &   71  &   3.2 \\
&   1553    &   50.0    &   263.4   &   116.2   &   6.7 &   292 &   113 &   6.2 \\
\hline
\end{tabular}
\end{center}
\end{table*}
The mineral densities have been taken from the webmineral
database\footnote{www.webmineral.com}. We determine the required ion
dose for complete amorphization of the dust grain material down to the
penetration depth of the incident ion by deriving the column density
of the lattice atoms. Further, we require a quantity for how many
displacements per lattice atom (dpa) are required to observe
amorphization. We set this number to be 2.5 based on the mineral
irradiation experiments for enstatite \citep{Jaeger:2003},
forsterite \citep{Bringa:2007} and olivine (\citealt{Demyk:2001};
\citealt{Carrez:2002}). However, this quantity remains inaccurate as
most of the above cited studies determined doses for He-irradiation
to fully amorphize the minerals varying within an order of
magnitude. \citet{Jaeger:2003} presented a dpa-value of 2.5 but the
lowest He-ion energy used was 50~keV. For protons, we assume the
same dpa ratio although this is hypothetical.

Table~\ref{si:SRIM} further shows that
$N_{\mathrm{Disp}}/\mathrm{ion}$ increases with increasing energy
but the required dose for amorphization increases as well. This can
be explained from the definition of the required dose. With higher
energies, the ions penetrate deeper into the dust grain and
consequently a thicker layer will be amorphized. This requires a
larger total number of displacements as more lattice atoms are
impacted.

We can therefore conclude from this calculation that the He-ions of
the present solar wind would amorphize the top $\sim$30~nm of the irradiated face of a
pyroxene dust grain and the required dose of
$\sim$2.3$\cdot10^{16}~\mathrm{ions}/\mathrm{cm}^2$ would be
accumulated within $\sim$50~years. To allow the full surface to be amorphized, the dust grain has to rotate with respect to the ion beam to allow for an isotropic irradiation, and the timescale has to be increased by a factor of a few. Obviously this indicates a very
efficient mechanism for amorphization and might be even more
efficient as we have ignored the irradiation by other ions.

On the other hand, this calculation assumes that all dust grains are
fully exposed to the radiation of the central object. This is obviously not true for
the majority of the observable grains in the disk atmosphere. The
ion flux gets extinct analogue to optical light due to absorption by
dust grains (self-shielding) and therefore, the amorphization
timescale depends strongly on the location of the grain within the
disk and the presence of vertical mixing of the disk material.
Therefore, the 50~years of irradiation time represent the timescale
for amorphization, based on the full flux. This number will increase
exponentially (or faster) the deeper the grains are located in the
disk. Consequently, our observations probe various timescales for
dust amorphization; conclusions depending on whether the presented
mechanisms are too efficient or too inefficient cannot be made at
the present time. Knowing the mean free path length along the
particle trajectories would enable more quantitative conclusions
about the amorphization timescale and potential with respect to
other dust processing mechanisms. Particles of higher energies could
also be considered by studying multiple scattering of ions with dust
grains. This would allow an amorphization of dust deeper in the
disk. For this purpose, dedicated modeling of the disk geometry and
grain size distribution is required. This will be addressed in
future studies.

For now, we can conclude that the dust amorphization by stellar wind
ions at the disk surface is sufficiently efficient so that this
mechanism might dominate the dust processing at the disk surface
layer.

\subsubsection{Optical properties and grain size dependency}

Since the dust grains are amorphized only at the surface layer, the
impact on the optical emissivity has to be investigated to
demonstrate the observability of the amorphization of circumstellar
dust by the low energetic stellar wind.

As an example, we calculate the optical mass absorption coefficients
of crystalline pyroxene, coated with a 30~nm thick layer of
amorphous pyroxene using Mie theory according to \citet{Bohren:1983}
for coated spheres. The calculation code in the appendix~B of
\citet{Bohren:1983} has been translated by \citet{Maetzler:2002}
into a Matlab script which we have used for this study. The optical
constants from \citet{Jaeger:1994} were used for crystalline
pyroxene while parameters for the amorphous pyroxene are taken from
\citet{Dorschner:1995}.

\begin{figure}[!ht]\begin{center}
\resizebox{\hsize}{!}{\includegraphics{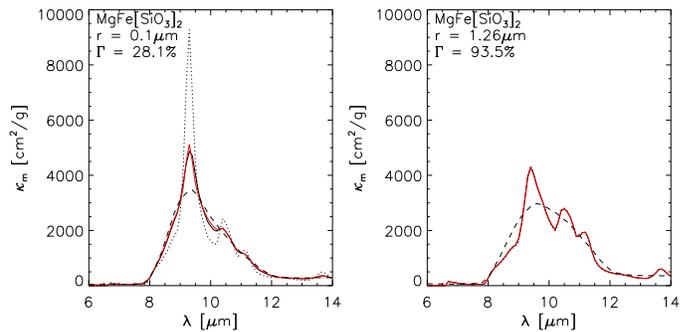}}
\caption{The thick, solid line corresponds to the mass absorption coefficient of a 2-layer spherical dust
grain with pyroxene stoichiometry and a radius of 0.1~$\mu$m (left) and 1.26~$\mu$m (right). The outer layer is amorphous with a thickness of 30~nm in both cases
and the core is crystalline. The dotted line corresponds to a pure
crystalline and the dashed line to a pure amorphous grain of the
same size and chemical composition. The red, thin solid line shows a
decomposition fit of the 2-layer absorption coefficient with a
purely amorphous and purely crystalline component and a crystalline
mass fraction of $\Gamma$=28.1\% and $\Gamma$=93.5\% is measured, respectively. Due to the high crystalline mass fraction for the larger grain, the dotted, red and thick lines overlap in the right panel.}\label{si:coated}
\end{center}\end{figure}

Figure~\ref{si:coated} shows the resulting mass absorption
coefficient within the wavelength range of interest for a dust grain of 0.1~$\mu$m and 1.26$\mu$m in radius, respectively.
Further, this plot indicates the two extremes of a purely
crystalline grain and a purely amorphous pyroxene grain. Finally, we calculate a linear combination of
the two extremes to fit the 2-layer case and get a contribution
ratio for the crystalline dust of $\Gamma$=28.1\% for the 0.1~$\mu$m grain and $\Gamma$=93.5\% for the 1.26~$\mu$m grain, where $\Gamma$
corresponds to the same definition as used for the decomposition
fits. This result reflects the mass ratio used for the coated sphere
calculation (29.1\% crystalline for the 0.1~$\mu$m grain and 93.0\% for the 1.26~$\mu$m grain) with relatively high accuracy.
Consequently, even if the dust grains are amorphized in the surface
layer only, the spectrally derived value for amorphization
corresponds to the mass ratio of crystalline and amorphous silicates
and conclusions about dust amorphization are valid even for
low-energy ions. 

However, the provided calculation assumes a purely crystalline dust grain. We expect to see a mixture of amorphous and crystalline dust regardless of the stellar wind and its impact on the dust structure. Consequently, the model described here delivers only  an additional component for the amorphous dust content and values for $\Gamma$ are expected to be lower than these limits.

Further, the results of the decompositional fits of our sample (see Table~\ref{si:fitresult}) suggest that a significant amount of the dust is present in larger grains and hence, the discussed mechanism might not be applicable. However, we repeated this study considering only small grains and no correlation between \Lx and $\Gamma$ was found. It is very likely that higher energetic ions are important and more abundant than in the present solar system. These ions  penetrate deeper into the dust grain and consequently, larger grains are amorphized. Therefore, the approach of using a monochromatic ion beam with solar wind properties is too elementary and has to be extended for more quantitative studies.

\section{Conclusions}\label{si:conclusions}

The previous calculations show that the amorphization of the surface layer of the
protoplanetary dust by the stellar wind is a viable scenario. We have
seen that a stellar wind with solar characteristics allows a very
fast amorphization of the upper layer of dust grains compared to the
longer timescales of dust processing during the star and disk
formation. For sub-micron grains, this process leads to an
observable structural change of the grain.

However, these processes are inefficient for micron-sized grains or
larger bodies as the penetration depth of the low-energetic ions are
too small to degrade the dust sufficiently. 

On the other hand, as
young stars are generally more active, we can assume that particle
fluxes and energies are generally higher than in the present solar
wind and consequently, the described processes may be even more
efficient. Further, the solar irradiation is not only composed of
the solar wind but a great variety of additional particles at
various energies and fluxes. Since our knowledge of these particle
fluxes ($E\la$10~MeV) in the early time of the solar history is very
poor, and since particle irradiation in other systems are
unmeasurable, the assessment of the amorphization efficiency remains
very inaccurate and difficult. Other irradiation sources than the
star, such as shocks from stellar winds in the protoplanetary gas
disk or from accretion or jets confuse the overall picture further.
Recrystallization due to thermal annealing of warmer dust grains
(see, e.g., \citealt{Djouadi:2005}) may inhibit the amorphization
process and needs to be taken into account for quantitative models.

Nonetheless, we observe a correlation for 20 objects between the X-ray luminosity as well as the X-ray luminosity multiplied with the X-ray hardness
and the crystalline mass fraction of the atmospheric dust of the
inner protoplanetary disk for objects within an age range of approx. 1 to 4.5~Myr. We interpret this as an indicator of
high-energy processes in the central object. We therefore postulate
degenerative processes for the crystalline structure of the dust by
ionic irradiation. Although we cannot observe the stellar wind
directly, its flux and/or speed might be related to the X-ray
luminosity, as shown for low X-ray fluxes by \citet{Wood:2005} by
comparing stellar mass loss rates with X-ray surface fluxes.

Self-shielding effects of the dust disk play an important role for
the time scale and the overall potential of amorphization.
Consequently, the disk geometry and the dust density distribution
have to be taken into account for studying evolutionary effects.
Further, self-shielding could explain why the correlation diminishes
if we include objects younger than \ap 1~Myr. These objects might
have well mixed disks even in the upper disk atmosphere where dust
settling has poorly progressed; consequently, the self-shielding is
more relevant.

It is less clear why the correlation worsens if we include objects
older than \ap 4.5~Myr. It could be related to a statistical problem as
our sample does not include many objects at such ages. Further
physical effects may also play a role on longer time scales.
Possibilities include radial dust mixing or the reproduction of
crystalline dust material in the inner disk region.

It would be interesting to verify if this correlation is present in
other star forming regions and for a larger sample of objects.
The extension of this study to Herbig Ae/Be systems could
provide different aspects of this process due to the shorter
evolutionary time scales of the central object.

\begin{acknowledgements}
The authors would like to thank Jean-Charles Augereau for reviewing this paper and providing many helpful comments and suggestions. This work is based in part on archival data obtained with the
Spitzer Space Telescope, which is operated by the Jet Propulsion
Laboratory, California Institute of Technology under a contract with
NASA. This research is based on observations obtained with
XMM-Newton, an ESA science mission with instruments and
contributions directly funded by ESA member states and the USA
(NASA). The OMNI data were obtained from the GSFC/SPDF OMNIWeb
interface at http://omniweb.gsfc.nasa.gov. The mineral data were
obtained from http://webmineral.com. M.A. and C.B.S. acknowledge
support from Swiss NSF grant PP002--110504. 
\end{acknowledgements}

\bibliographystyle{aa}
\bibliography{references}
\begin{appendix}
\section{Resulting fit parameters}\label{si:fitresult}
\begin{landscape}
\begin{center}
\begin{table}[!h]
\caption{Resulting fit parameters for the decomposition analysis of
the 10~$\mu$m silicate feature. Dashes indicate that the lower boundary
has been reached and the component proportion has been set to 0.}
\begin{tabular}{l@{   }c@{ }c@{ }c@{ }c@{ }c@{ }c@{ }c@{ }c@{ }c@{ }c@{ }c} \hline \hline
Name        &   \multicolumn{1}{c}{$q_\mathrm{disk}$}                   &   \multicolumn{2}{c}{MgFeSiO$_4$}                                         &   \multicolumn{2}{c}{MgFe[SiO$_3$]$_2$}                                           &   \multicolumn{2}{c}{Mg$_2$SiO$_4$}                                           &   \multicolumn{2}{c}{MgSiO$_3$}                                           &   \multicolumn{2}{c}{SiO$_2$}                                         \\
    &   \multicolumn{1}{c}{$\times 10^{-1}$}                    &   Small [\%]                  &   \multicolumn{1}{c}{Large [\%]}                  &   \multicolumn{1}{c}{Small [\%]}                  &   \multicolumn{1}{c}{Large [\%]}                  &   \multicolumn{1}{c}{Small [\%]}                  &   \multicolumn{1}{c}{Large [\%]}                  &   \multicolumn{1}{c}{Small [\%]}                  &   \multicolumn{1}{c}{Large [\%]}                  &   \multicolumn{1}{c}{Small [\%]}                  &   \multicolumn{1}{c}{Large [\%]}                  \\
\hline \\
04187+1927  &$  -6.32   _{  -0.01   }^{+    0.01    }$&$    -                   $&$ -                   $&$ 12.8    _{- 2.0 }^{+    0.1 }$&$    53.7    _{- 0.1 }^{+    3.1 }$&$    -                   $&$ 10.4    _{- 0.5 }^{+    0.1 }$&$    1.9 _{- 0.3 }^{+    0.0 }$&$    15.7    _{- 0.1 }^{+    0.1 }$&$    4.8 _{- 0.0 }^{+    0.1 }$&$    0.7 _{- 0.0 }^{+    0.1 }$\\
04303+2240  &$  -5.84   _{  -0.08   }^{+    0.09    }$&$    -                   $&$ 66.4    _{- 0.8 }^{+    0.8 }$&$    -                   $&$ -                   $&$ 7.1 _{- 0.2 }^{+    0.2 }$&$    18.2    _{- 0.6 }^{+    0.6 }$&$    -                   $&$ -                   $&$ -                   $&$ 8.4 _{- 0.0 }^{+    0.0 }$\\
04385+2550  &$  -4.35   _{  -0.08   }^{+    0.07    }$&$    32.4    _{- 0.2 }^{+    0.2 }$&$    47.6    _{- 3.6 }^{+    4.6 }$&$    -                   $&$ 18.3    _{- 4.0 }^{+    3.7 }$&$    0.0 _{- 0.0 }^{+    0.0 }$&$    1.7 _{- 0.1 }^{+    0.2 }$&$    -                   $&$ -                   $&$ -                   $&$ -                   $\\
AA Tau      &$  -6.07   _{  -0.21   }^{+    0.16    }$&$    29.0    _{- 1.4 }^{+    1.5 }$&$    -                   $&$ -                   $&$ 55.4    _{- 3.3 }^{+    2.6 }$&$    0.0 _{- 0.0 }^{+    0.0 }$&$    5.9 _{- 0.7 }^{+    0.7 }$&$    -                   $&$ 7.3 _{- 0.6 }^{+    0.8 }$&$    -                   $&$ 2.4 _{- 0.1 }^{+    0.1 }$\\
BP Tau      &$  -5.50   _{  -0.36   }^{+    0.19    }$&$    44.5    _{- 2.1 }^{+    1.2 }$&$    0.0 _{- 0.0 }^{+    5.3 }$&$    9.9 _{- 0.9 }^{+    1.6 }$&$    37.0    _{- 6.6 }^{+    1.1 }$&$    1.2 _{- 0.0 }^{+    0.0 }$&$    4.4 _{- 0.3 }^{+    0.2 }$&$    2.6 _{- 0.4 }^{+    0.5 }$&$    -                   $&$ -                   $&$ 0.6 _{- 0.5 }^{+    0.2 }$\\
CI Tau      &$  -5.29   _{  -0.11   }^{+    0.11    }$&$    21.0    _{- 0.6 }^{+    0.7 }$&$    -                   $&$ -                   $&$ 63.7    _{- 1.7 }^{+    1.7 }$&$    2.1 _{- 0.1 }^{+    0.1 }$&$    6.1 _{- 0.5 }^{+    0.5 }$&$    -                   $&$ 3.4 _{- 0.4 }^{+    0.4 }$&$    -                   $&$ 3.8 _{- 0.0 }^{+    0.0 }$\\
CoKu Tau/3  &$  -5.51   _{  -0.17   }^{+    0.15    }$&$    -                   $&$ -                   $&$ -                   $&$ 77.3    _{- 1.4 }^{+    1.3 }$&$    3.4 _{- 0.1 }^{+    0.1 }$&$    9.6 _{- 0.7 }^{+    0.8 }$&$    -                   $&$ 6.7 _{- 0.7 }^{+    0.8 }$&$    -                   $&$ 3.0 _{- 0.2 }^{+    0.2 }$\\
CW Tau      &$  -7.50   _{  -0.14   }^{+    0.05    }$&$    0.0 _{- 0.0 }^{+    21.7    }$&$    -                   $&$ -                   $&$ 79.3    _{- 27.5    }^{+    0.6 }$&$    3.8 _{- 0.0 }^{+    0.2 }$&$    4.8 _{- 0.3 }^{+    2.6 }$&$    -                   $&$ 7.3 _{- 0.3 }^{+    3.7 }$&$    -                   $&$ 4.8 _{- 0.8 }^{+    0.0 }$\\
CY Tau      &$  -5.43   _{  -0.60   }^{+    0.49    }$&$    -                   $&$ -                   $&$ 40.9    _{- 0.3 }^{+    1.9 }$&$    22.3    _{- 8.2 }^{+    4.6 }$&$    3.4 _{- 0.0 }^{+    0.1 }$&$    14.0    _{- 1.4 }^{+    2.2 }$&$    0.0 _{- 0.0 }^{+    0.5 }$&$    11.0    _{- 2.8 }^{+    3.3 }$&$    0.4 _{- 0.3 }^{+    0.2 }$&$    8.2 _{- 0.2 }^{+    0.2 }$\\
CZ Tau      &$  -5.10   _{  -0.02   }^{+    0.01    }$&$    15.3    _{- 0.6 }^{+    1.0 }$&$    -                   $&$ 41.7    _{- 1.0 }^{+    0.6 }$&$    40.8    _{- 0.0 }^{+    0.1 }$&$    0.8 _{- 0.1 }^{+    0.0 }$&$    -                   $&$ 1.4 _{- 0.1 }^{+    0.1 }$&$    -                   $&$ -                   $&$ 0.0 _{- 0.0 }^{+    0.0 }$\\
DD Tau      &$  -5.59   _{  -0.29   }^{+    0.20    }$&$    0.0 _{- 0.0 }^{+    7.6 }$&$    0.0 _{- 0.0 }^{+    24.4    }$&$    4.9 _{- 2.3 }^{+    8.6 }$&$    71.2    _{- 25.4    }^{+    7.1 }$&$    1.5 _{- 0.0 }^{+    0.1 }$&$    5.5 _{- 0.2 }^{+    0.8 }$&$    1.7 _{- 1.2 }^{+    0.1 }$&$    7.2 _{- 0.2 }^{+    1.1 }$&$    -                   $&$ 1.9 _{- 1.3 }^{+    0.1 }$\\
DK Tau      &$  -4.78   _{  -0.08   }^{+    0.09    }$&$    -                   $&$ 0.0 _{- 0.0 }^{+    1.4 }$&$    1.7 _{- 0.7 }^{+    1.4 }$&$    80.6    _{- 1.1 }^{+    0.5 }$&$    3.5 _{- 0.1 }^{+    0.1 }$&$    3.4 _{- 0.2 }^{+    0.1 }$&$    7.1 _{- 0.3 }^{+    0.2 }$&$    -                   $&$ 0.3 _{- 0.0 }^{+    0.0 }$&$    3.0 _{- 0.1 }^{+    0.0 }$\\
DN Tau      &$  -5.74   _{  -0.14   }^{+    0.20    }$&$    -                   $&$ -                   $&$ 12.1    _{- 0.6 }^{+    0.6 }$&$    43.8    _{- 3.1 }^{+    4.3 }$&$    3.0 _{- 0.0 }^{+    0.0 }$&$    14.7    _{- 0.9 }^{+    0.7 }$&$    -                   $&$ 17.7    _{- 2.3 }^{+    2.2 }$&$    1.1 _{- 0.0 }^{+    0.1 }$&$    7.8 _{- 0.0 }^{+    0.0 }$\\
FM Tau      &$  -4.81   _{  -0.14   }^{+    0.17    }$&$    57.1    _{- 0.5 }^{+    0.4 }$&$    13.1    _{- 3.6 }^{+    3.5 }$&$    -                   $&$ 28.8    _{- 3.8 }^{+    3.7 }$&$    0.8 _{- 0.0 }^{+    0.0 }$&$    -                   $&$ 0.0 _{- 0.0 }^{+    0.3 }$&$    -                   $&$ -                   $&$ 0.2 _{- 0.2 }^{+    0.4 }$\\
FO Tau      &$  -5.65   _{  -0.76   }^{+    0.60    }$&$    -                   $&$ -                   $&$ 25.2    _{- 0.5 }^{+    4.6 }$&$    50.3    _{- 13.5    }^{+    5.9 }$&$    2.9 _{- 0.1 }^{+    0.1 }$&$    2.8 _{- 0.8 }^{+    1.8 }$&$    0.0 _{- 0.0 }^{+    0.4 }$&$    17.2    _{- 5.9 }^{+    8.3 }$&$    1.2 _{- 0.4 }^{+    0.1 }$&$    0.7 _{- 0.7 }^{+    1.0 }$\\
FQ Tau      &$  -8.11   _{  -0.06   }^{+    0.04    }$&$    -                   $&$ -                   $&$ -                   $&$ 49.3    _{- 1.6 }^{+    0.6 }$&$    -                   $&$ 17.7    _{- 0.4 }^{+    1.1 }$&$    2.5 _{- 0.0 }^{+    0.1 }$&$    24.8    _{- 0.2 }^{+    0.3 }$&$    0.1 _{- 0.1 }^{+    0.0 }$&$    5.6 _{- 0.1 }^{+    0.3 }$\\
FS Tau      &$  -5.21   _{  -0.06   }^{+    0.08    }$&$    20.2    _{- 0.5 }^{+    0.4 }$&$    56.4    _{- 1.8 }^{+    1.4 }$&$    -                   $&$ 22.3    _{- 1.9 }^{+    2.5 }$&$    1.1 _{- 0.1 }^{+    0.1 }$&$    -                   $&$ 0.0 _{- 0.0 }^{+    0.1 }$&$    -                   $&$ -                   $&$ 0.0 _{- 0.0 }^{+    0.0 }$\\
FV Tau      &$  -4.31   _{  -0.06   }^{+    0.04    }$&$    -                   $&$ 98.5    _{- 0.5 }^{+    0.4 }$&$    -                   $&$ -                   $&$ 0.3 _{- 0.0 }^{+    0.0 }$&$    0.5 _{- 0.4 }^{+    0.6 }$&$    -                   $&$ -                   $&$ -                   $&$ 0.7 _{- 0.1 }^{+    0.1 }$\\
FX Tau      &$  -4.71   _{  -0.09   }^{+    0.08    }$&$    40.5    _{- 0.5 }^{+    0.6 }$&$    -                   $&$ -                   $&$ 54.9    _{- 1.4 }^{+    1.2 }$&$    2.4 _{- 0.1 }^{+    0.1 }$&$    0.6 _{- 0.4 }^{+    0.6 }$&$    1.5 _{- 0.2 }^{+    0.3 }$&$    -                   $&$ -                   $&$ 0.2 _{- 0.1 }^{+    0.1 }$\\
FZ Tau      &$  -4.94   _{  -0.15   }^{+    0.15    }$&$    -                   $&$ 48.9    _{- 7.6 }^{+    6.0 }$&$    -                   $&$ 8.0 _{- 8.0 }^{+    8.6 }$&$    4.2 _{- 0.0 }^{+    0.0 }$&$    11.9    _{- 0.0 }^{+    0.1 }$&$    13.2    _{- 0.3 }^{+    0.2 }$&$    6.8 _{- 1.0 }^{+    1.0 }$&$    5.0 _{- 0.2 }^{+    0.2 }$&$    2.1 _{- 0.5 }^{+    0.5 }$\\
GH Tau      &$  -6.64   _{  -0.41   }^{+    0.37    }$&$    -                   $&$ -                   $&$ 0.0 _{- 0.0 }^{+    1.1 }$&$    65.1    _{- 3.7 }^{+    2.2 }$&$    0.1 _{- 0.1 }^{+    0.1 }$&$    12.3    _{- 1.0 }^{+    1.5 }$&$    -                   $&$ 20.6    _{- 2.5 }^{+    3.1 }$&$    1.2 _{- 0.0 }^{+    0.0 }$&$    0.8 _{- 0.3 }^{+    0.2 }$\\
GI Tau      &$  -5.73   _{  -0.19   }^{+    0.13    }$&$    45.1    _{- 1.0 }^{+    1.3 }$&$    -                   $&$ 0.0 _{- 0.0 }^{+    1.1 }$&$    44.5    _{- 2.3 }^{+    0.3 }$&$    1.4 _{- 0.0 }^{+    0.1 }$&$    2.9 _{- 0.1 }^{+    0.6 }$&$    0.1 _{- 0.1 }^{+    0.0 }$&$    3.9 _{- 0.2 }^{+    0.6 }$&$    -                   $&$ 2.1 _{- 0.2 }^{+    0.1 }$\\
GK Tau      &$  -4.61   _{  -0.10   }^{+    0.09    }$&$    43.2    _{- 0.6 }^{+    0.7 }$&$    -                   $&$ 15.2    _{- 1.6 }^{+    1.4 }$&$    34.0    _{- 0.7 }^{+    0.8 }$&$    3.2 _{- 0.0 }^{+    0.0 }$&$    0.4 _{- 0.0 }^{+    0.0 }$&$    2.6 _{- 0.1 }^{+    0.1 }$&$    -                   $&$ -                   $&$ 1.4 _{- 0.0 }^{+    0.0 }$\\
GN Tau      &$  -5.11   _{  -0.18   }^{+    0.10    }$&$    -                   $&$ -                   $&$ -                   $&$ 82.8    _{- 1.7 }^{+    0.9 }$&$    3.5 _{- 0.1 }^{+    0.2 }$&$    3.0 _{- 0.4 }^{+    0.8 }$&$    -                   $&$ 5.5 _{- 0.4 }^{+    0.8 }$&$    2.7 _{- 0.1 }^{+    0.0 }$&$    2.5 _{- 0.0 }^{+    0.0 }$\\
GO Tau      &$  -5.40   _{  -0.28   }^{+    0.36    }$&$    -                   $&$ -                   $&$ 0.3 _{- 0.3 }^{+    4.3 }$&$    82.1    _{- 2.3 }^{+    1.4 }$&$    0.8 _{- 0.1 }^{+    0.1 }$&$    1.7 _{- 0.6 }^{+    1.4 }$&$    -                   $&$ 13.7    _{- 2.2 }^{+    1.9 }$&$    -                   $&$ -                   $\\
Haro 6-13   &$  -3.81   _{  -0.03   }^{+    0.03    }$&$    11.8    _{- 0.0 }^{+    0.0 }$&$    61.7    _{- 1.6 }^{+    1.9 }$&$    -                   $&$ 25.0    _{- 2.0 }^{+    1.7 }$&$    1.4 _{- 0.1 }^{+    0.1 }$&$    -                   $&$ -                   $&$ -                   $&$ -                   $&$ -                   $\\
Haro 6-28   &$  -6.97   _{  -0.06   }^{+    0.03    }$&$    -                   $&$ -                   $&$ -                   $&$ 69.7    _{- 1.0 }^{+    0.2 }$&$    5.6 _{- 0.0 }^{+    0.3 }$&$    -                   $&$ 15.4    _{- 0.1 }^{+    0.5 }$&$    8.0 _{- 0.1 }^{+    0.0 }$&$    0.5 _{- 0.3 }^{+    0.0 }$&$    0.7 _{- 0.1 }^{+    0.6 }$\\
HK Tau      &$  -4.75   _{- 0.00    }^{+    0.58    }$&$    8.3 _{- 0.0 }^{+    2.9 }$&$    34.1    _{- 34.1    }^{+    0.0 }$&$    -                   $&$ 16.8    _{- 0.0 }^{+    39.3    }$&$    1.2 _{- 0.0 }^{+    1.0 }$&$    17.2    _{- 0.0 }^{+    2.4 }$&$    13.7    _{- 0.0 }^{+    0.2 }$&$    8.8 _{- 8.8 }^{+    0.0 }$&$    -                   $&$ 0.0 _{- 0.0 }^{+    1.8 }$\\
HO Tau      &$  -4.38   _{  -0.38   }^{+    0.18    }$&$    34.9    _{- 0.7 }^{+    0.2 }$&$    0.0 _{- 0.0 }^{+    11.1    }$&$    -                   $&$ 59.7    _{- 9.3 }^{+    0.7 }$&$    2.1 _{- 0.3 }^{+    0.2 }$&$    0.9 _{- 0.4 }^{+    0.5 }$&$    -                   $&$ -                   $&$ 0.2 _{- 0.1 }^{+    0.1 }$&$    2.0 _{- 0.7 }^{+    0.1 }$\\
HP Tau      &$  -4.82   _{  -0.08   }^{+    0.11    }$&$    47.1    _{- 0.2 }^{+    0.4 }$&$    3.5 _{- 3.5 }^{+    2.2 }$&$    -                   $&$ 45.6    _{- 2.3 }^{+    3.3 }$&$    2.5 _{- 0.0 }^{+    0.1 }$&$    -                   $&$ -                   $&$ -                   $&$ -                   $&$ 1.7 _{- 0.1 }^{+    0.2 }$\\
IQ Tau      &$  -5.41   _{  -0.28   }^{+    0.33    }$&$    12.0    _{- 0.7 }^{+    1.0 }$&$    60.4    _{- 7.5 }^{+    5.8 }$&$    -                   $&$ 20.8    _{- 5.2 }^{+    7.0 }$&$    2.6 _{- 0.0 }^{+    0.0 }$&$    -                   $&$ -                   $&$ 1.3 _{- 0.7 }^{+    0.6 }$&$    -                   $&$ 2.9 _{- 0.3 }^{+    0.3 }$\\
IS Tau      &$  -4.64   _{  -0.19   }^{+    0.18    }$&$    -                   $&$ -                   $&$ -                   $&$ 71.0    _{- 2.6 }^{+    2.6 }$&$    4.8 _{- 0.3 }^{+    0.3 }$&$    8.3 _{- 1.1 }^{+    1.2 }$&$    3.9 _{- 0.4 }^{+    0.4 }$&$    4.4 _{- 0.5 }^{+    0.6 }$&$    2.8 _{- 0.1 }^{+    0.1 }$&$    4.9 _{- 0.2 }^{+    0.2 }$\\
IT Tau      &$  -5.58   _{  -0.29   }^{+    0.28    }$&$    -                   $&$ -                   $&$ -                   $&$ 67.7    _{- 7.4 }^{+    4.7 }$&$    -                   $&$ 3.1 _{- 0.5 }^{+    1.8 }$&$    1.0 _{- 1.0 }^{+    0.3 }$&$    28.0    _{- 2.8 }^{+    6.1 }$&$    0.0 _{- 0.0 }^{+    0.4 }$&$    -                   $\\
MHO-3       &$  -3.43   _{  -0.02   }^{+    0.02    }$&$    39.9    _{- 0.0 }^{+    0.0 }$&$    37.4    _{- 1.1 }^{+    1.1 }$&$    -                   $&$ 21.2    _{- 1.1 }^{+    1.2 }$&$    1.1 _{- 0.1 }^{+    0.1 }$&$    0.4 _{- 0.0 }^{+    0.0 }$&$    -                   $&$ -                   $&$ -                   $&$ -                   $\\
RY Tau      &$  -5.40   _{  -0.11   }^{+    0.05    }$&$    47.5    _{- 0.4 }^{+    0.5 }$&$    -                   $&$ -                   $&$ 45.5    _{- 1.5 }^{+    0.7 }$&$    2.2 _{- 0.1 }^{+    0.1 }$&$    2.7 _{- 0.2 }^{+    0.3 }$&$    0.9 _{- 0.1 }^{+    0.1 }$&$    -                   $&$ -                   $&$ 1.2 _{- 0.0 }^{+    0.0 }$\\
UZ Tau/e    &$  -6.30   _{  -1.31   }^{+    0.16    }$&$    11.3    _{- 11.3    }^{+    0.4 }$&$    44.4    _{- 2.1 }^{+    37.5    }$&$    0.0 _{- 0.0 }^{+    8.1 }$&$    32.8    _{- 32.8    }^{+    2.5 }$&$    0.9 _{- 0.1 }^{+    0.0 }$&$    2.8 _{- 2.0 }^{+    0.1 }$&$    -                   $&$ 4.7 _{- 0.3 }^{+    0.9 }$&$    -                   $&$ 3.1 _{- 1.1 }^{+    0.1 }$\\
V410 Anon 13&$  -6.93   _{  -0.09   }^{+    0.35    }$&$    -                   $&$ -                   $&$ 11.8    _{- 2.7 }^{+    0.9 }$&$    30.0    _{- 3.9 }^{+    5.6 }$&$    14.6    _{- 0.8 }^{+    0.5 }$&$    16.4    _{- 1.3 }^{+    0.8 }$&$    23.3    _{- 2.2 }^{+    1.4 }$&$    3.2 _{- 0.1 }^{+    0.7 }$&$    1.2 _{- 0.2 }^{+    0.2 }$&$    -                   $\\
V710 Tau    &$  -3.98   _{  -1.02   }^{+    0.17    }$&$    -                   $&$ 0.0 _{- 0.0 }^{+    40.5    }$&$    23.3    _{- 2.7 }^{+    3.0 }$&$    61.1    _{- 46.9    }^{+    1.8 }$&$    0.4 _{- 0.4 }^{+    0.1 }$&$    4.3 _{- 0.3 }^{+    0.2 }$&$    0.0 _{- 0.0 }^{+    0.8 }$&$    6.6 _{- 1.0 }^{+    4.2 }$&$    -                   $&$ 4.9 _{- 2.1 }^{+    0.2 }$\\
V773 Tau    &$  -4.56   _{  -0.22   }^{+    0.00    }$&$    -                   $&$ 69.8    _{- 0.0 }^{+    2.4 }$&$    -                   $&$ 7.9 _{- 7.9 }^{+    0.0 }$&$    5.3 _{- 0.0 }^{+    0.7 }$&$    6.3 _{- 0.0 }^{+    0.9 }$&$    -                   $&$ 10.8    _{- 0.0 }^{+    2.0 }$&$    -                   $&$ -                   $\\
V807 Tau    &$  -7.26   _{  -0.39   }^{+    0.47    }$&$    -                   $&$ -                   $&$ -                   $&$ 67.9    _{- 4.6 }^{+    3.6 }$&$    5.3 _{- 0.3 }^{+    0.4 }$&$    10.4    _{- 1.3 }^{+    1.9 }$&$    -                   $&$ 16.3    _{- 3.1 }^{+    2.8 }$&$    -                   $&$ 0.0 _{- 0.0 }^{+    1.3 }$\\
V955 Tau    &$  -5.53   _{  -0.17   }^{+    0.20    }$&$    -                   $&$ -                   $&$ -                   $&$ 63.5    _{- 2.2 }^{+    2.7 }$&$    7.7 _{- 0.3 }^{+    0.2 }$&$    8.5 _{- 0.8 }^{+    0.6 }$&$    2.5 _{- 0.0 }^{+    0.0 }$&$    8.2 _{- 1.5 }^{+    1.3 }$&$    3.1 _{- 0.0 }^{+    0.0 }$&$    6.4 _{- 0.0 }^{+    0.0 }$\\
XZ Tau      &$  -5.27   _{  -0.04   }^{+    0.05    }$&$    -                   $&$ 76.5    _{- 1.8 }^{+    1.6 }$&$    -                   $&$ -                   $&$ 4.9 _{- 0.2 }^{+    0.4 }$&$    13.4    _{- 1.1 }^{+    1.1 }$&$    5.2 _{- 0.6 }^{+    0.5 }$&$    -                   $&$ 0.0 _{- 0.0 }^{+    0.2 }$&$    0.0 _{- 0.0 }^{+    0.2 }$\\
\hline
\end{tabular}
\end{table}
\end{center}
\end{landscape}
\end{appendix}

\begin{appendix}
\onecolumn
\section{Spectra of the complete Data Sample}
\begin{longtable}{c c c c}
\caption{IRS spectra (black line) and fit (red line) including the
corresponding uncertainties (gray lines) of the full dataset. The
continuum background used for the fit function is shown in blue. In
the lower part of the figures, the resulting $\chi^2$, multiplied by
the sign of the deviation, are shown.}\\

\includegraphics[width=4.2cm]{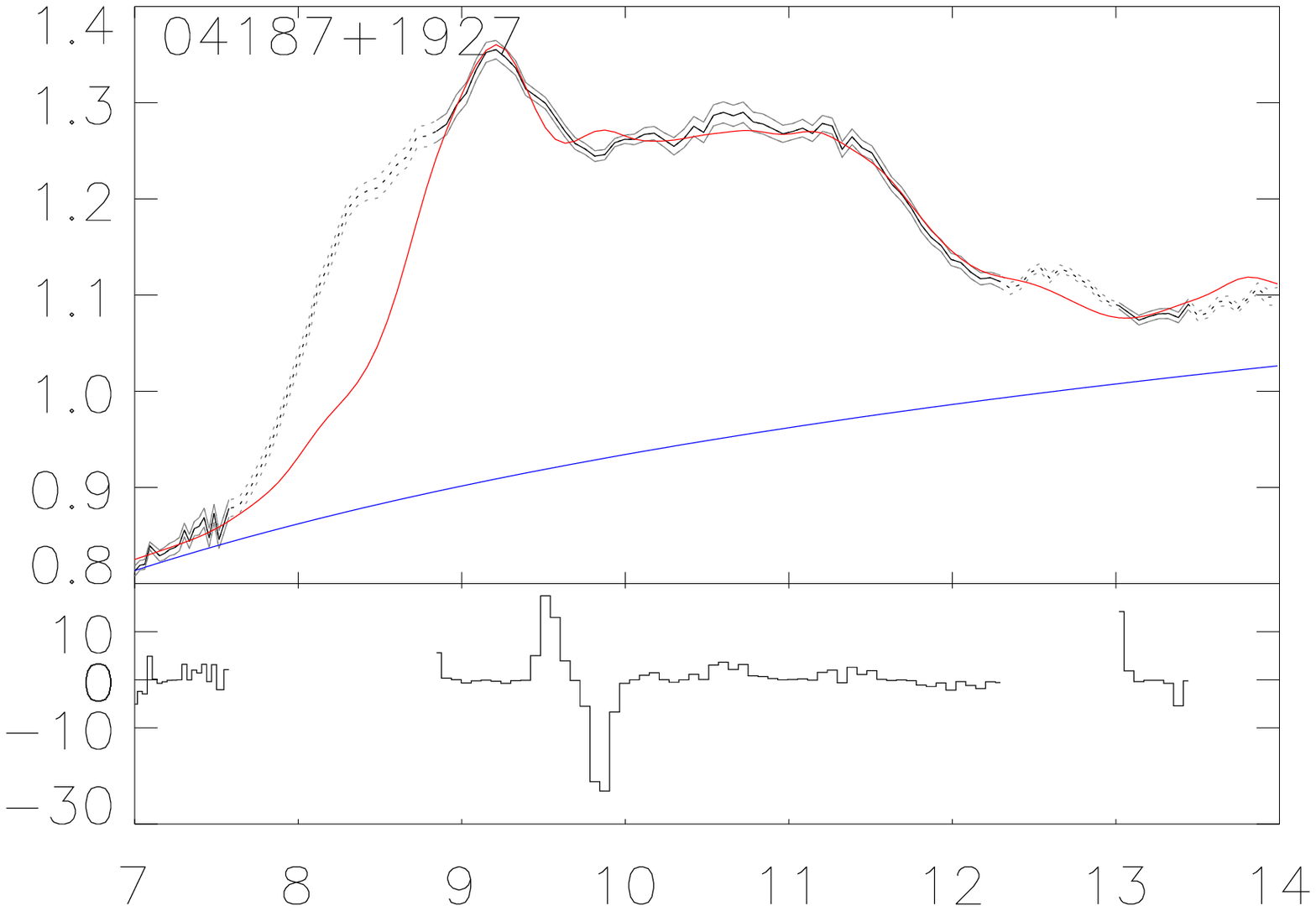}&\includegraphics[width=4.2cm]{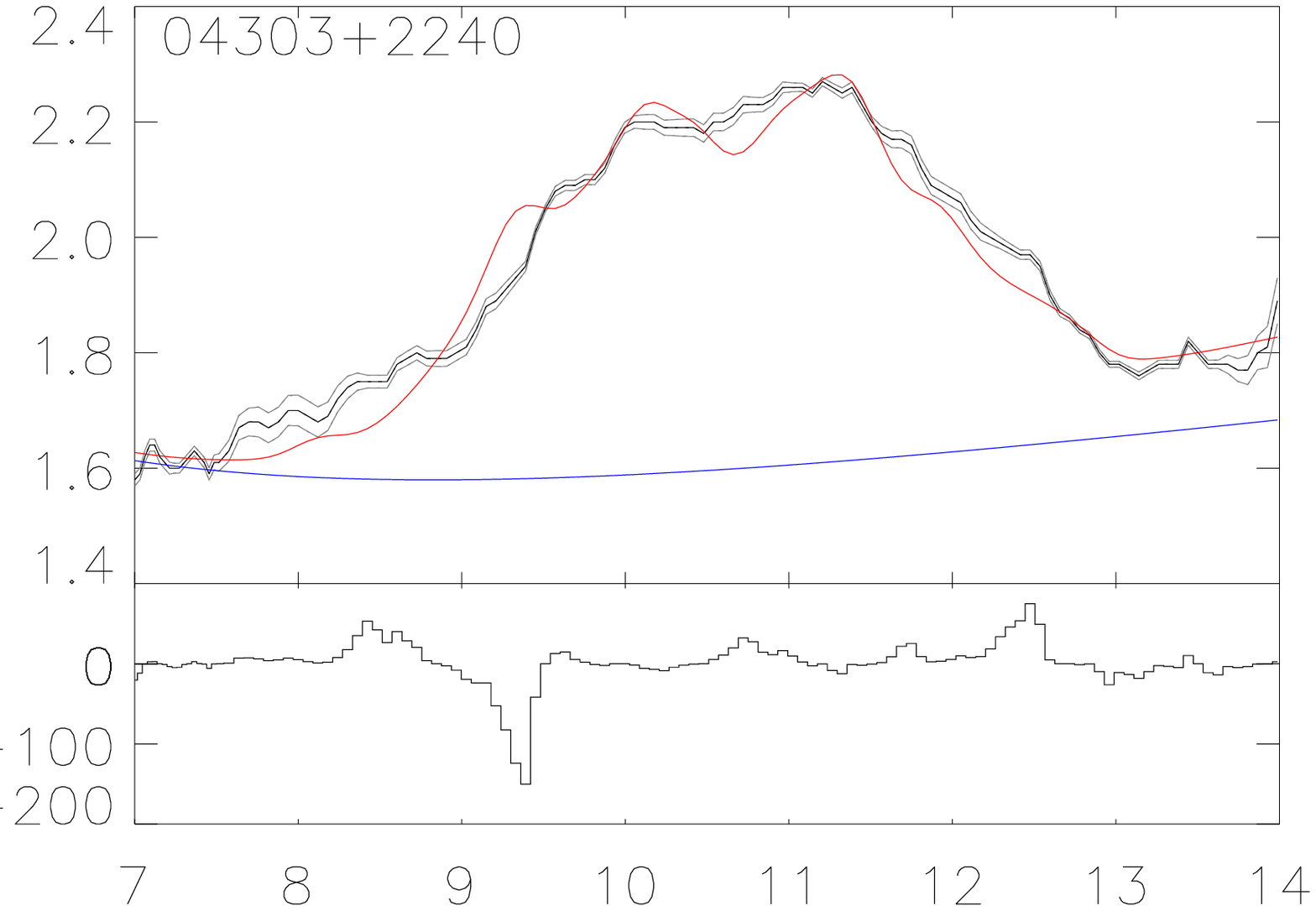}&
\includegraphics[width=4.2cm]{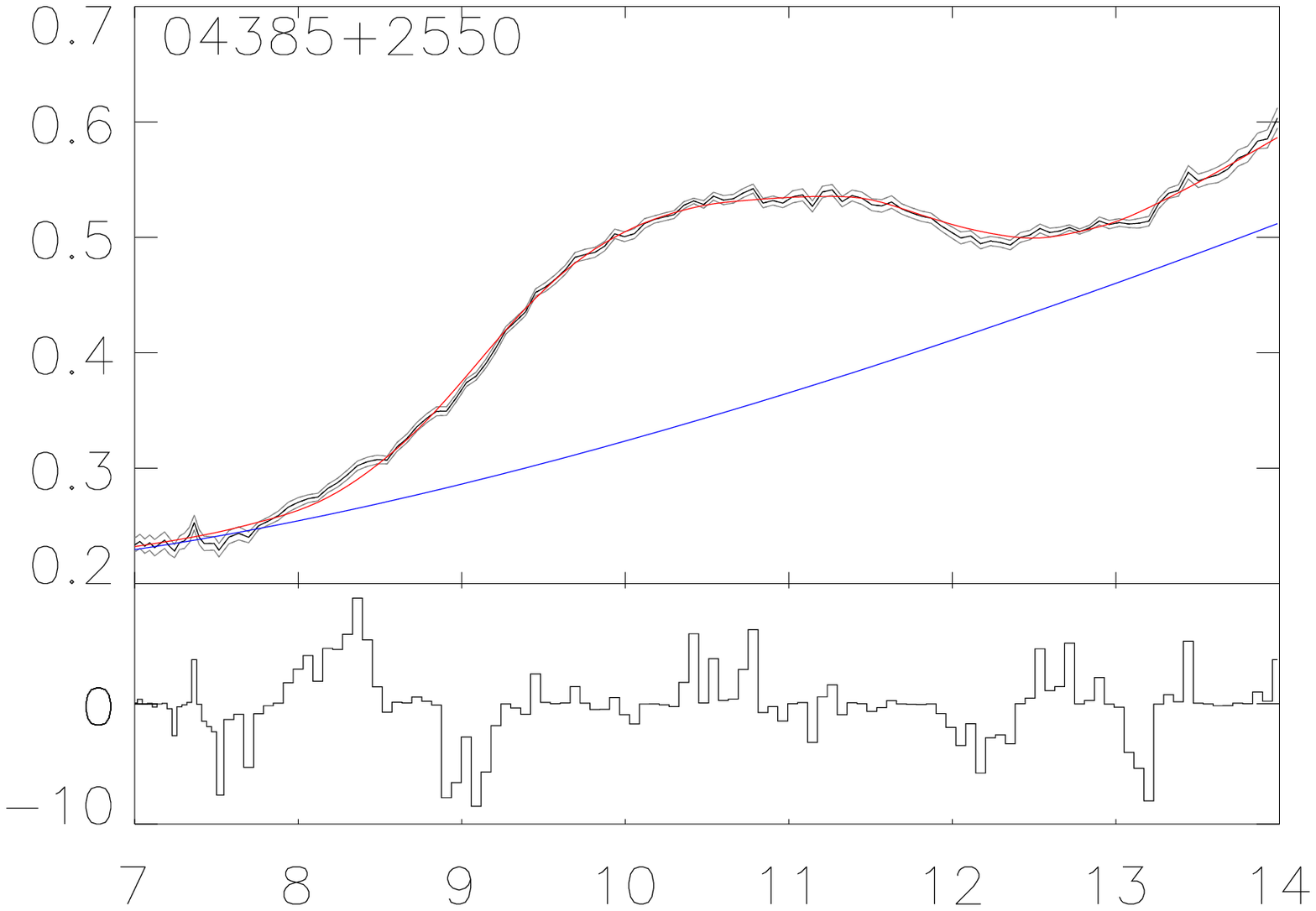}&\includegraphics[width=4.2cm]{sifitres09.eps}\\
\\
\includegraphics[width=4.2cm]{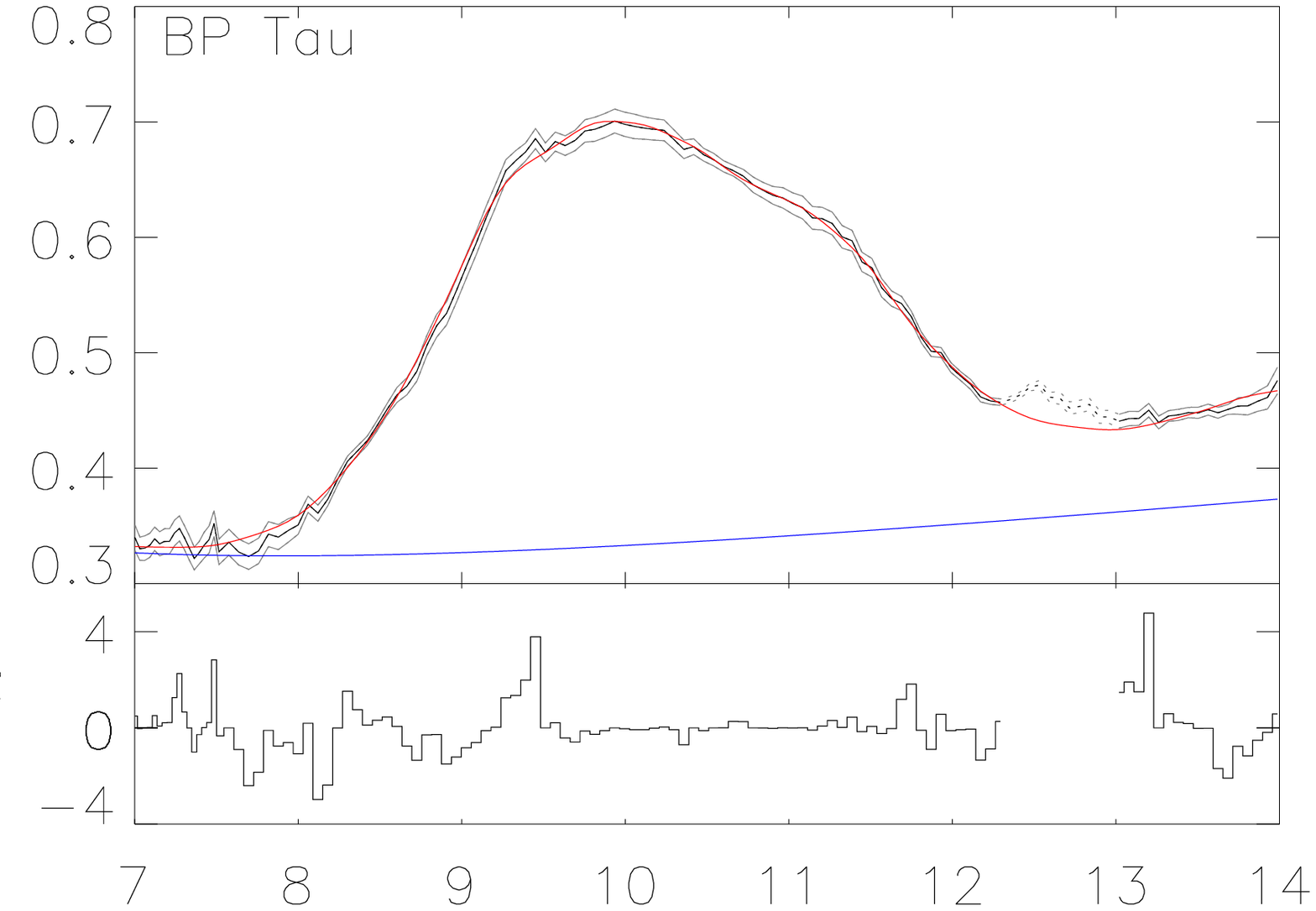}&\includegraphics[width=4.2cm]{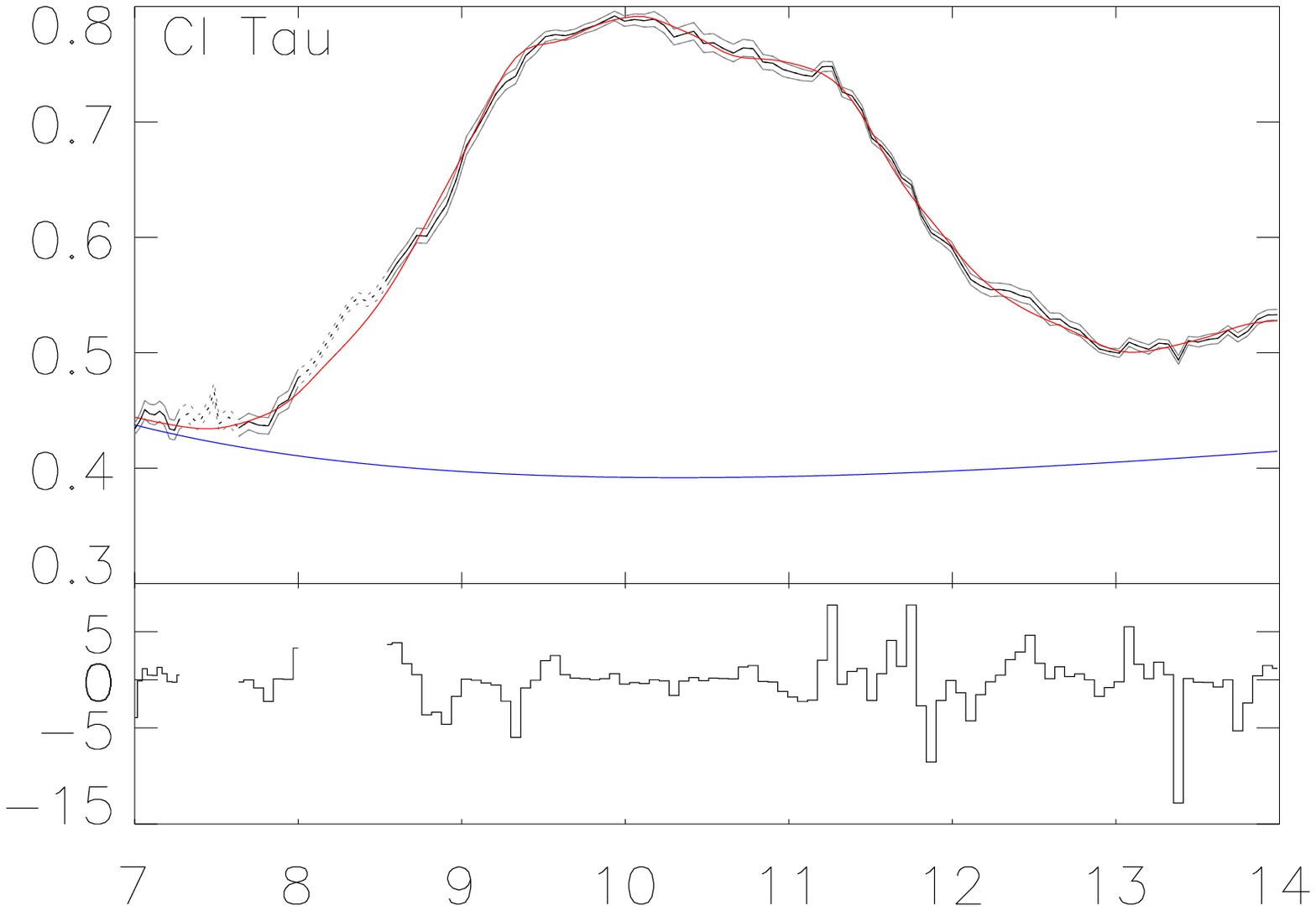}&
\includegraphics[width=4.2cm]{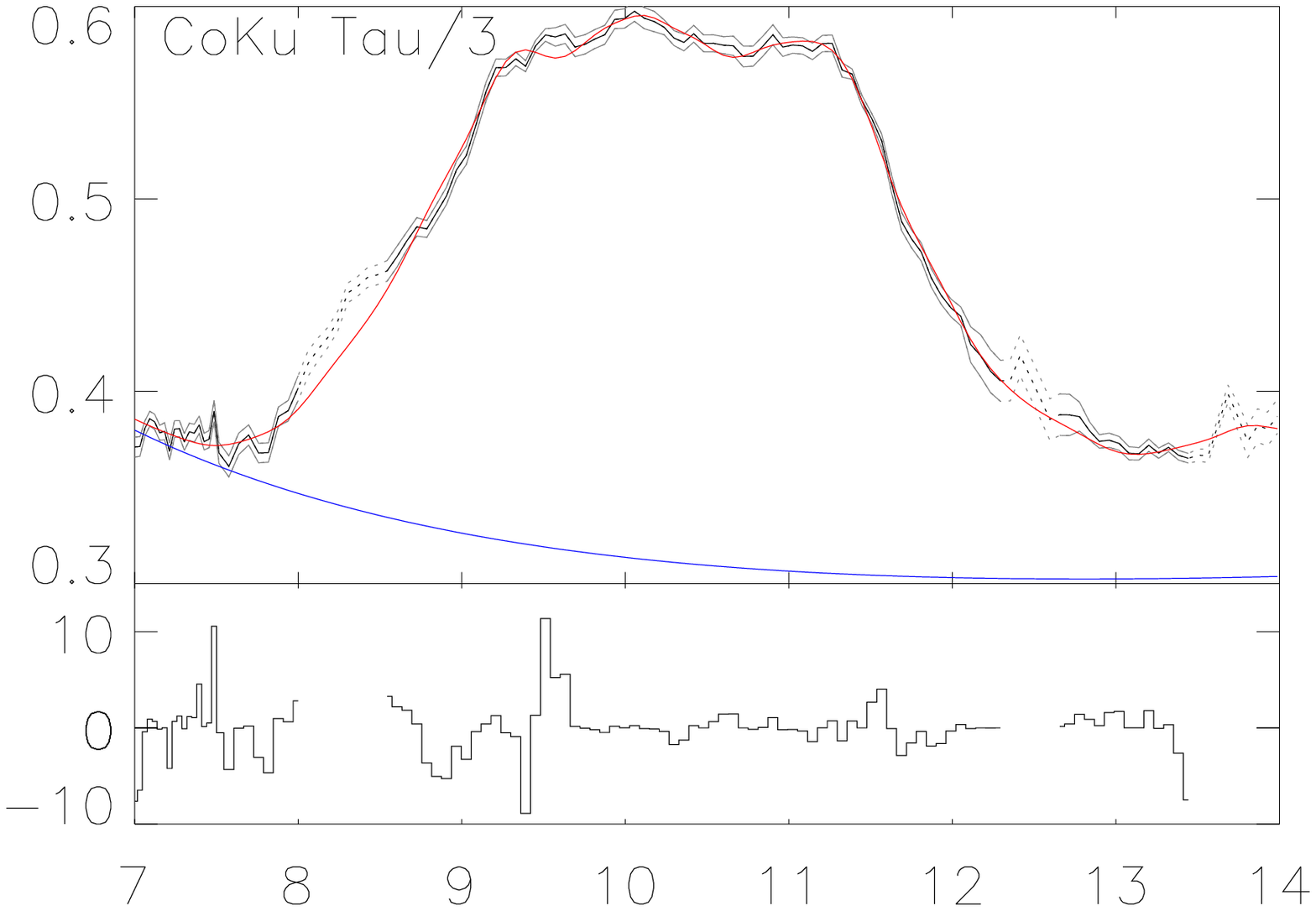}&\includegraphics[width=4.2cm]{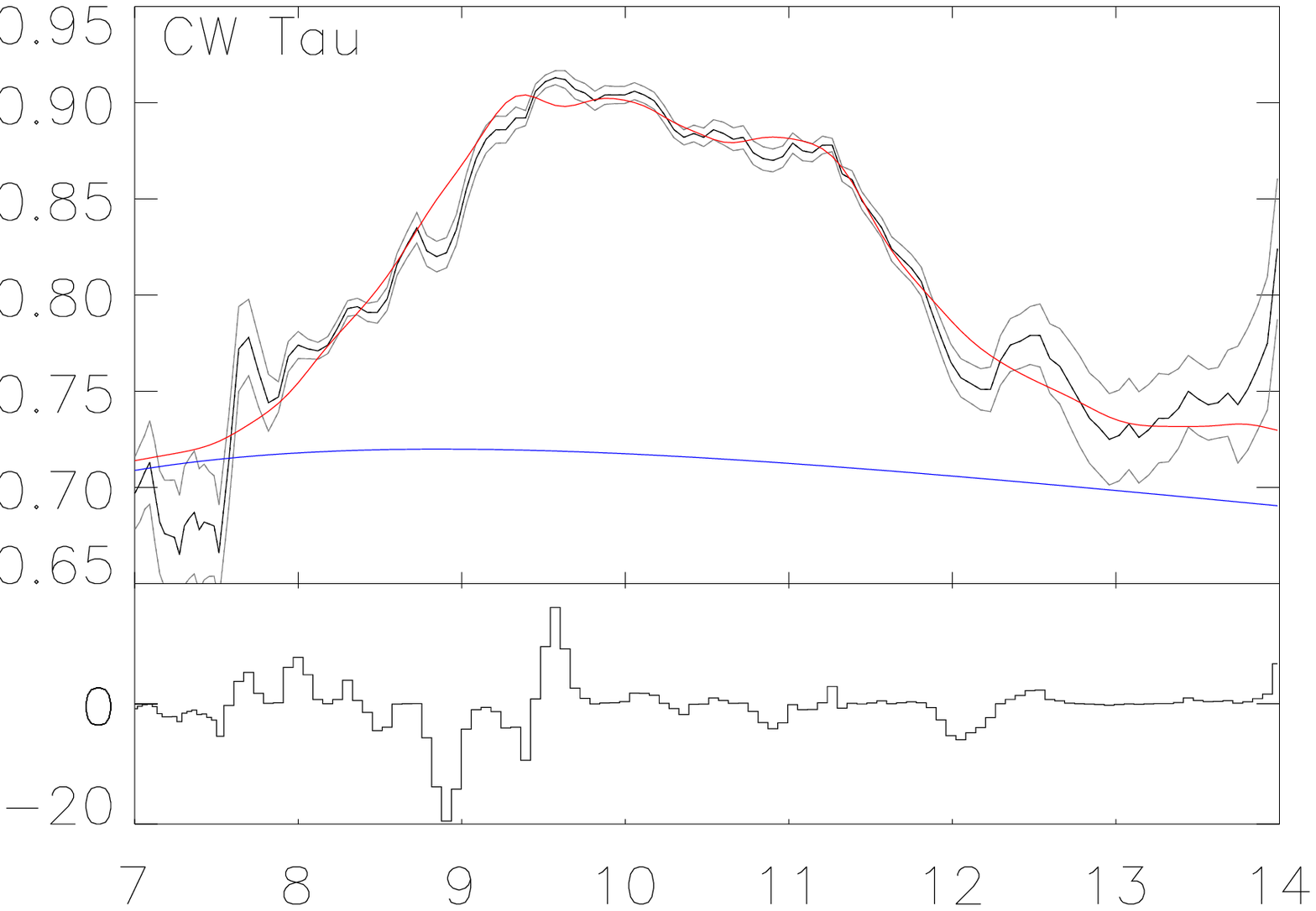}\\
\\
\includegraphics[width=4.2cm]{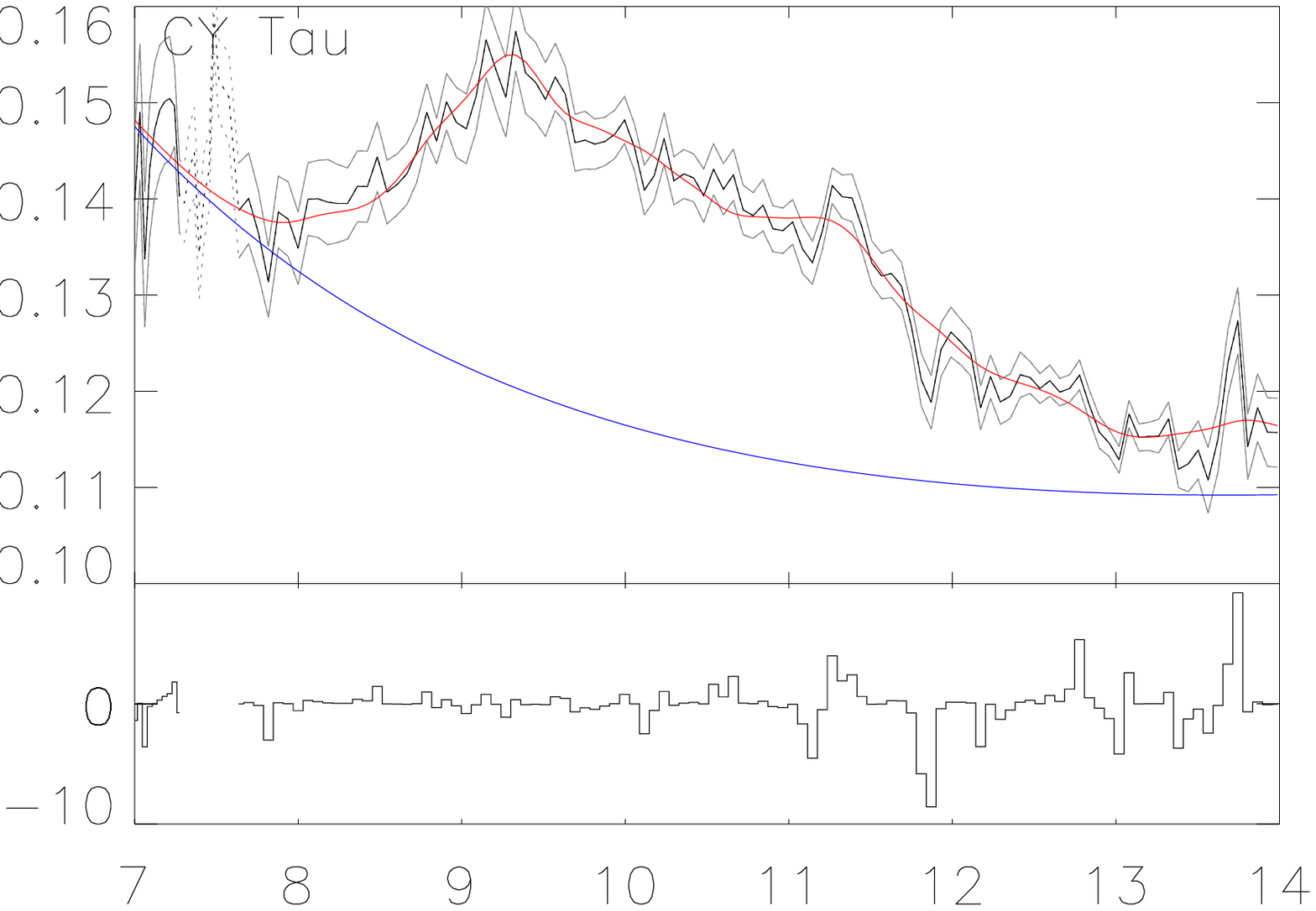}&\includegraphics[width=4.2cm]{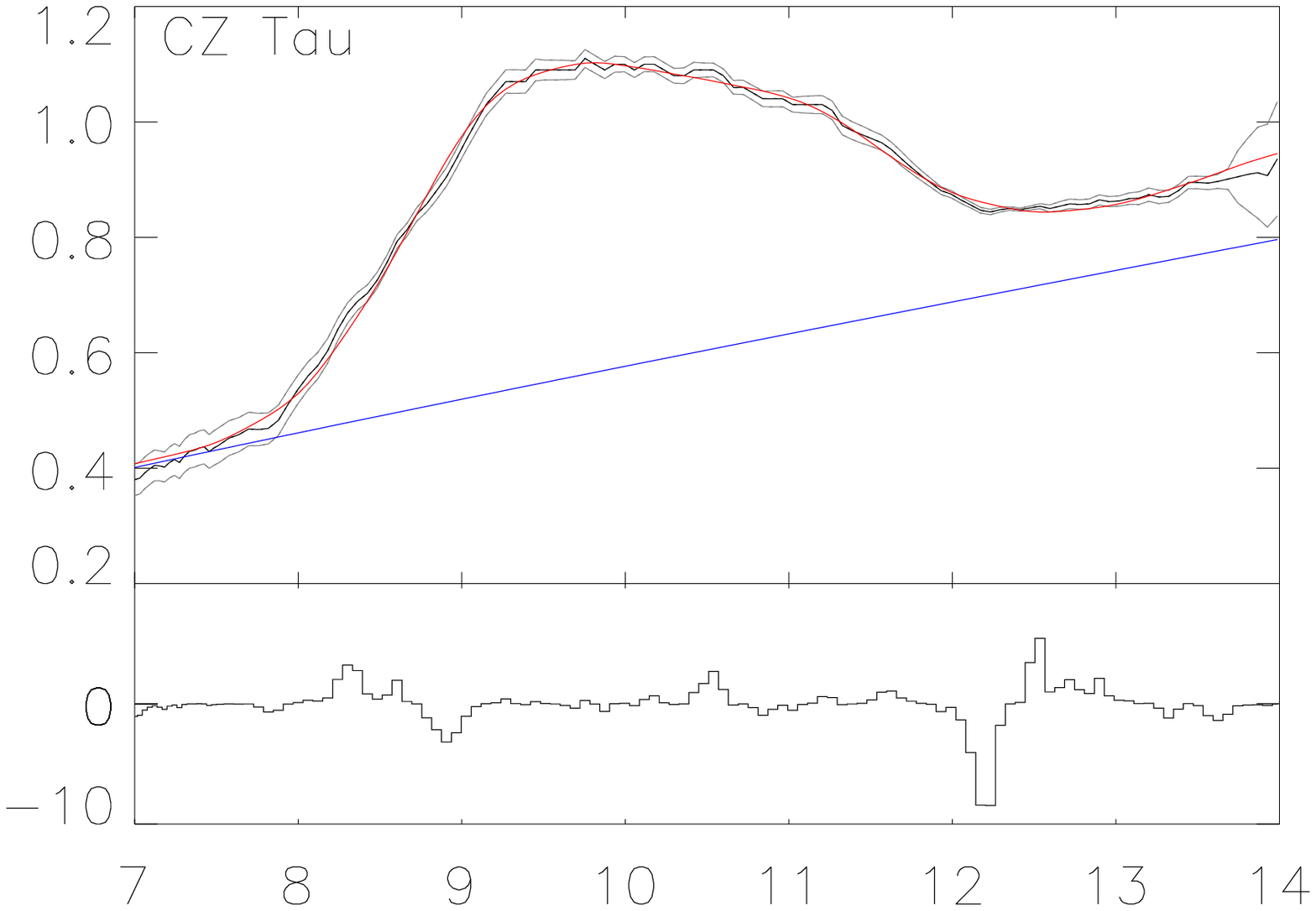}&
\includegraphics[width=4.2cm]{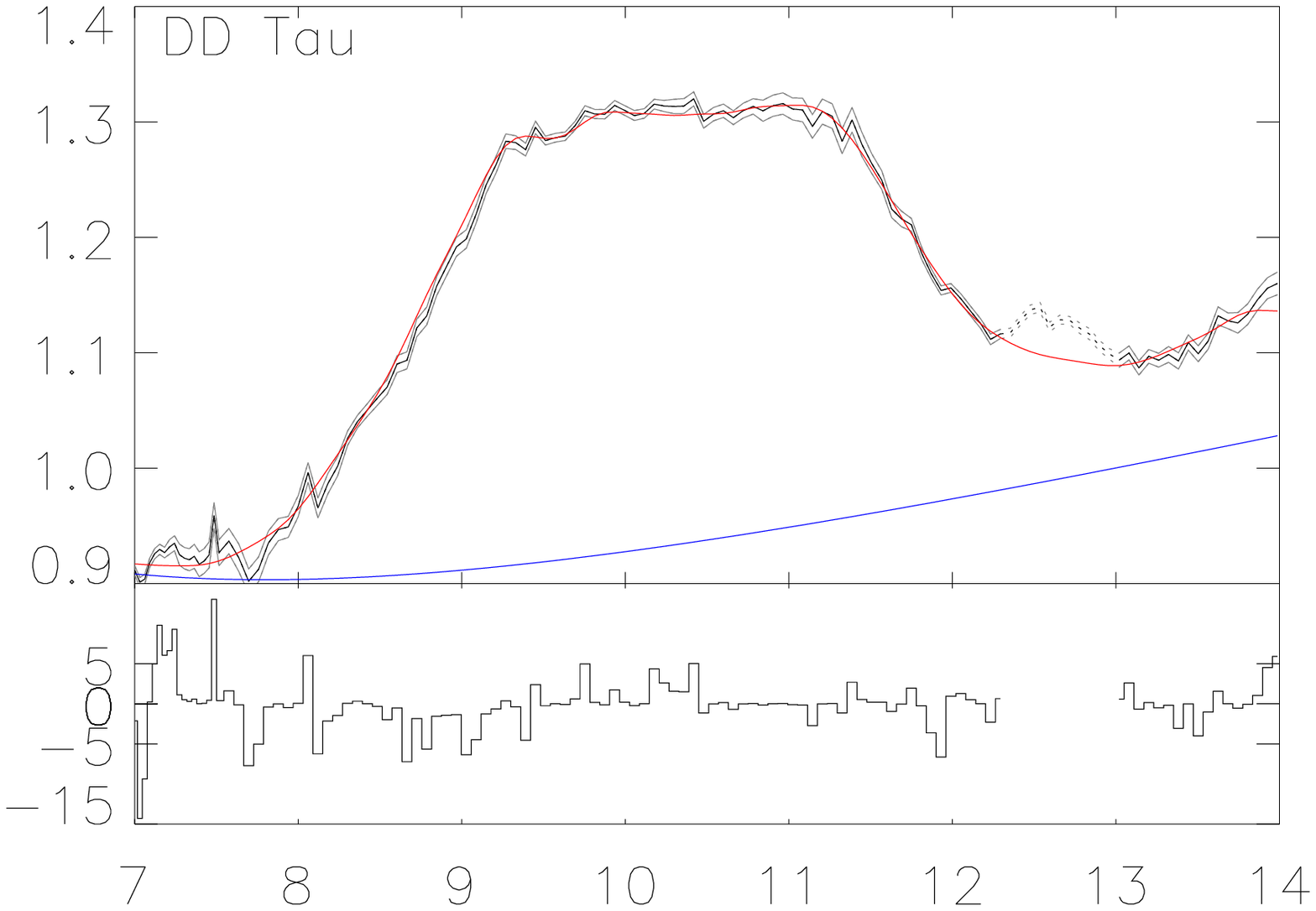}&\includegraphics[width=4.2cm]{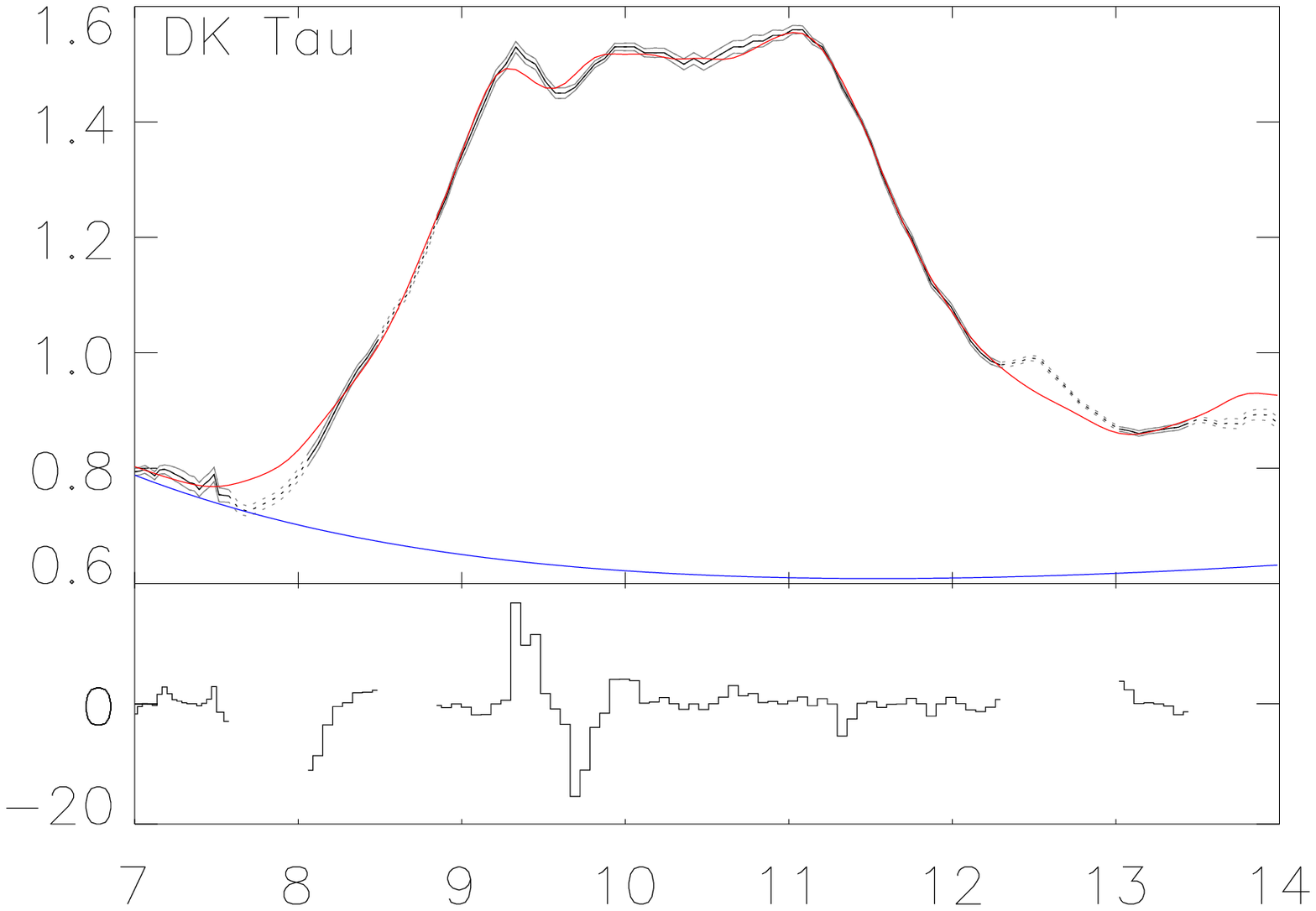}\\
\\
\includegraphics[width=4.2cm]{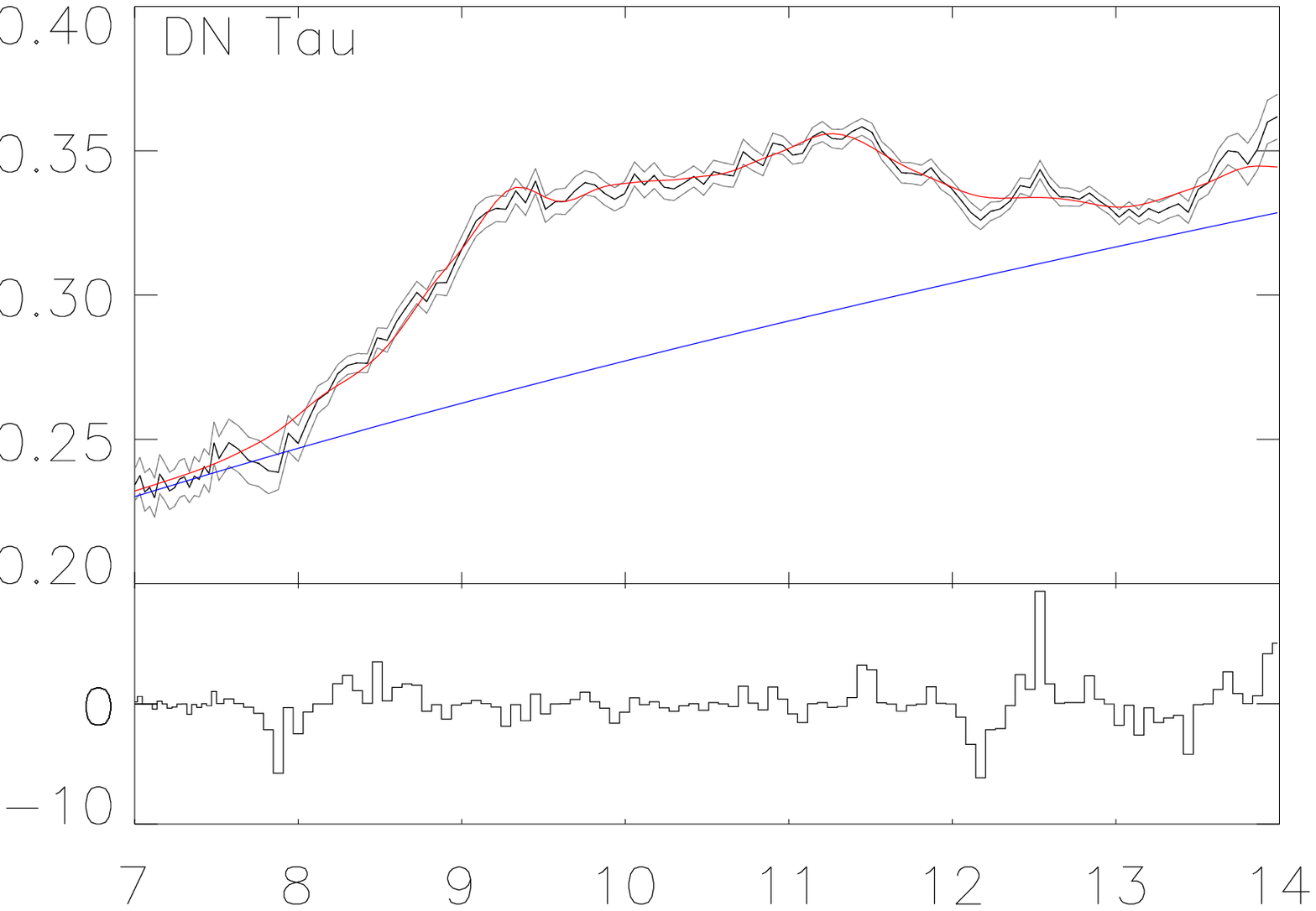}&\includegraphics[width=4.2cm]{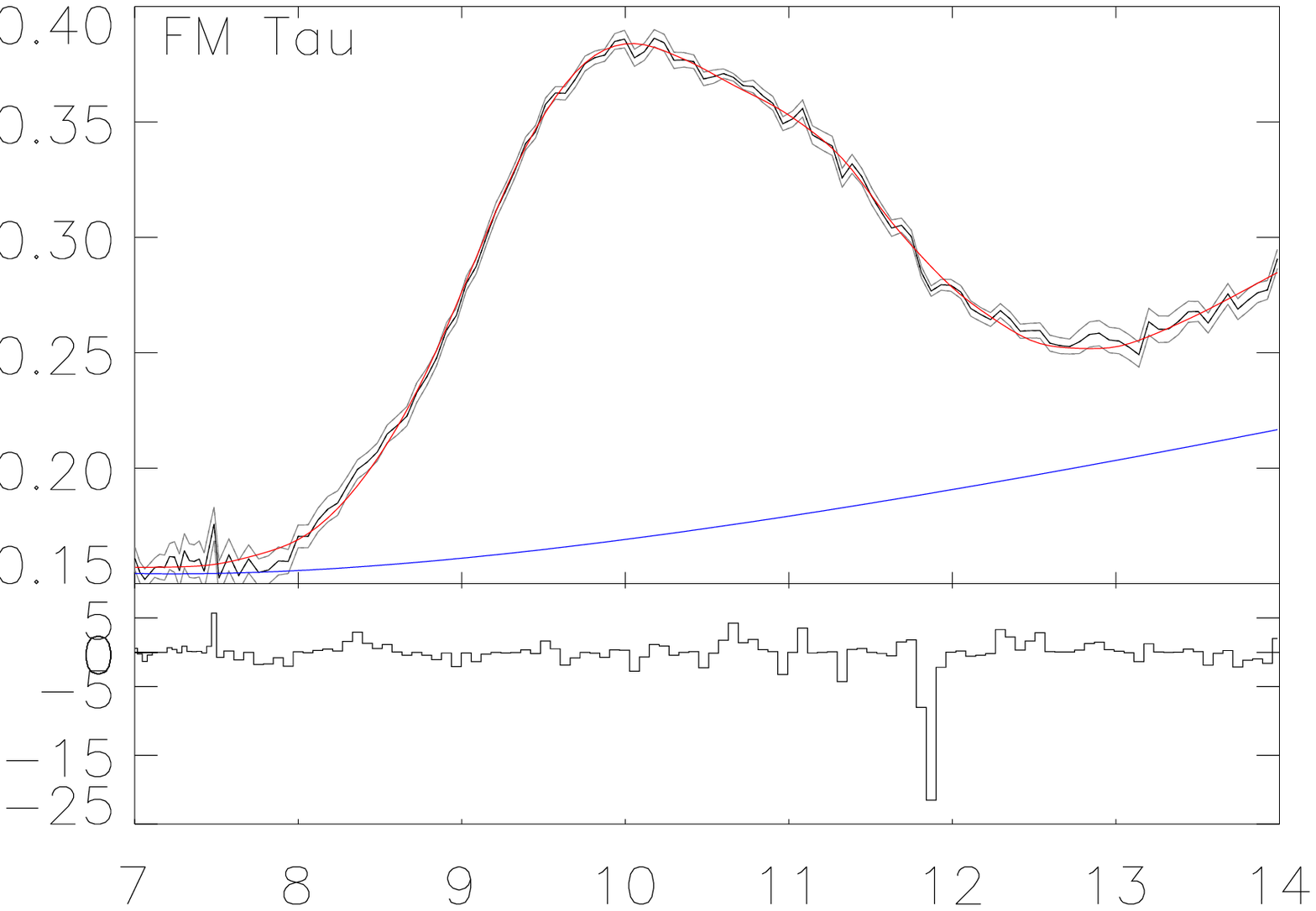}&
\includegraphics[width=4.2cm]{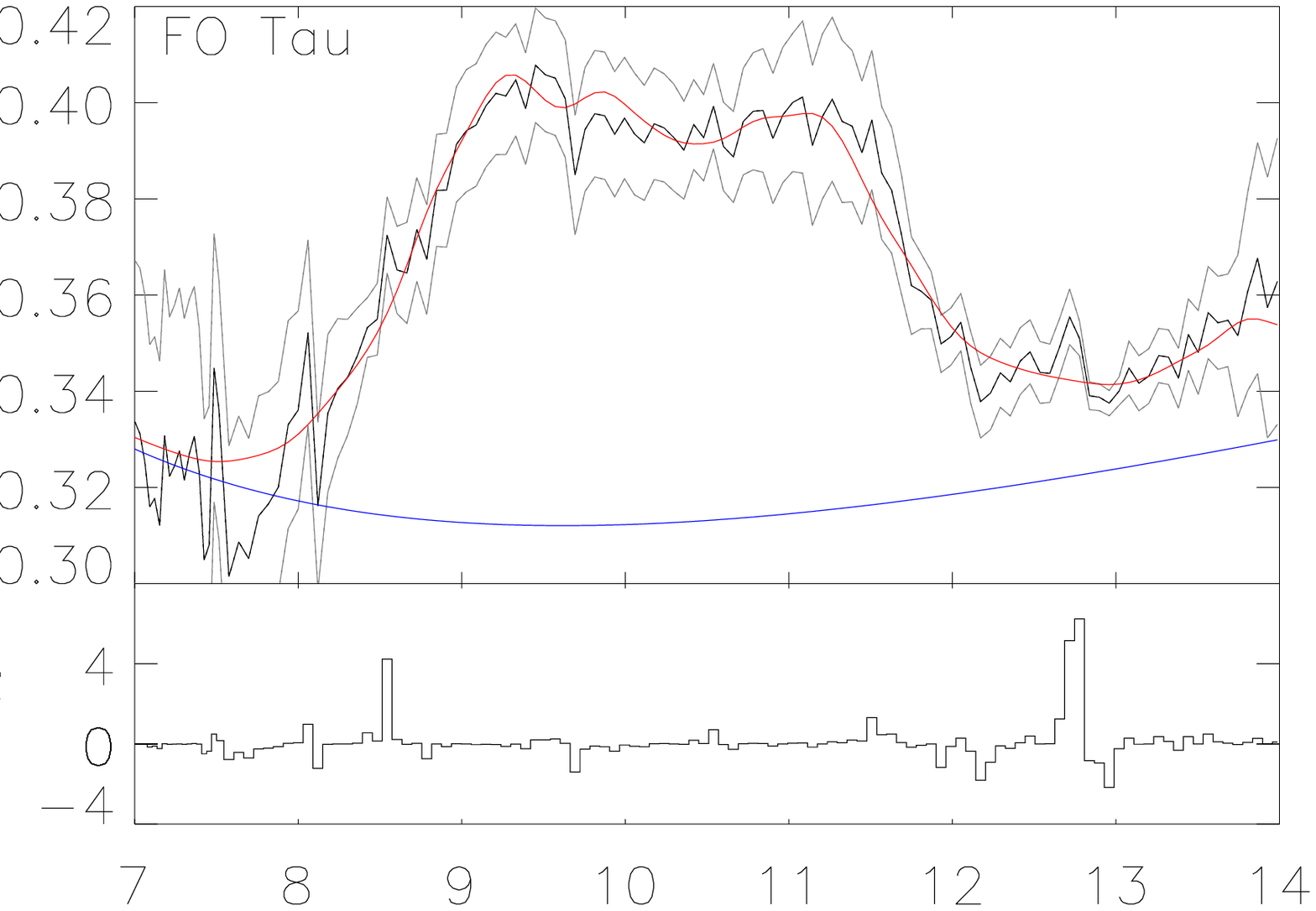}&\includegraphics[width=4.2cm]{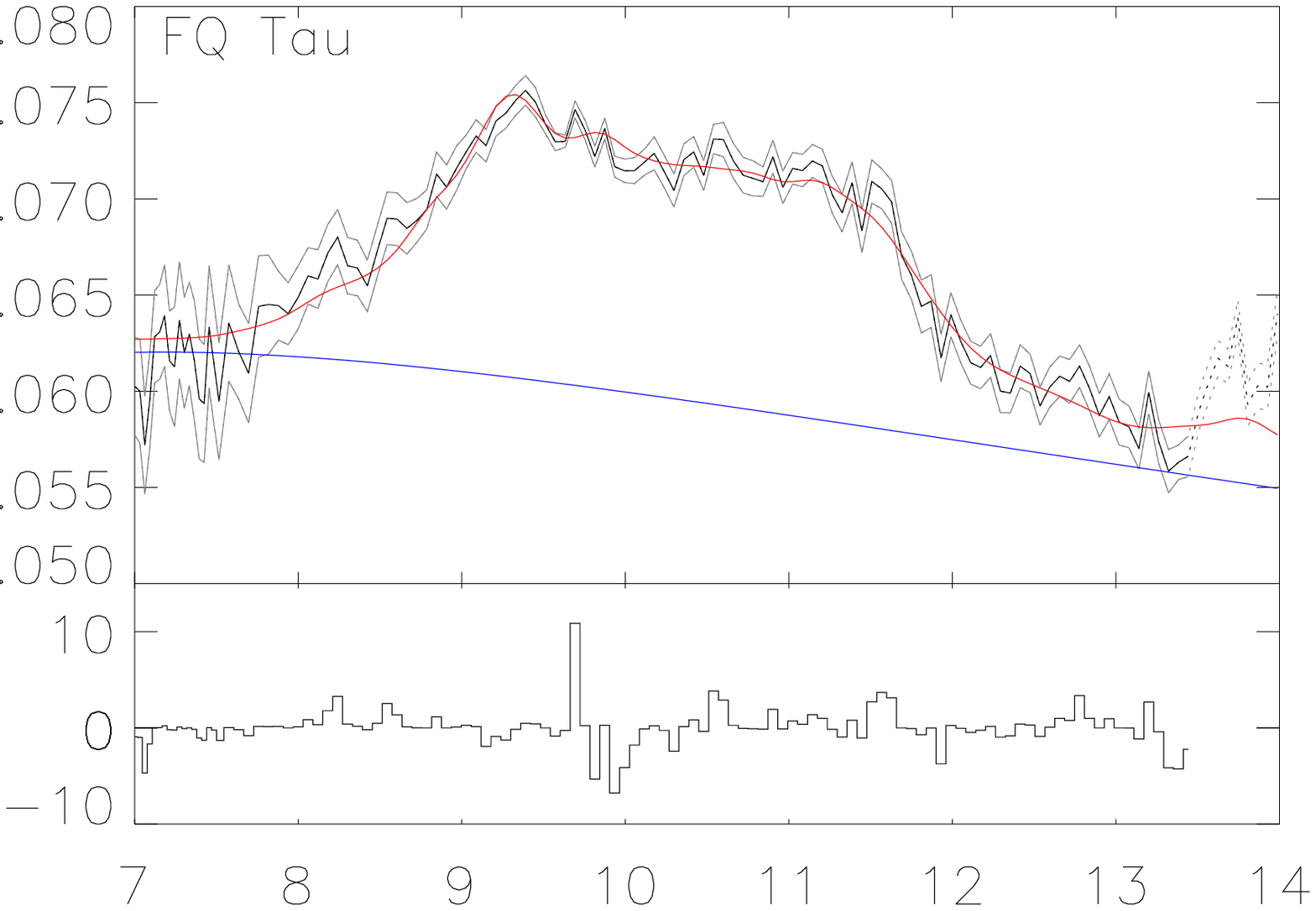}\\
\\
\includegraphics[width=4.2cm]{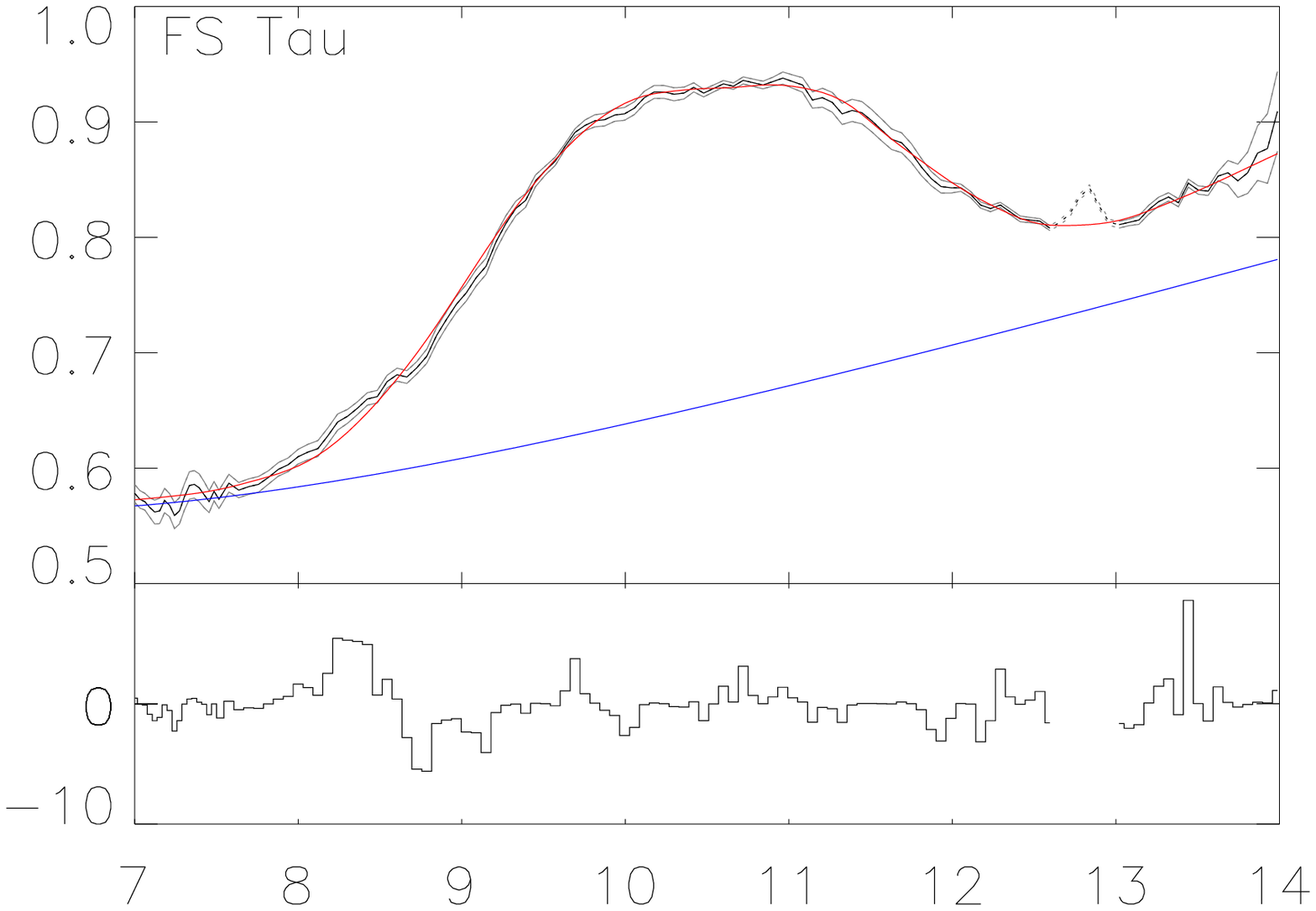}&\includegraphics[width=4.2cm]{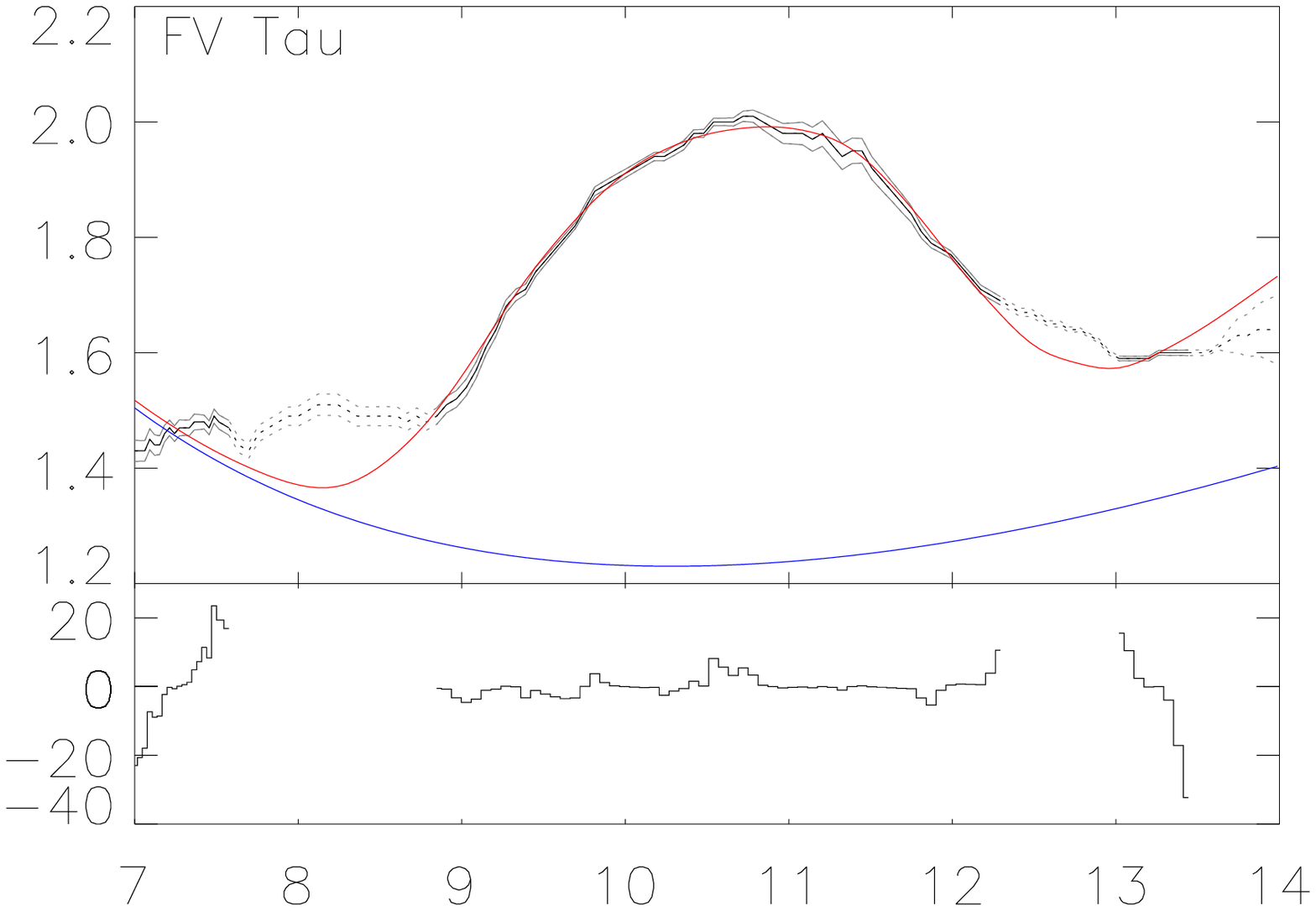}&
\includegraphics[width=4.2cm]{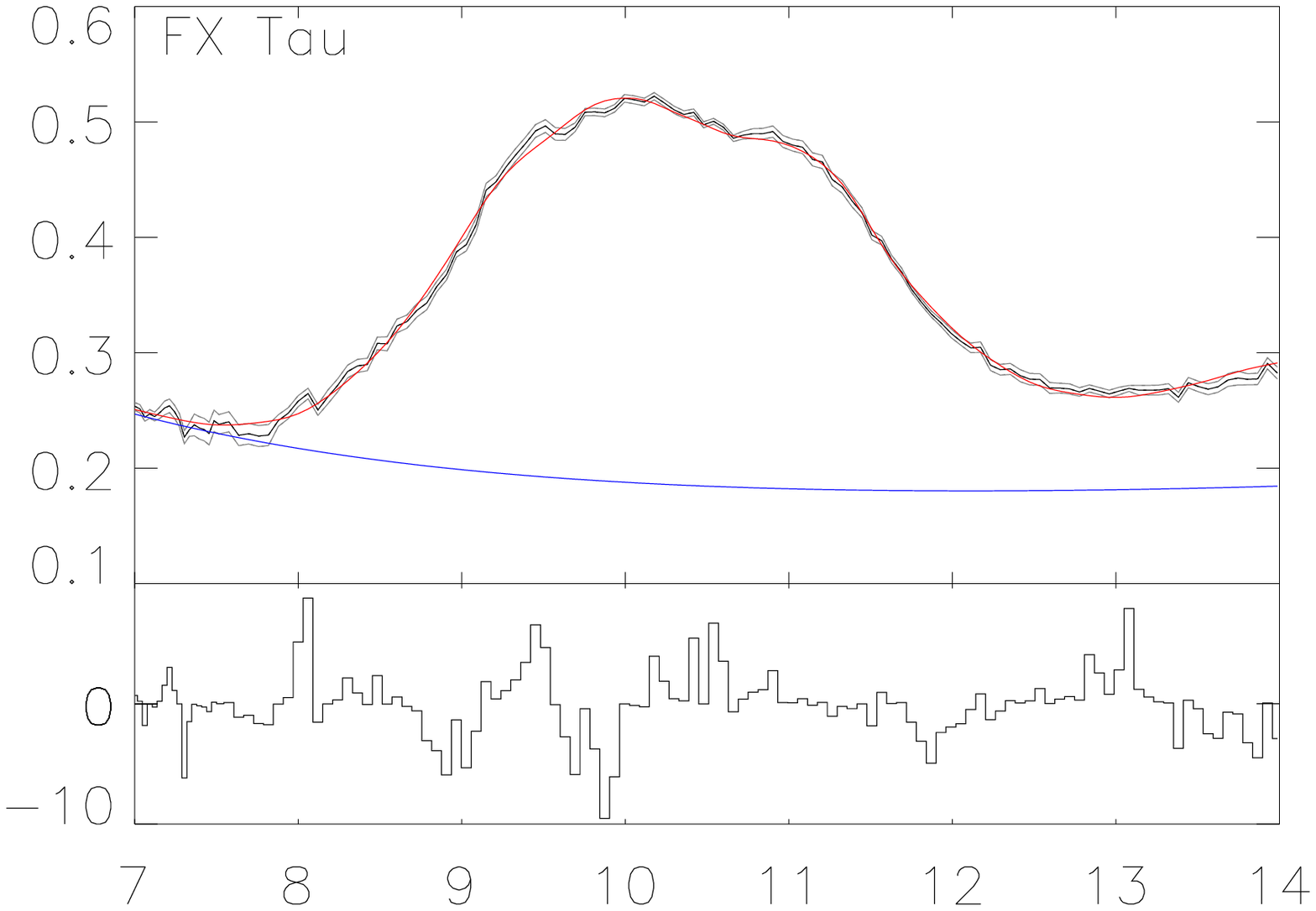}&\includegraphics[width=4.2cm]{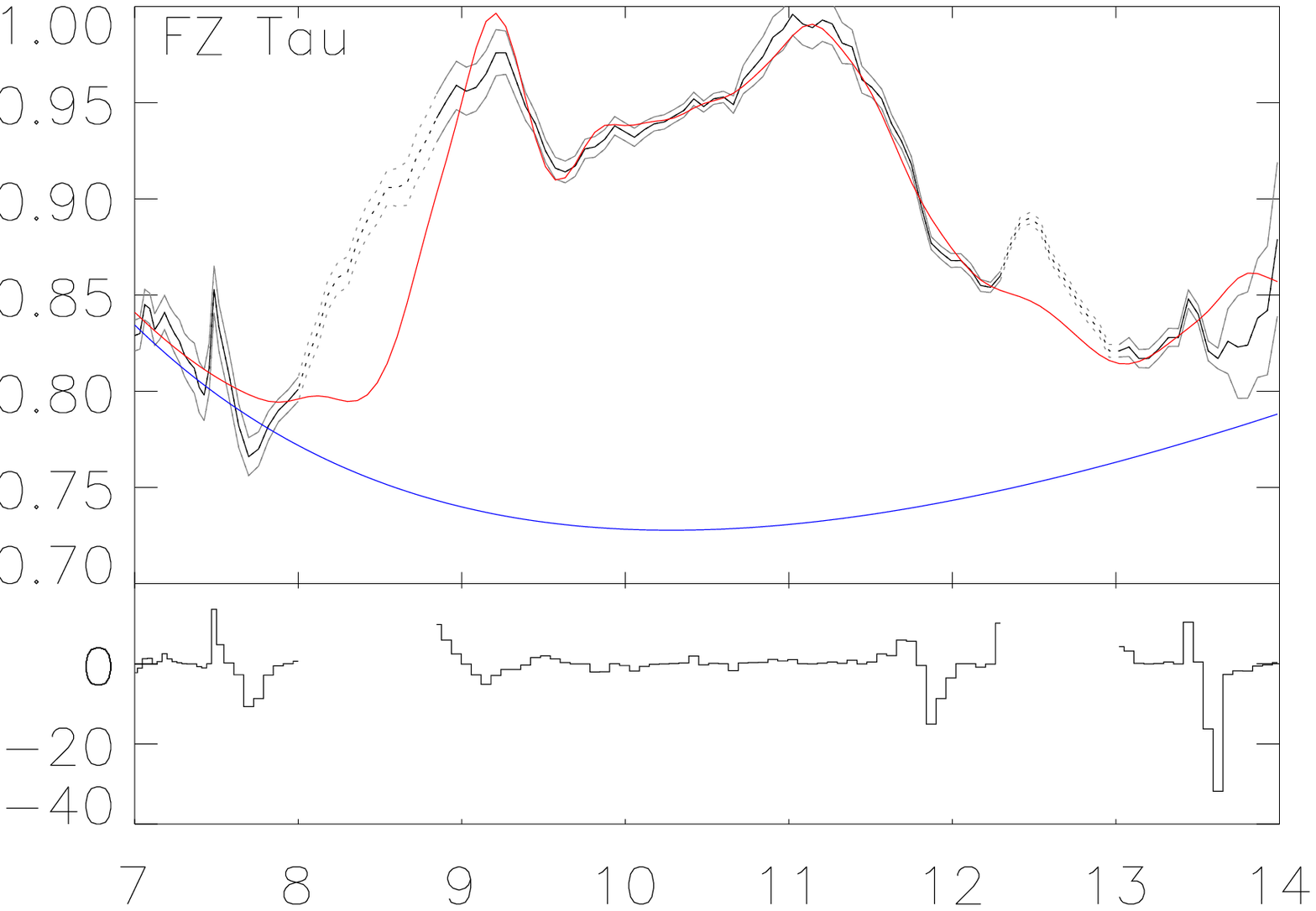}\\
\\
\includegraphics[width=4.2cm]{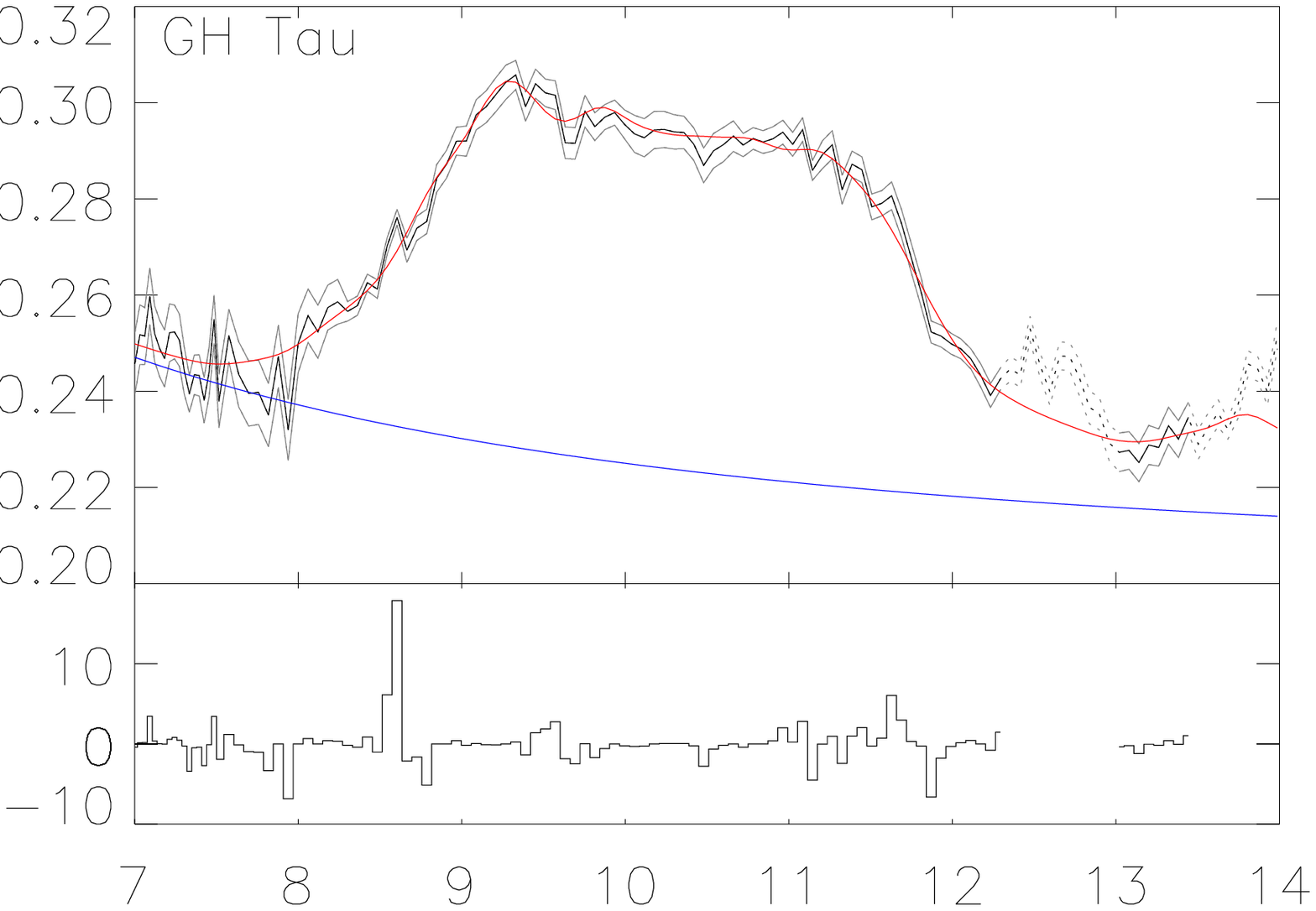}&\includegraphics[width=4.2cm]{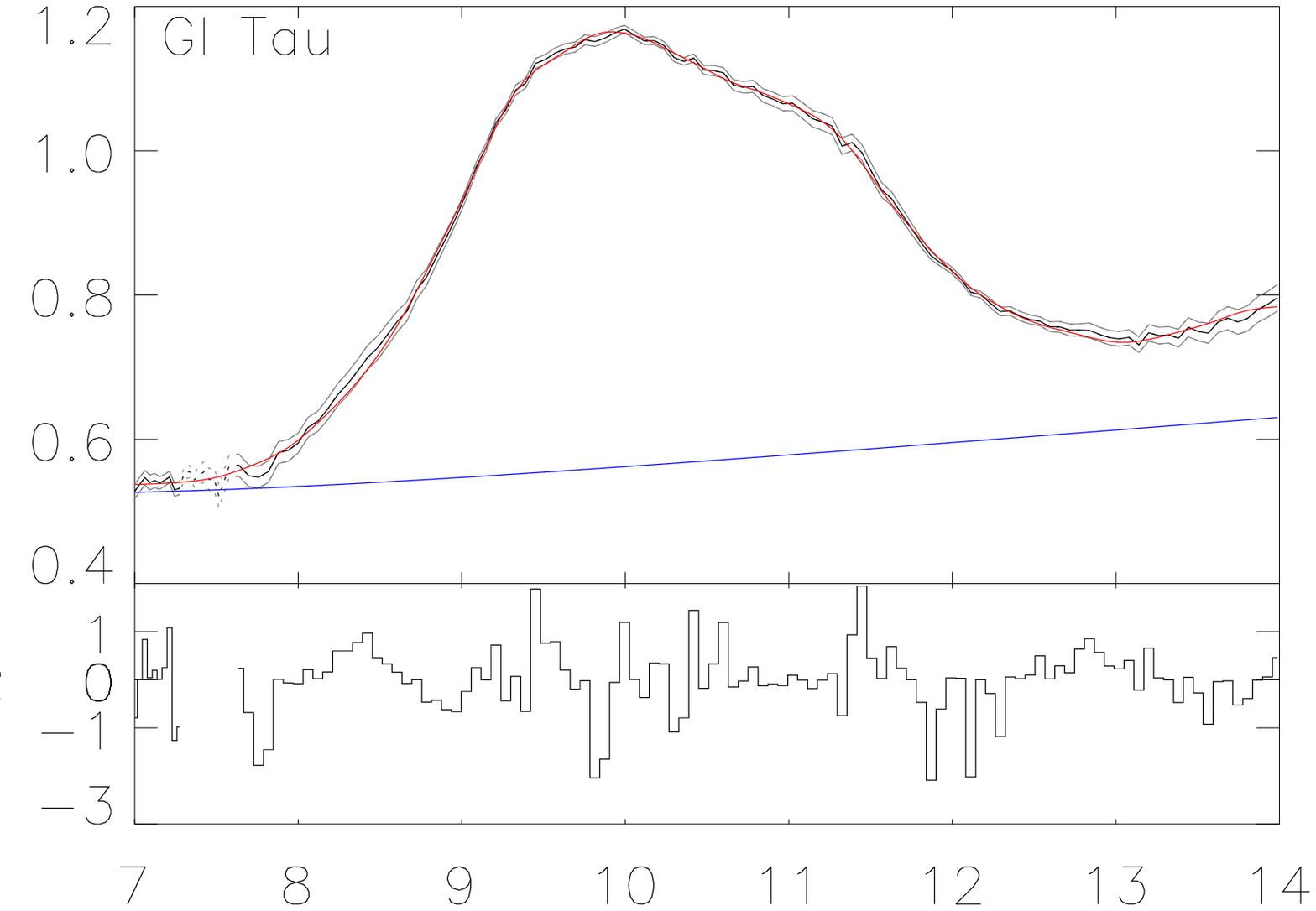}&
\includegraphics[width=4.2cm]{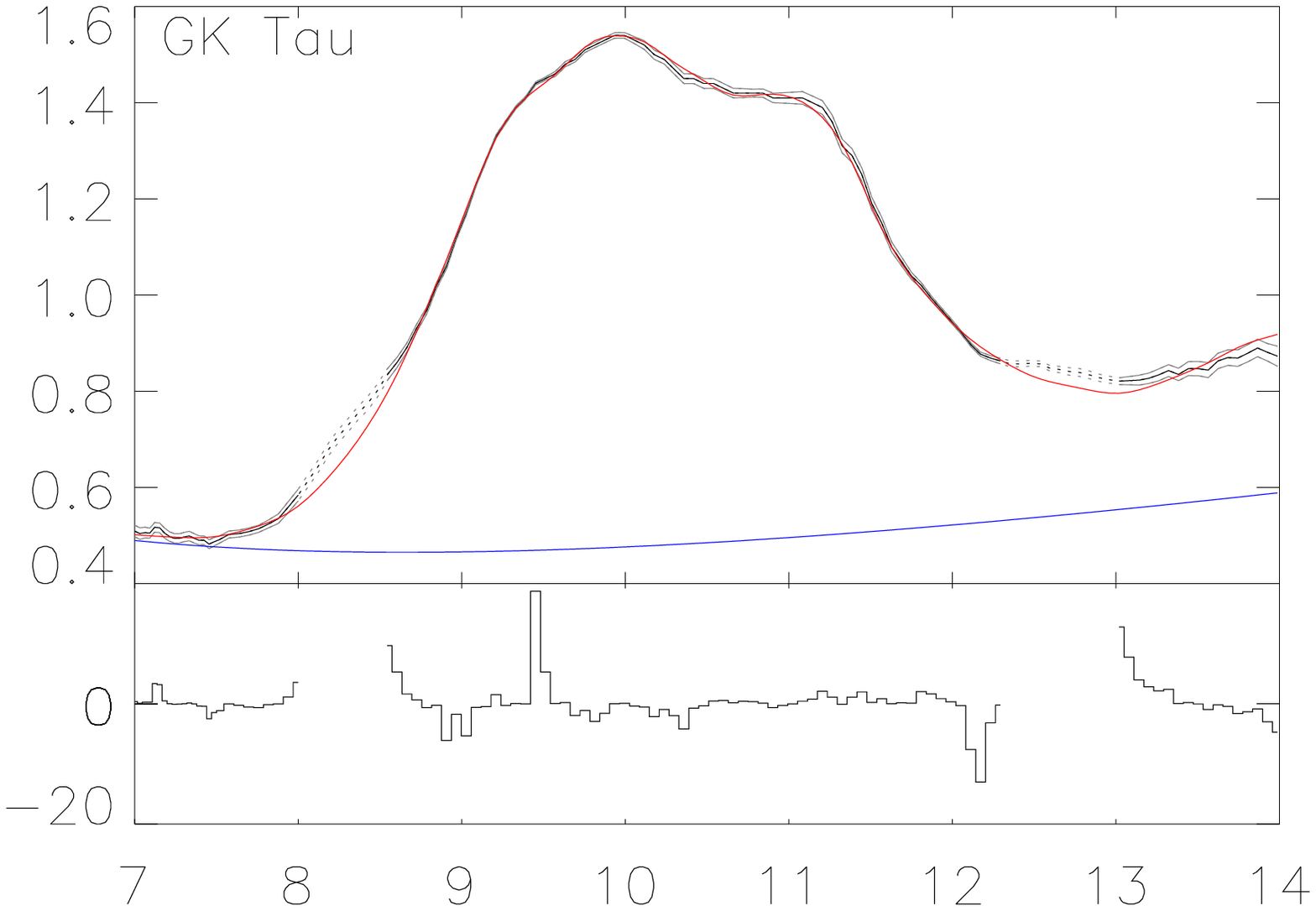}&\includegraphics[width=4.2cm]{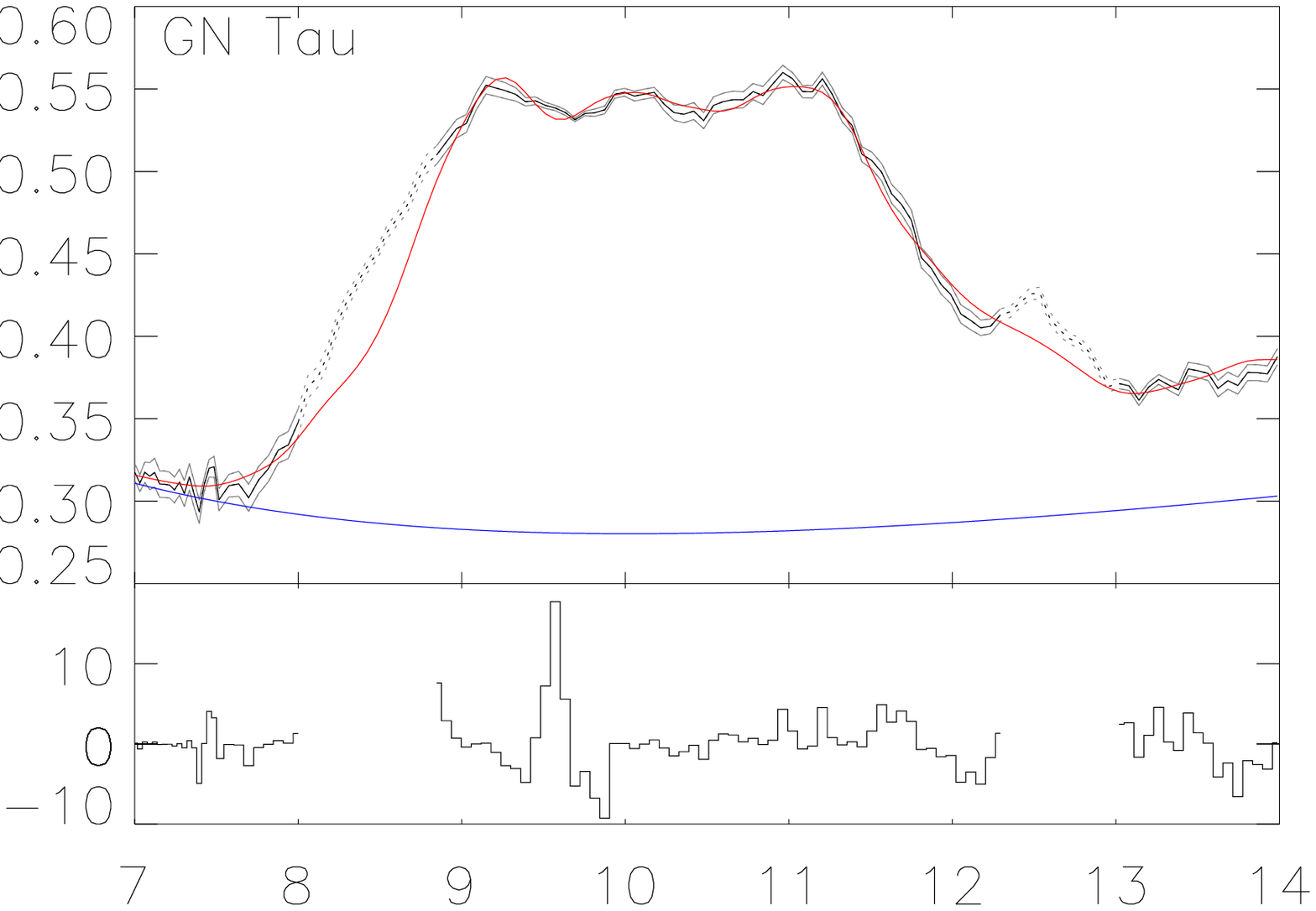}\\
\\
\includegraphics[width=4.2cm]{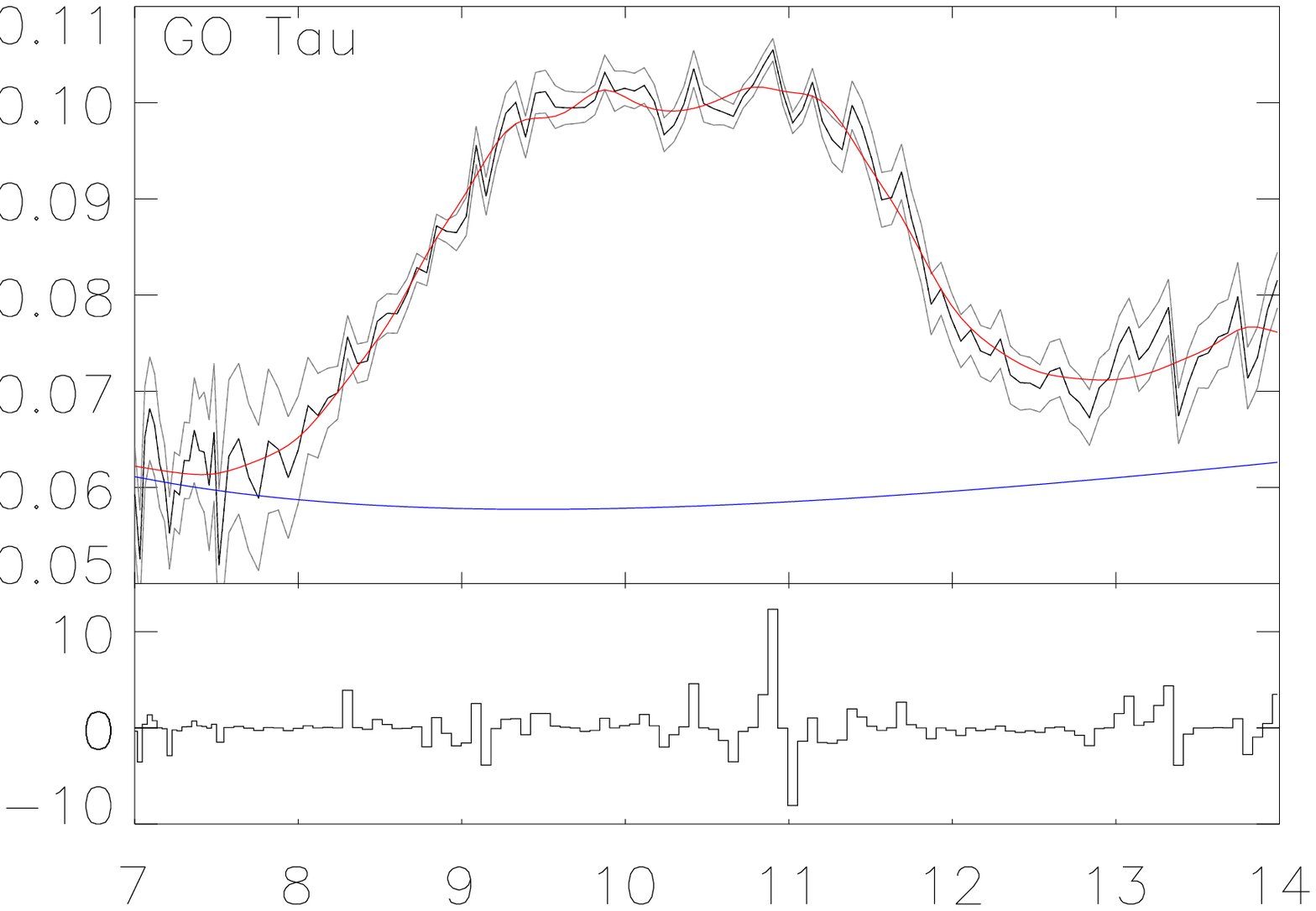}&\includegraphics[width=4.2cm]{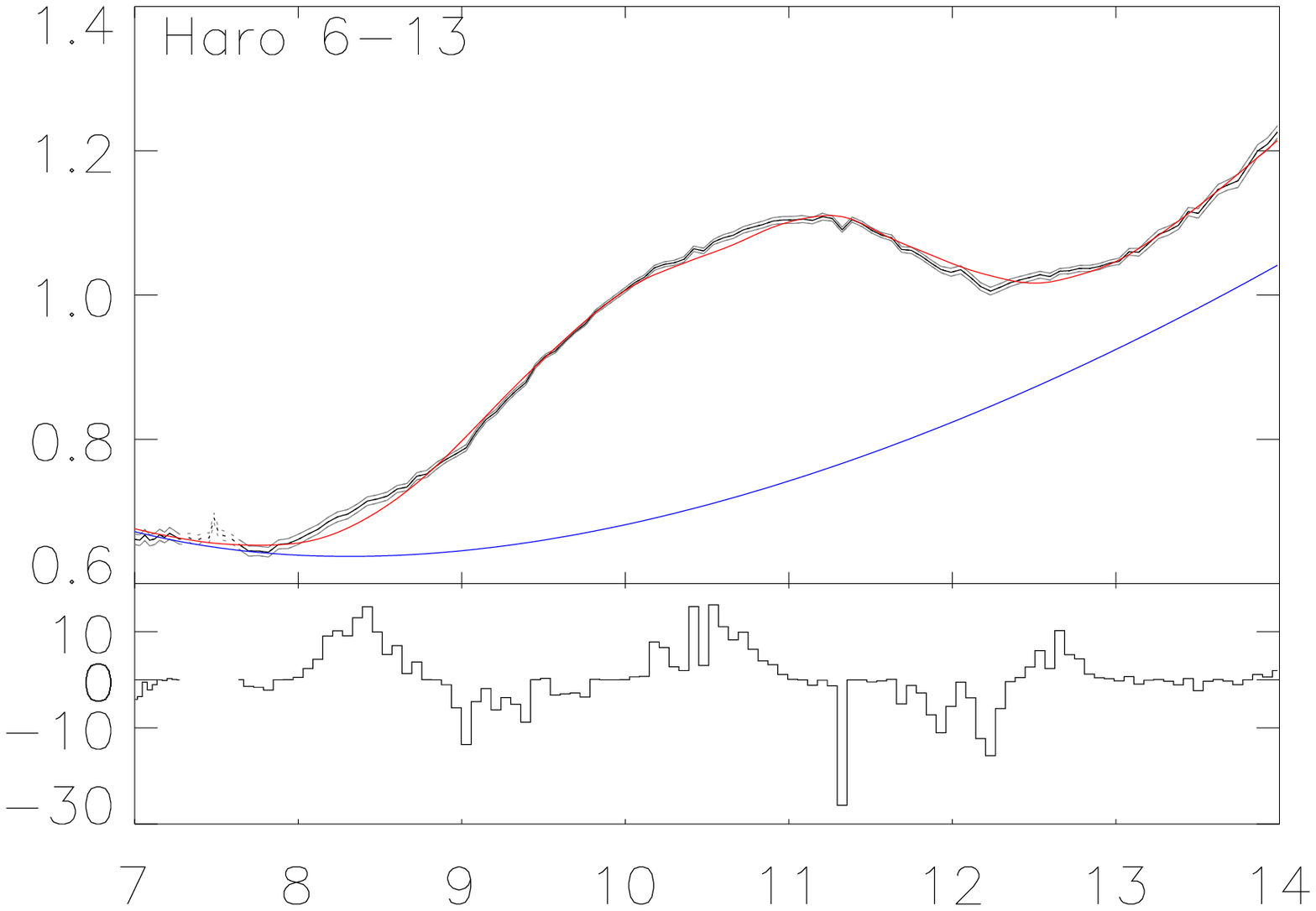}&
\includegraphics[width=4.2cm]{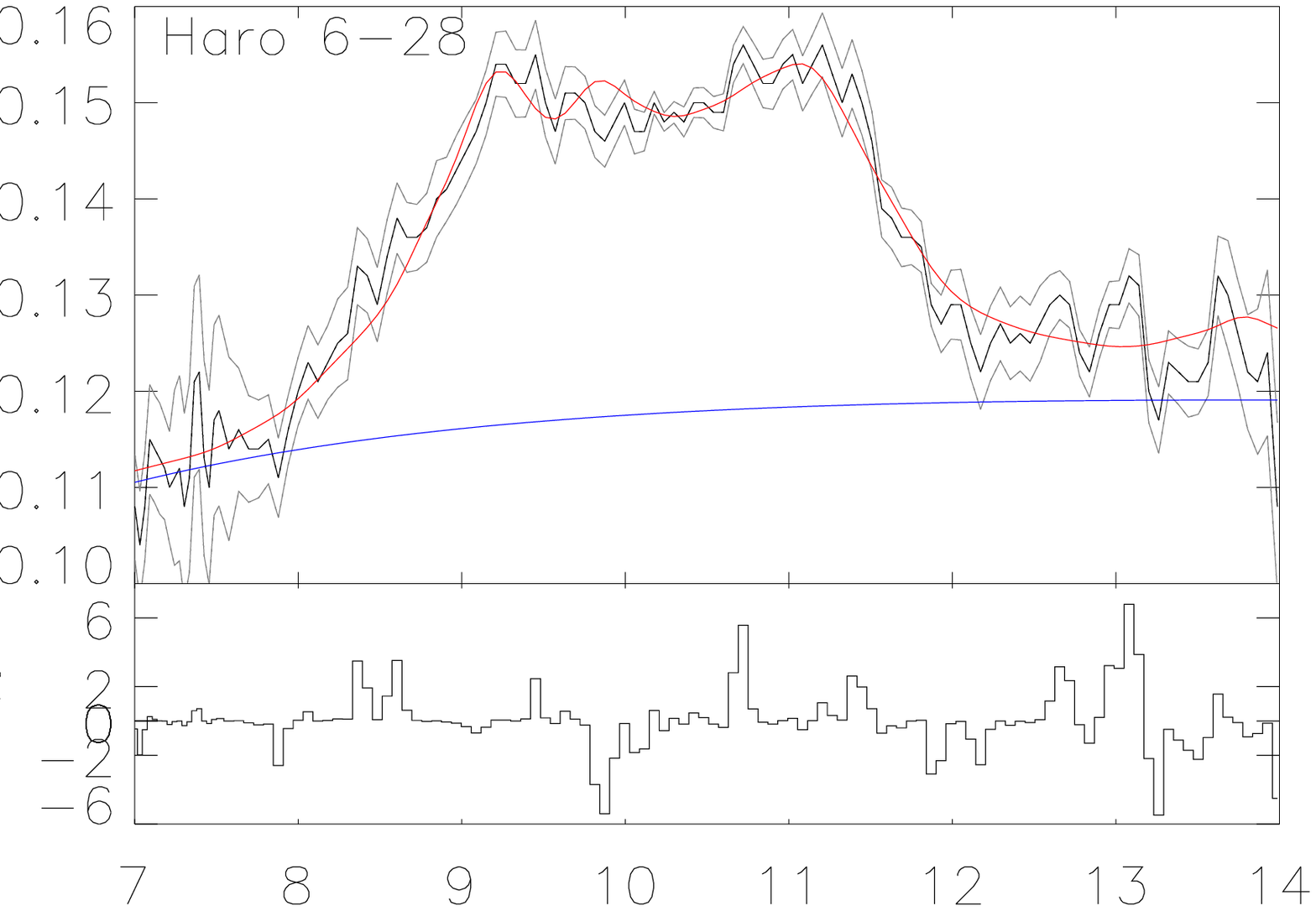}&\includegraphics[width=4.2cm]{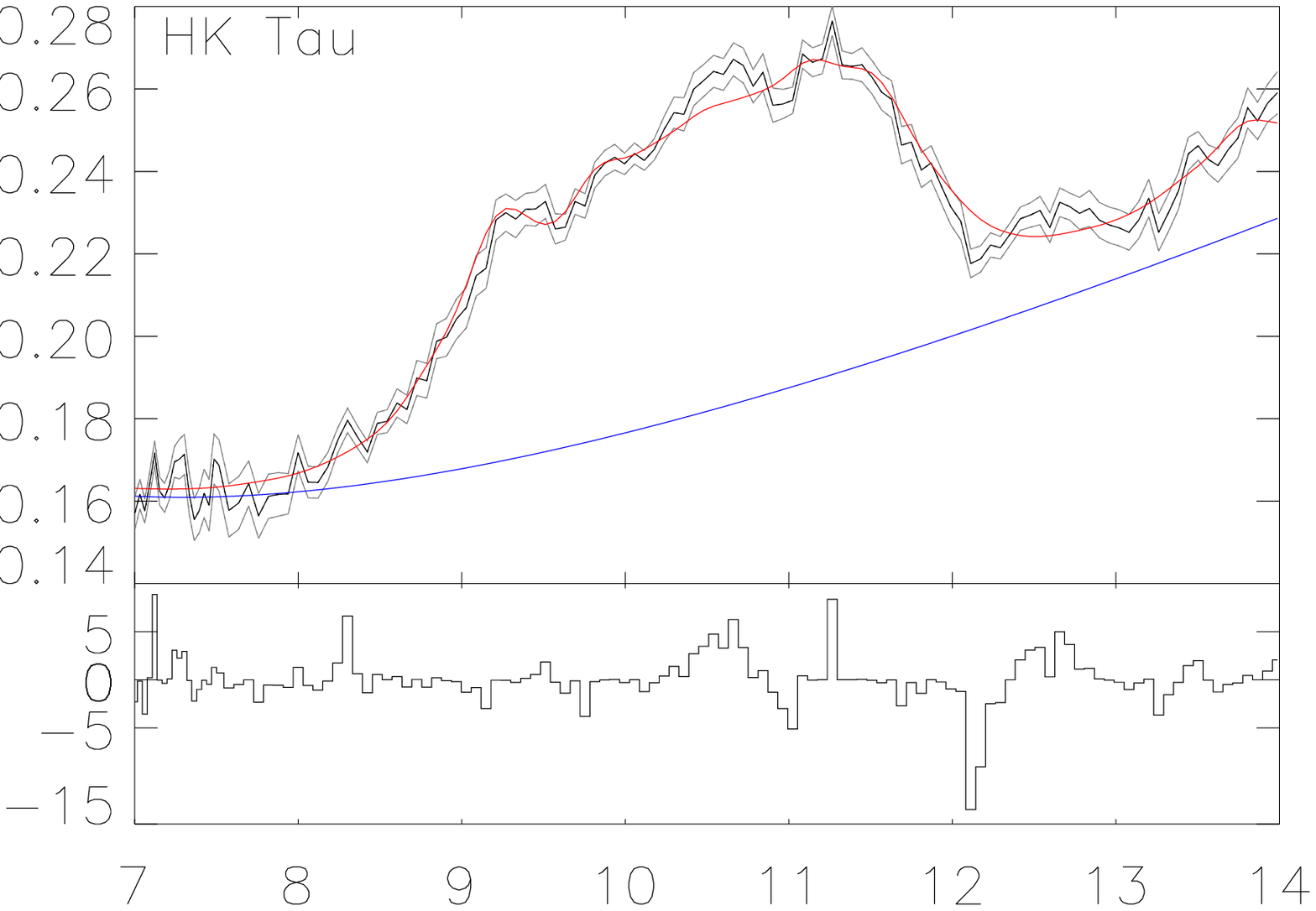}\\
\\
\includegraphics[width=4.2cm]{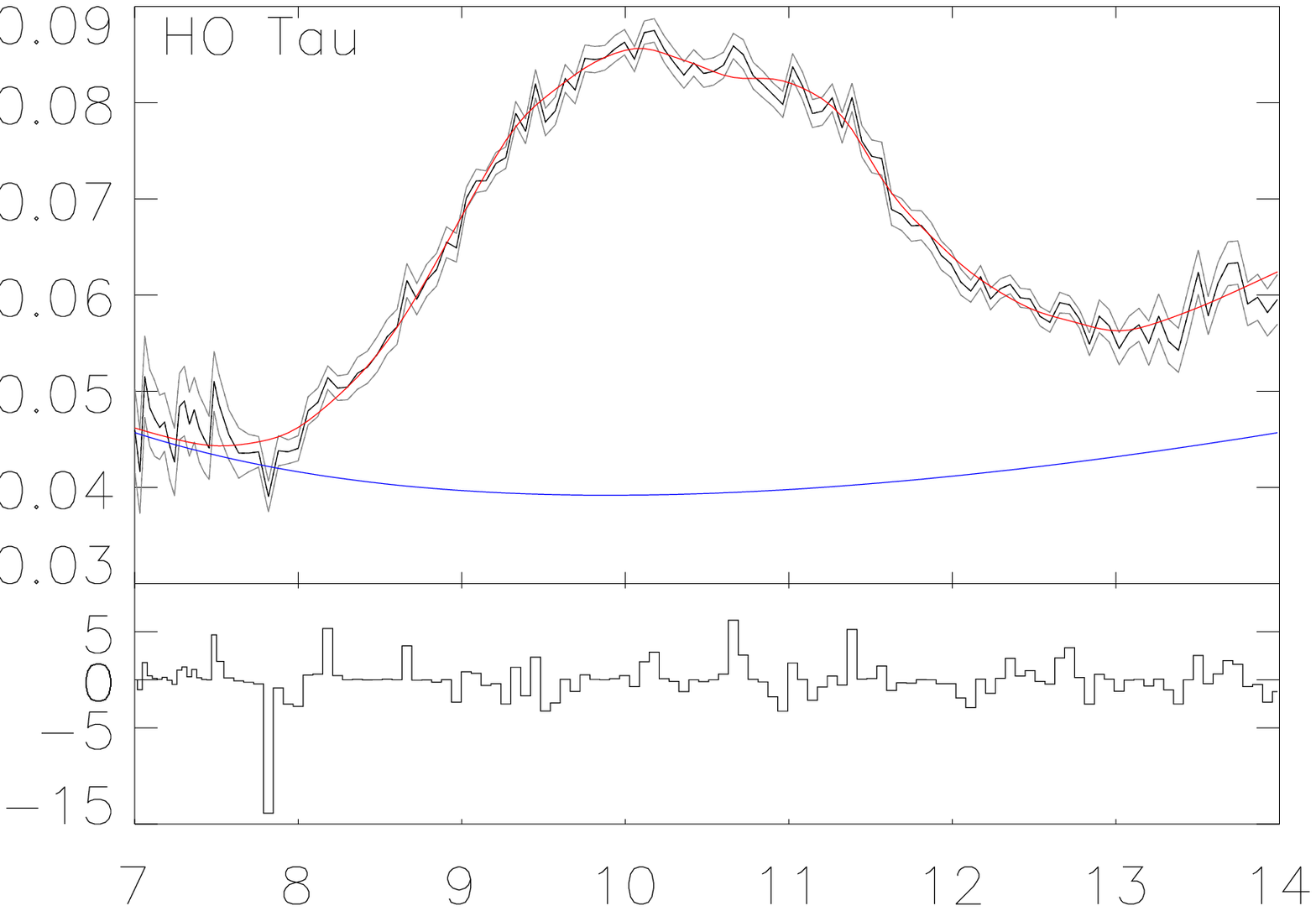}&\includegraphics[width=4.2cm]{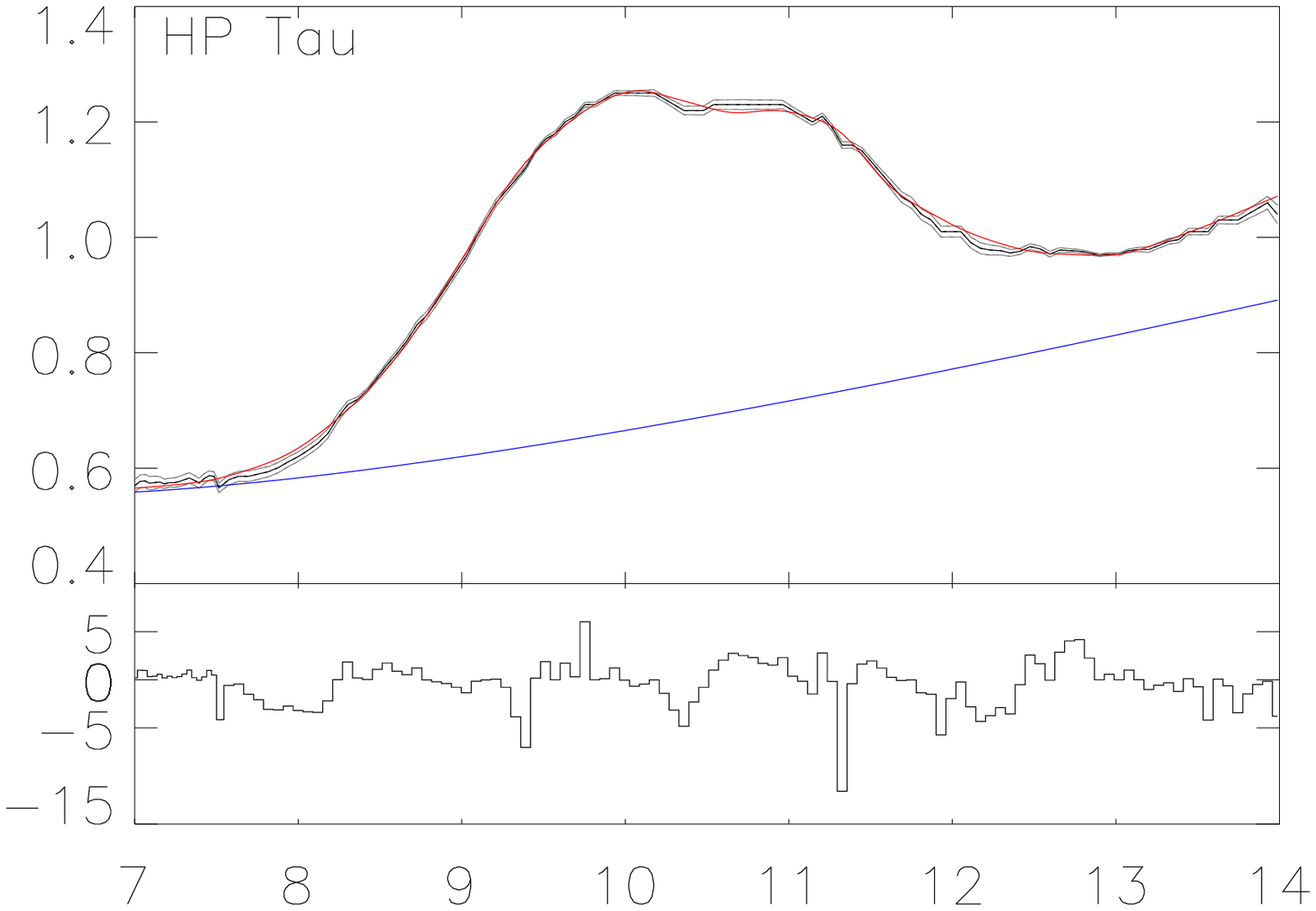}&
\includegraphics[width=4.2cm]{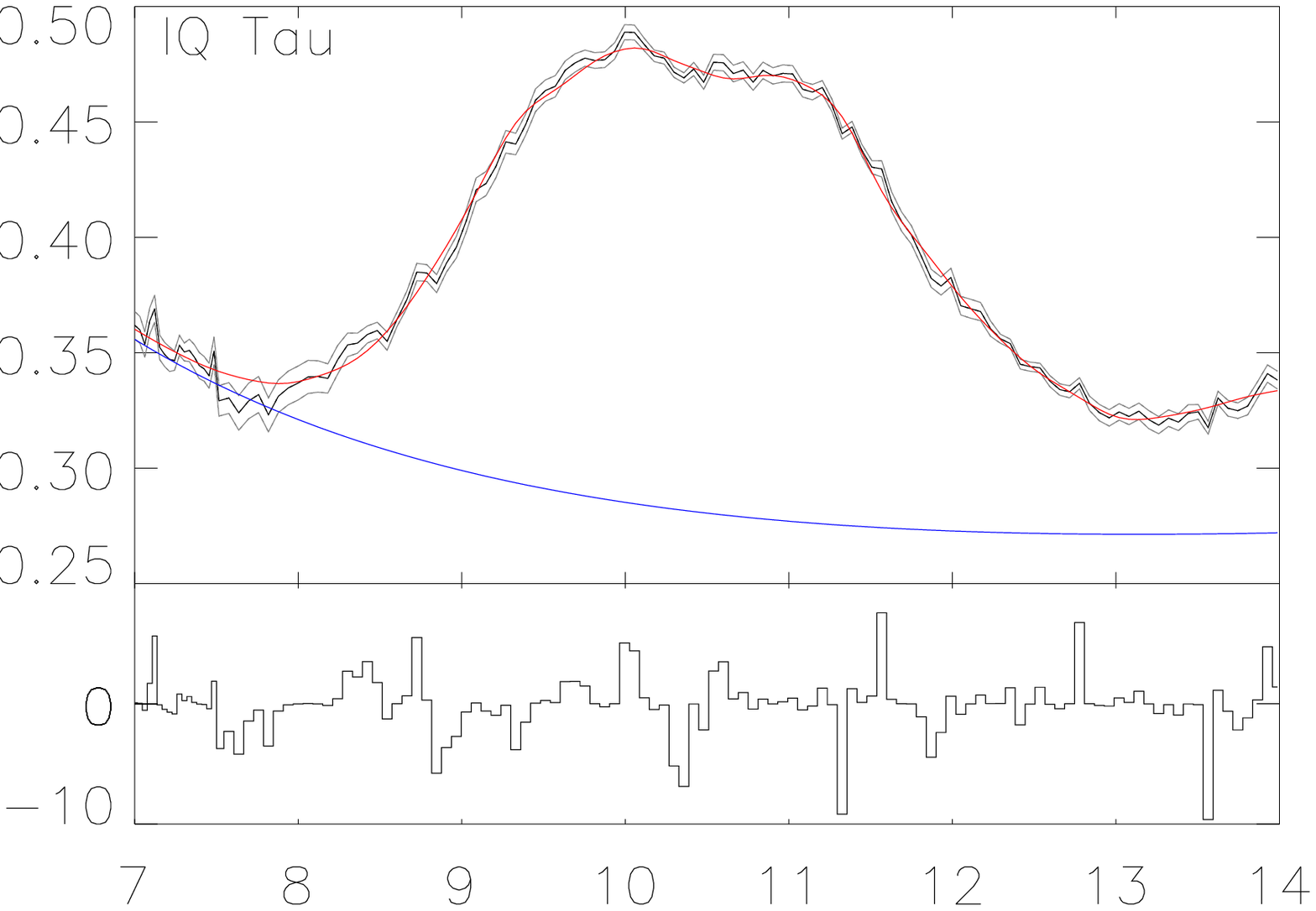}&\includegraphics[width=4.2cm]{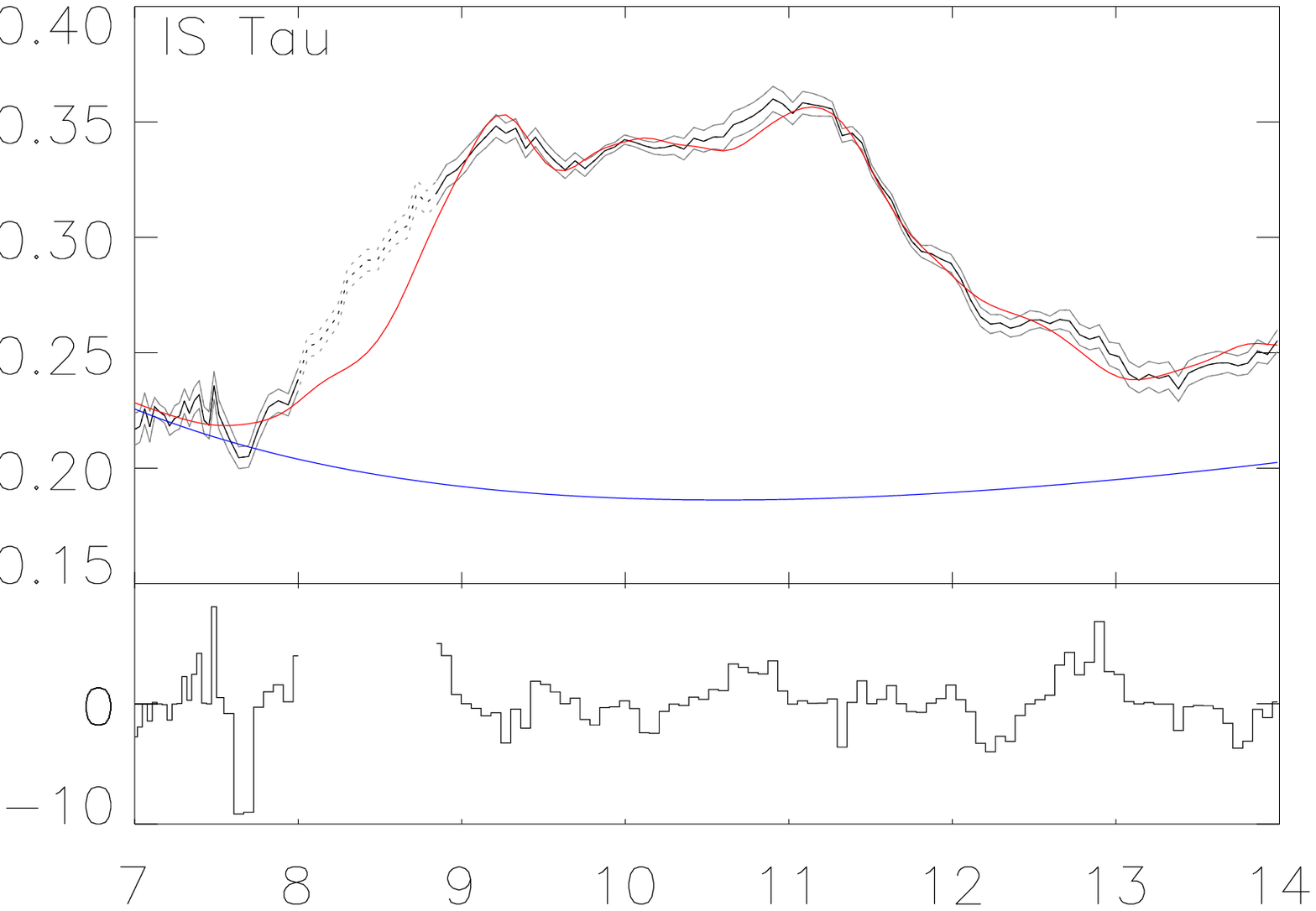}\\
\\
\includegraphics[width=4.2cm]{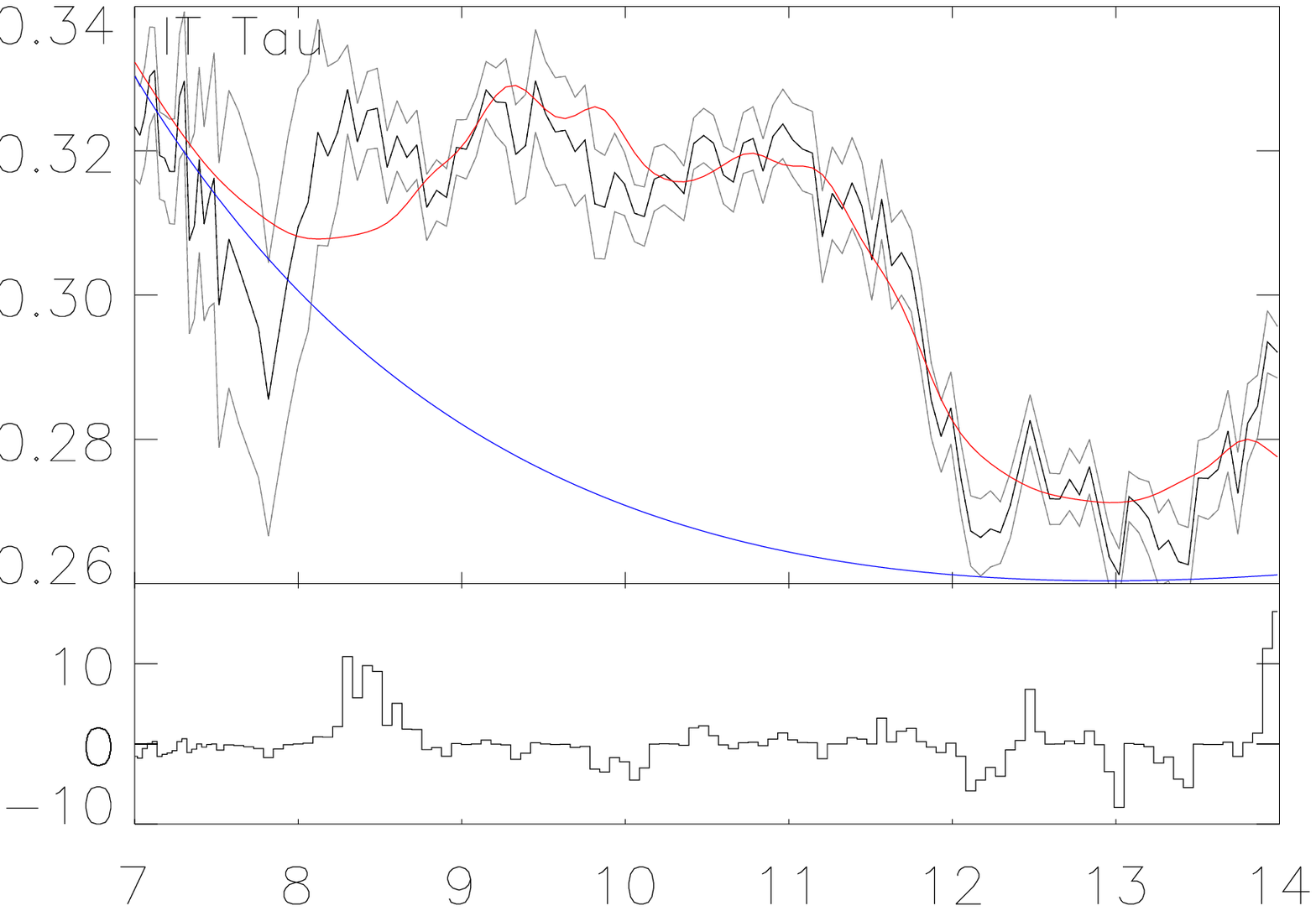}&\includegraphics[width=4.2cm]{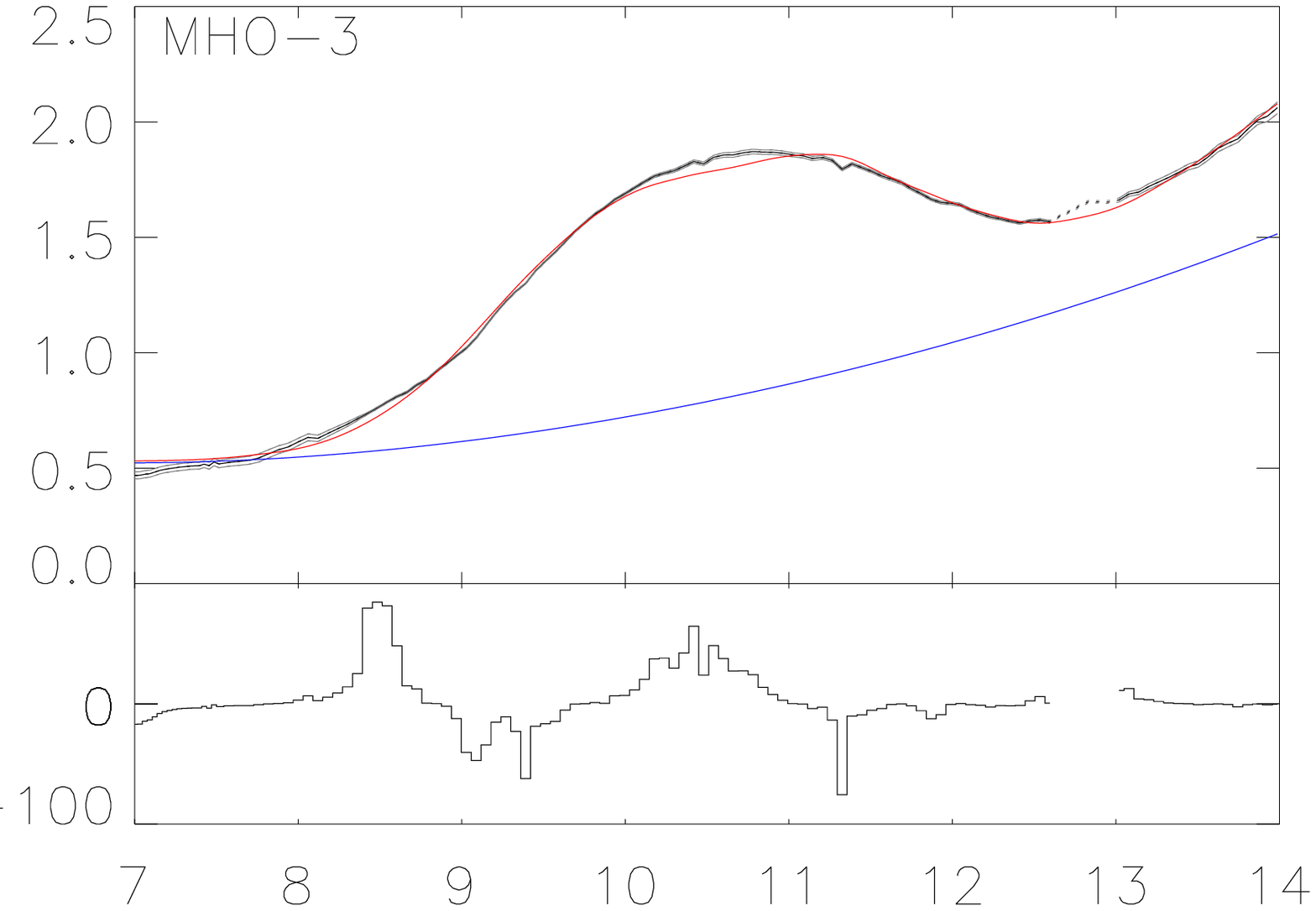}&
\includegraphics[width=4.2cm]{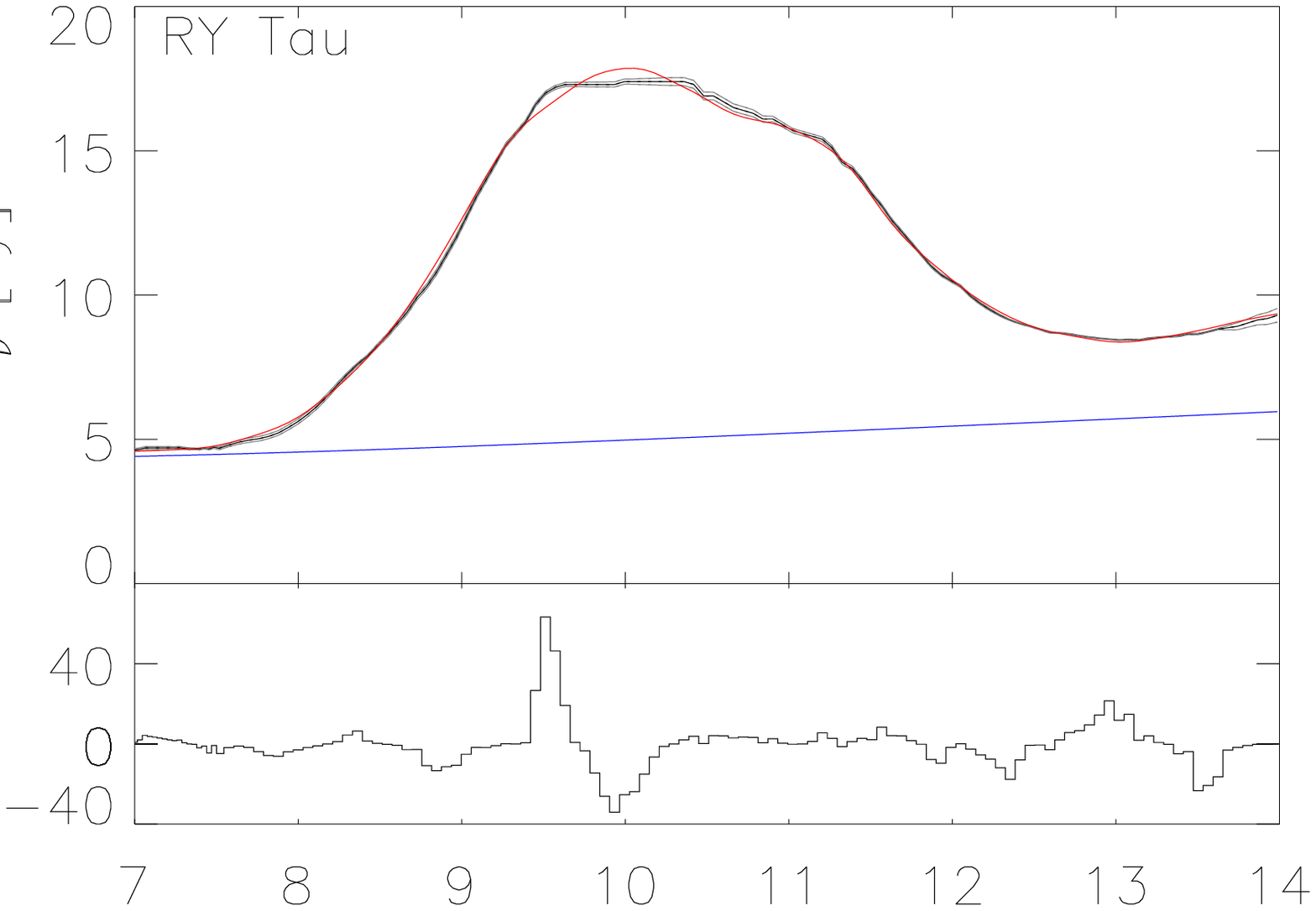}&\includegraphics[width=4.2cm]{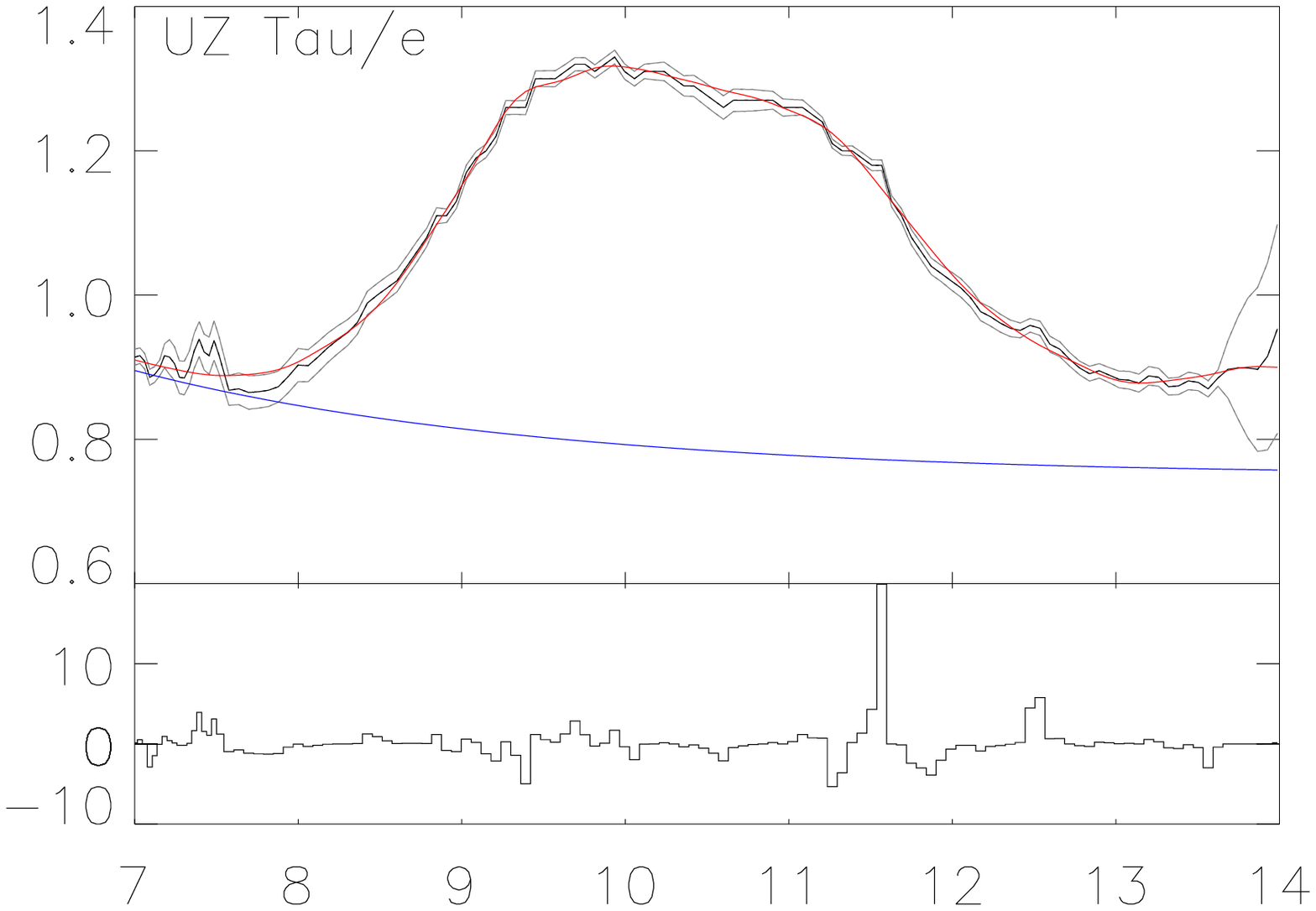}\\
\\
\includegraphics[width=4.2cm]{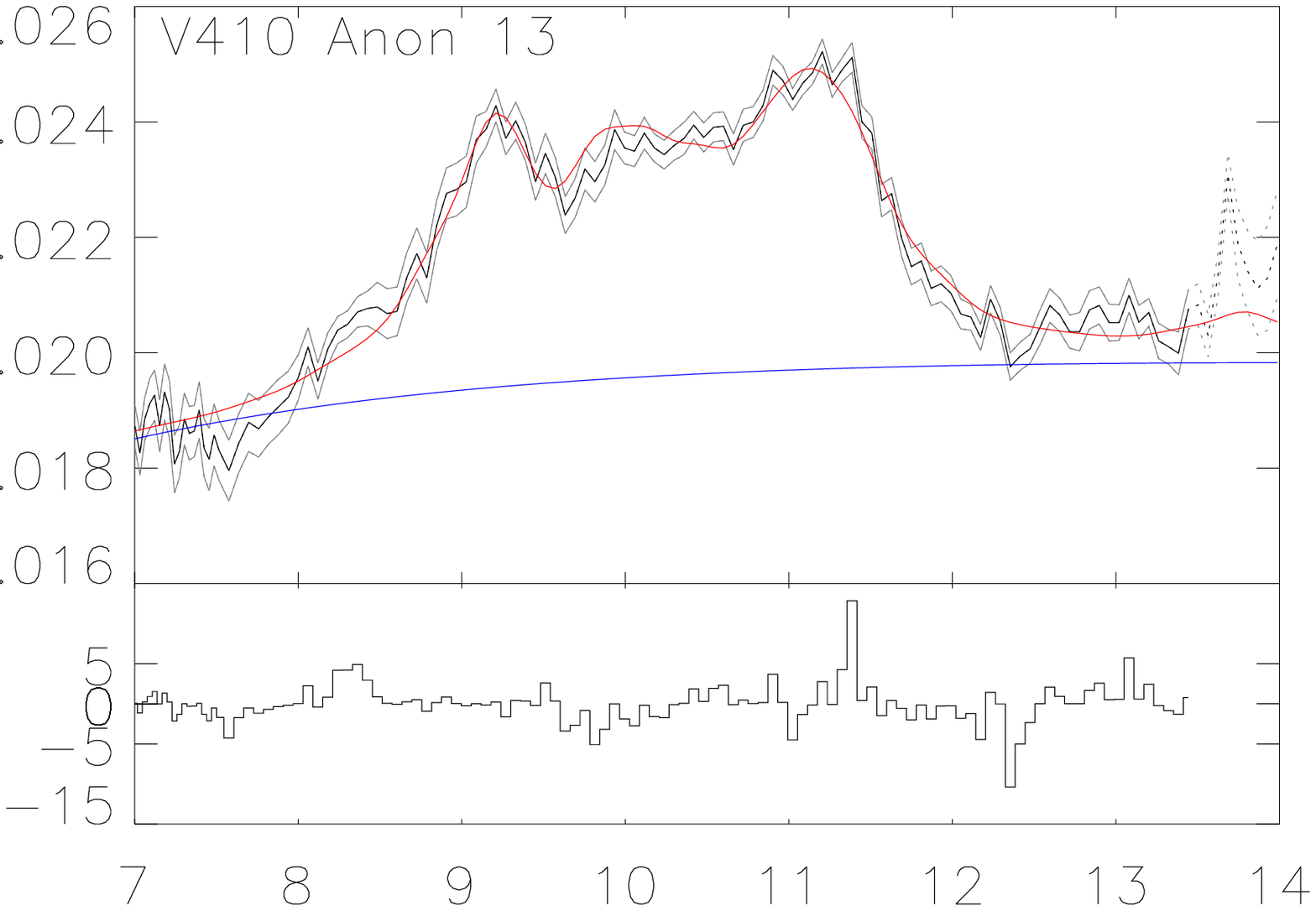}&\includegraphics[width=4.2cm]{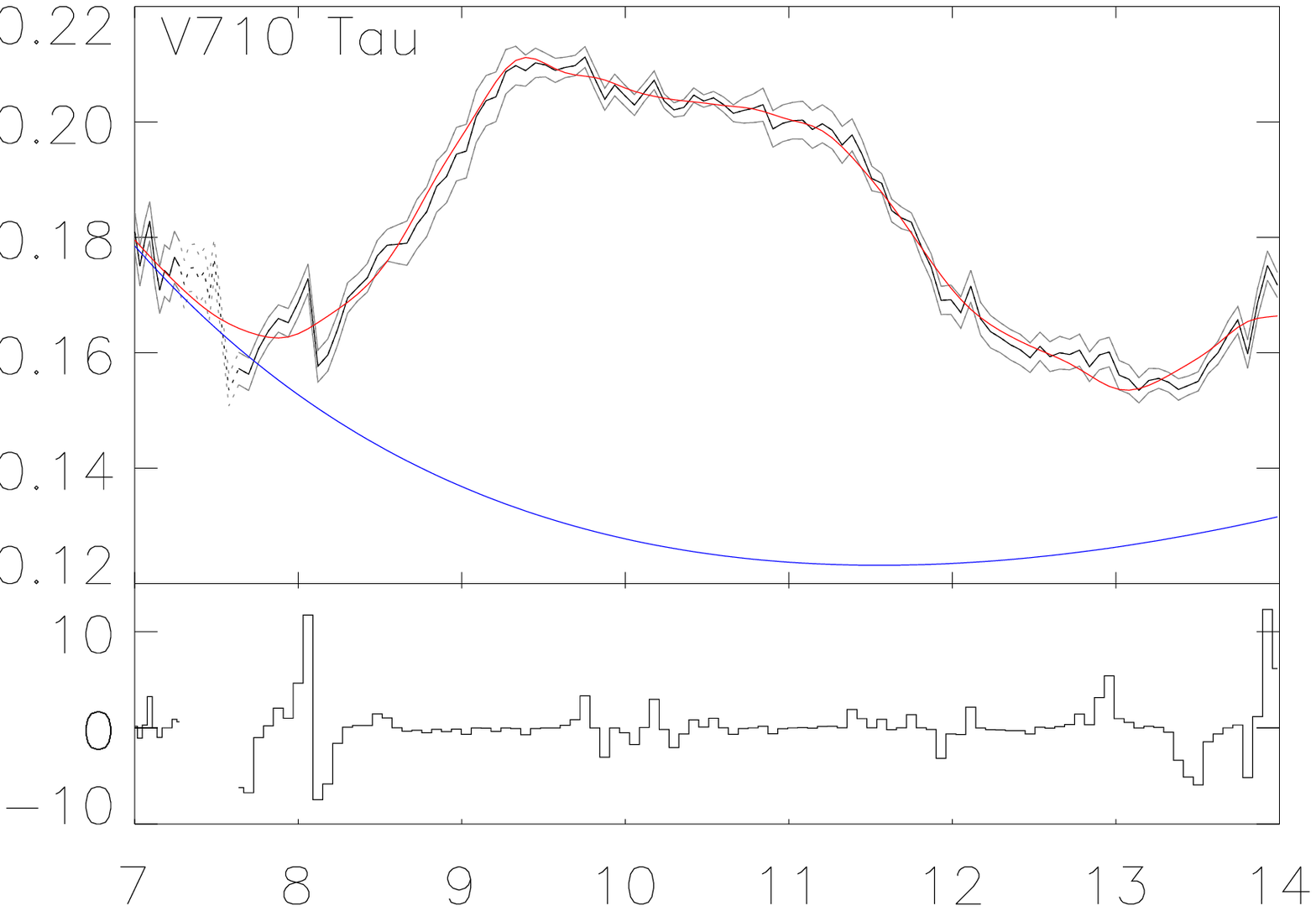}&
\includegraphics[width=4.2cm]{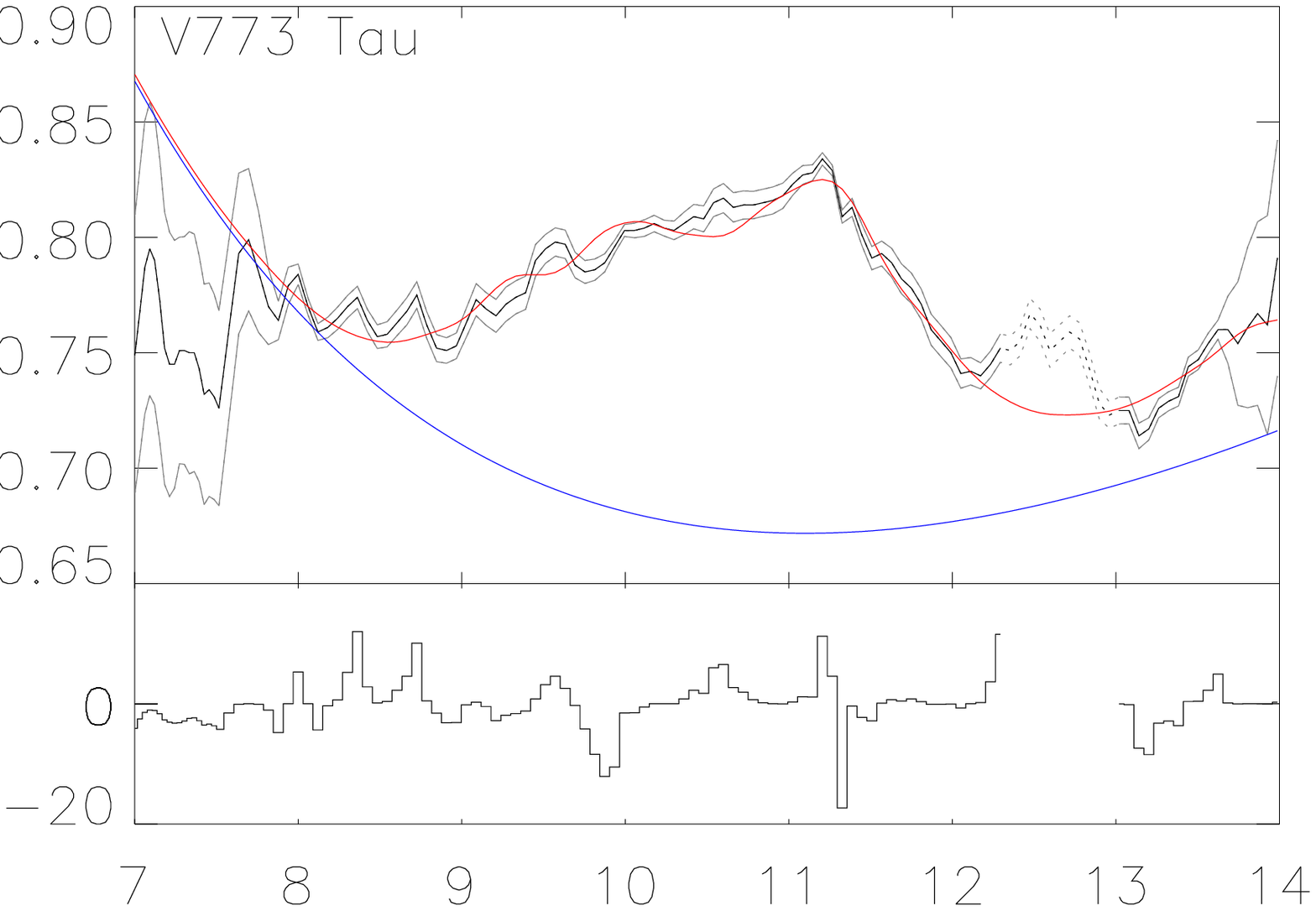}&\includegraphics[width=4.2cm]{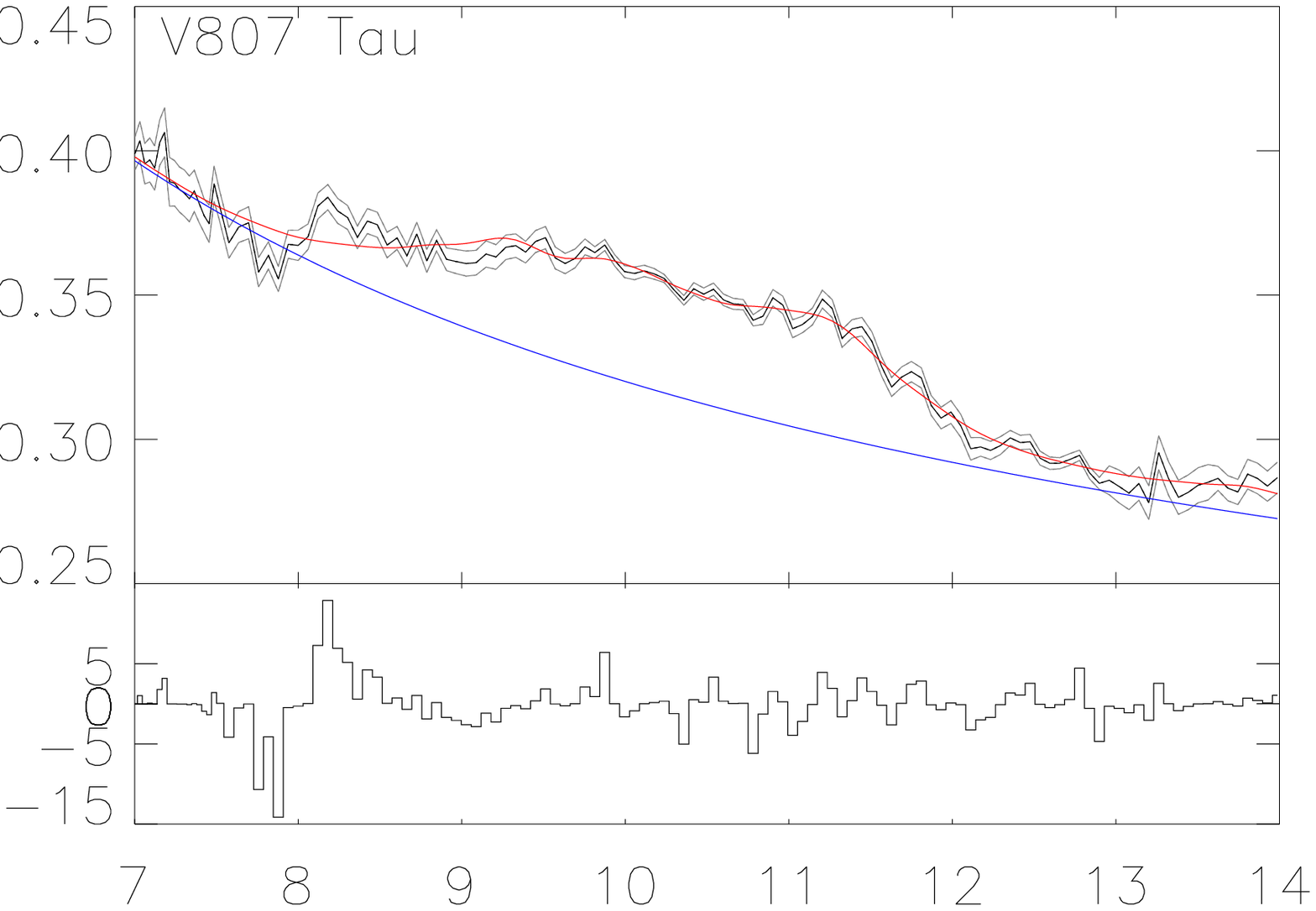}\\
\\
\includegraphics[width=4.2cm]{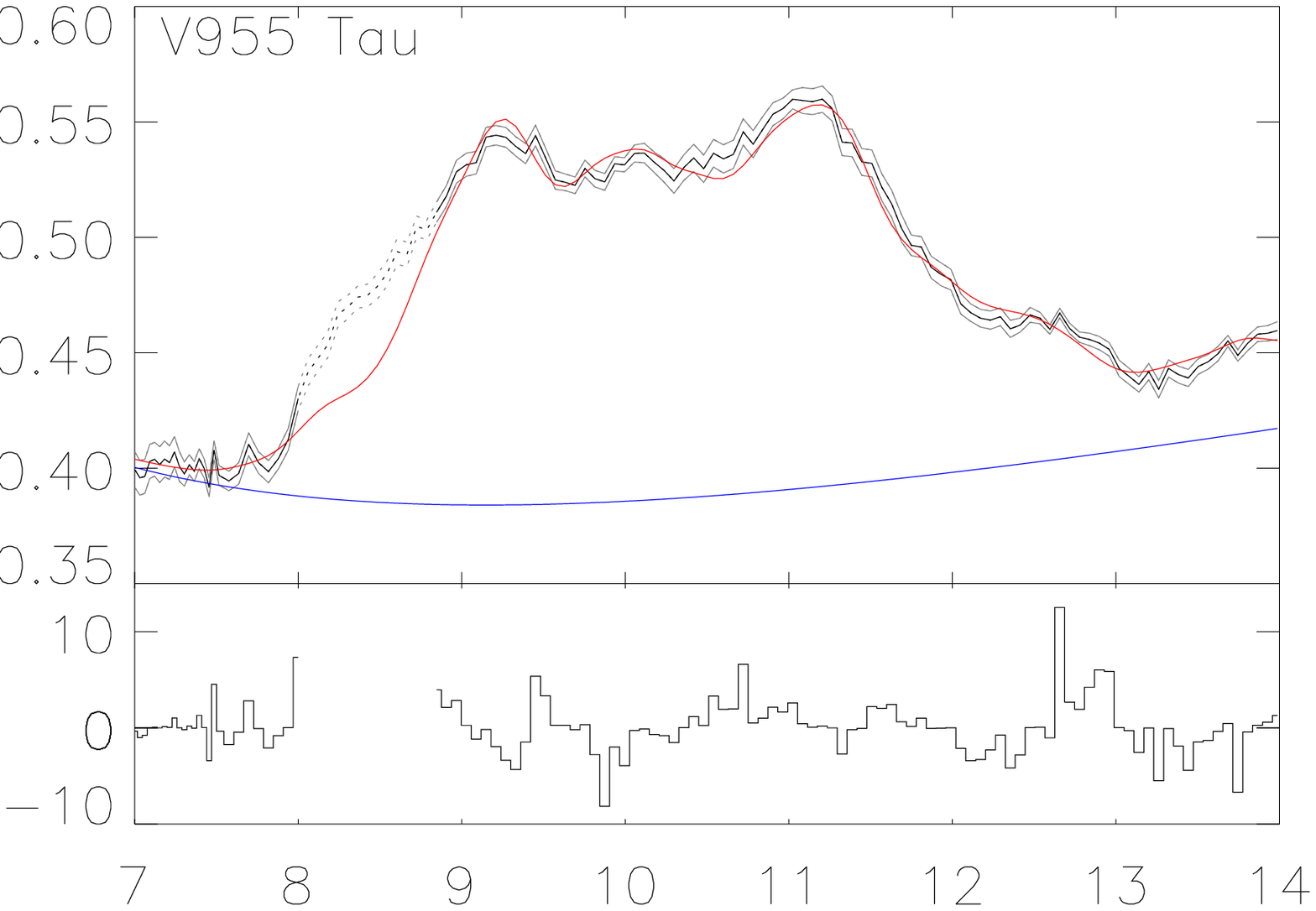}&\includegraphics[width=4.2cm]{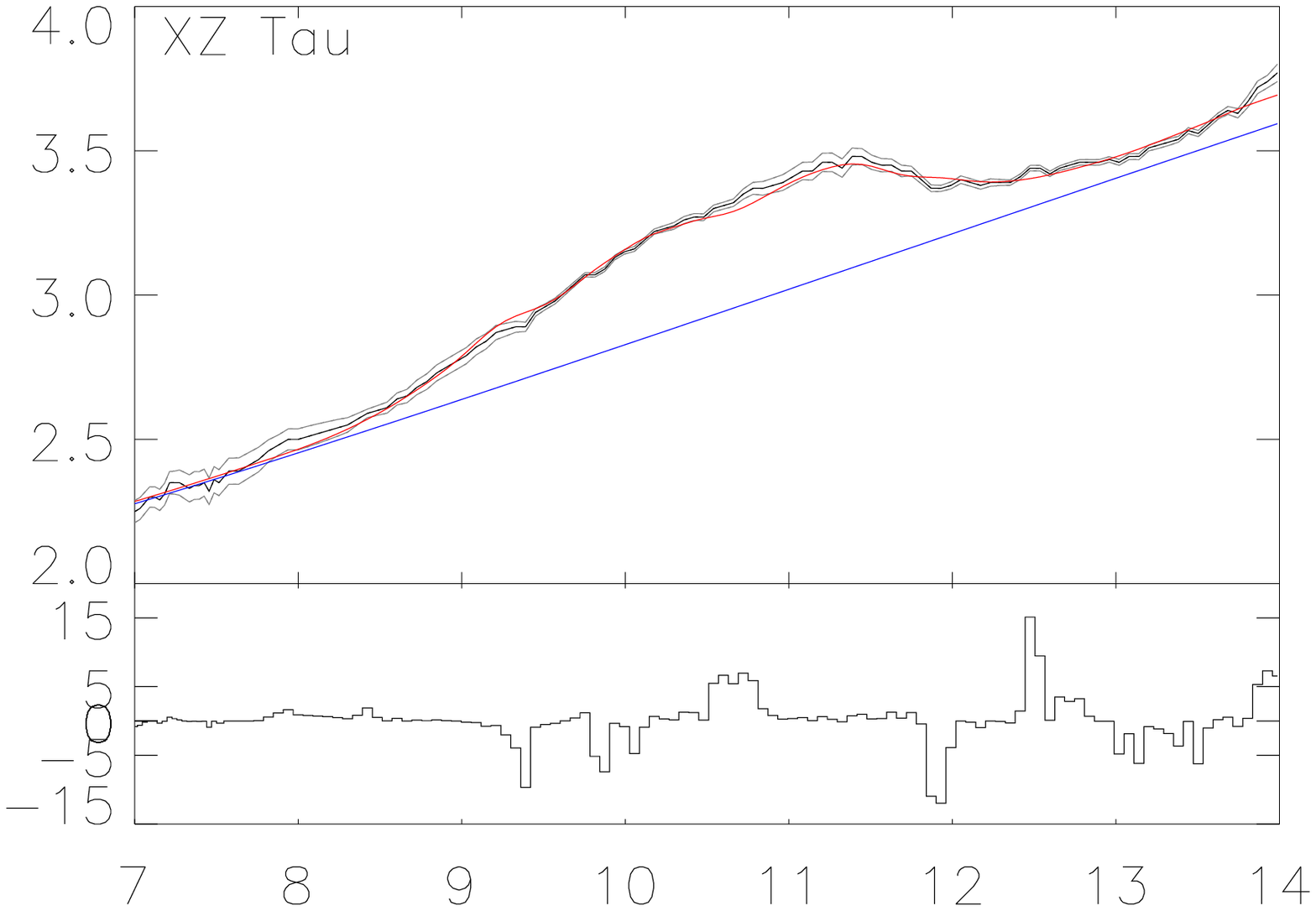}\\
\end{longtable}

\end{appendix}

\end{document}